\documentclass[%
 reprint, 
superscriptaddress,
 amsmath,amssymb,
 aps, physrev,
floatfix,
]{revtex4-2}

\usepackage{graphicx}
\usepackage{dcolumn}
\usepackage{bm}
\usepackage{braket}
\usepackage{xcolor}
\usepackage[colorlinks=true,citecolor=blue,linkcolor=blue,urlcolor=blue]{hyperref}

\begin{document}

\preprint{APS/123-QED}

\title{\textbf{Spectroscopic factors as a probe of nuclear shape in $^{44}$S via one-neutron knockout reaction} 
}%

\author{Ranojit Barman}
 \email{Contact author: ranojit@ribf.riken.jp}
\affiliation{Department of Physics, Indian Institute of Technology Roorkee, Roorkee 247 667, India}
\affiliation{RIKEN Nishina Center, Wako, Saitama 351-0198, Japan}

\author{Masaaki Kimura}%
\affiliation{RIKEN Nishina Center, Wako, Saitama 351-0198, Japan}
 
\author{Yoshiki Chazono}
\affiliation{Department of Physics, Kyushu University, Fukuoka 819-0395, Japan}

\author{Kazuki Yoshida}
\affiliation{Research Center for Nuclear Physics (RCNP), The University of Osaka, Ibaraki 567-0047, Japan}
\affiliation{Interdisciplinary Theoretical and Mathematical Sciences Program (iTHEMS), RIKEN, Wako 351-0198, Japan}

\author{Kazuyuki Ogata}
\affiliation{Department of Physics, Kyushu University, Fukuoka 819-0395, Japan}

\author{Rajdeep Chatterjee}%
\affiliation{Department of Physics, Indian Institute of Technology Roorkee, Roorkee 247 667, India}

\date{\today}

\begin{abstract}
\noindent
\textbf{Background:}
Neutron-rich nucleus $^{44}$S lies in the region where traditional $N=28$ shell closure weakens, leading to the emergence of shape coexistence and large-amplitude collective motion (LACM). Understanding the nature and degree of shape mixing in this nucleus remains an important and fascinating problem.\\
\textbf{Purpose:}
We investigate the manifestation of shape fluctuations in $^{44}$S and examine how the electric transitions and the spectroscopic factors from one-neutron knockout reactions can serve as probes of shapes mixing.\\
\textbf{Method:}
The antisymmetrized molecular dynamics combined with the generator coordinate method (AMD+GCM) is used to study the structure of $^{44}$S and $^{43}$S. Calculations are performed by using Gogny effective interactions with two different parameter sets, D1S and D1M, to explore the interaction dependence of shape mixing. Monopole and quadrupole transition strengths and spectroscopic factors are evaluated. The cross sections for the $^{44}$S$(p,pn)^{43}$S reaction are calculated within the distorted wave impulse approximation (DWIA).\\
\textbf{Results:} 
The calculations reveal a strong interaction dependence of shape fluctuation in $^{44}$S. The structural differences obtained from D1S and D1M interactions produce distinct patterns of the electric transitions, the spectroscopic factors, and the cross sections for $^{44}$S$(p,pn)^{43}$S knockout reaction.\\
\textbf{Conclusion}
The population of $3/2^-$ and $7/2^-$ states of $^{43}$S is particularly sensitive to the underlying shape fluctuation in $^{44}$S. Thus, the measurement of $^{44}$S$(p,pn)^{43}$S reaction can provide a direct experimental probe.
\end{abstract}

\maketitle


\section{INTRODUCTION}
 
The evolution of nuclear shell structure far from stability often leads to the weakening of traditional magic numbers, and has become a major theme in nuclear structure studies~\cite{thibault_PhysRevC.12.644, CAMPI1975193, detraz_PhysRevC.19.164, ioi_PhysRevC.41.1147, MOTOBAYASHI19959, IWASAKI2001227, YANAGISAWA200384, Rodriguez_2002, Otsuka_2005, SORLIN2008602}. One prominent example is the breakdown of $N=28$ shell closure in neutron-rich nuclei, which results in the emergence of diverse nuclear shapes~\cite{GLASMACHER1997163, Gaudefroy_2006, Bastin_2007, Sorlin_2013}. This evolution is driven by enhanced quadrupole correlations induced by the reduction of the neutron shell gap between the $0f_{7/2}$ and $1p_{3/2}$ orbitals, a gap originally formed through spin-orbit splitting~\cite{Goeppert_1949, Haxel_1949}. As a result, a progressive onset of deformation and various shapes is found as one moves away from the doubly magic $^{48}$Ca. 

In this region, the structure of $^{44}$S has been widely studied. Experimental studies identified many non-yrast states, establishing the coexistence and mixing of different shapes in this nucleus~\cite{GLASMACHER1997163,Caceres_PhysRevC.85.024311,Force_2010, Riley_2025}. For example, measurements at GANIL identified a low-lying isomeric $0_2^+$ state at 1365 keV, whose $E0$ and $E2$ branches to the $0_1^+$ and $2_1^+$ states suggest a prolate-spherical shape coexistence~\cite{Force_2010}. Furthermore, a two-neutron knockout experiment from $^{46}$Ar revealed additional excited states with a strongly deformed $4^+$ state, proposing a triple configuration coexistence in $^{44}$S~\cite{Santiago-Gonzalez_PhysRevC.83.061305}. On the theoretical side, several studies predicted a pronounced shape mixing in its ground and $0_2^+$ state~\cite{Peru2000, Li_PhysRevC.84.054304,Chevrier_2014,Rodriguez_PhysRevC.84.051307}. The study performed within a symmetry-conserving configuration-mixing framework showed large quadrupole fluctuations~\cite{Rodriguez_PhysRevC.84.051307}. Similar to this picture, antisymmetrized molecular dynamics (AMD) calculations combined with the generator coordinate method (GCM) revealed that the energy surface of $^{44}$S is remarkably soft in the triaxial degree of freedom, leading to dynamical shape fluctuation characterized by large amplitude collective motion (LACM)~\cite{suzuki_kimura_2021, suzuki_2022}. Therefore, the detailed nature and degree of shape mixing in the ground state of $^{44}$S remain to be unambiguously established through further experimental investigations.

Several observables can probe complex shape dynamics in nuclei, including the energies of low-lying excited states, electromagnetic transition strengths, and quadrupole moments~\cite{Heyde_1988, WOOD1992101, Heyde_2011, Otsuka_2020, GARRETT2022103931}. Notably, electric monopole and quadrupole transitions provide important signatures of shape coexistence and configuration mixing~\cite{Heyde_1988, Heyde_2011}. Large $E0$ transition strengths and enhanced $B(E2)$ for interband transitions signal mixing between coexisting shapes. In addition to these quantities, spectroscopic factors for nucleon removal or knockout reactions also provide insight into nuclear shapes. They are determined by the overlap between the wave functions of the target and residual nuclei and represent the probability of removing a nucleon from a specific single-particle orbit. Thus, they are sensitive to nuclear shapes and can serve as a useful experimental probe.

With these in mind, we employ AMD with GCM to investigate the emergence of LACM in the ground state of $^{44}$S and explore its signature in experimentally accessible observables. In particular, we analyze electric monopole and quadrupole transition strengths, which are sensitive to shape fluctuations and configuration mixing among low-lying $0^+$ and $2^+$ states. As a complementary approach, we study spectroscopic factors associated with one neutron knockout reactions to examine how LACM is reflected in spectroscopic factors. We further calculate cross sections for the proton-induced one-neutron knockout reaction within the distorted wave impulse approximation (DWIA)~\cite{NSChant77, NSChant83, TWakasa17, KOgata24} to investigate how these structural features can be probed via reaction observables. As the structures in such nuclei are also often found to be strongly dependent on the choice of effective interactions~\cite{Barman_PhysRevC.111.064305}, we perform the calculations using two different Gogny interactions, D1S and D1M. We will see that these interactions lead to different manifestations of LACM in $^{44}$S, reflecting different patterns of shape mixing.

In the following section, we briefly outline the theoretical framework, including AMD with GCM and the reaction formalism based on DWIA. In Sec.~\ref{results}, we present the structural properties of $^{44}$S obtained with the two Gogny interactions and discuss its shape fluctuation through monopole and quadrupole transition strengths. We then investigate the low-lying states of $^{43}$S relevant to the $^{44}\mathrm{S}(p,pn)^{43}\mathrm{S}$ reaction and analyze the corresponding spectroscopic factors as probes of the shape of $^{44}$S. Finally, we examine signatures of shape mixing through reaction observables for $^{44}\mathrm{S}(p,pn)^{43}\mathrm{S}$ reaction, and summarize the results in Sec.~\ref{summary}.

\section{FORMALISM}
\subsection{Framework of AMD+GCM}\label{sec:AMD}
 At first, the microscopic Hamiltonian is defined as,
 \begin{align}
   \hat{H} = \sum_{i=1}^{A} \hat{t}_i - \hat{T}_{\rm cm} + \sum_{i<j}^{A} \hat{v}^{nn}_{ij} + \sum_{i<j}^{A} \hat{v}^{\rm C}_{ij}.
 \end{align}
 Here, $\hat{t}_i$ denotes the single-particle kinetic energy operator, $\hat{T}_{\rm cm}$ is the kinetic energy of the center-of-mass motion, and $\hat{v}^{nn}_{ij}$ is the effective two-body interaction. In this work, the Gogny interaction with D1S~\cite{BERGER1991365} and D1M~\cite{Goriely_2009} parameterizations have been used. The Coulomb interaction $\hat{v}^{\rm C}_{ij}$ is approximated by a sum of seven Gaussians. The intrinsic AMD wave function is expressed as a Slater determinant of single-particle wave packets,
  \begin{align}
   \Phi_{\rm int} &= \frac{1}{\sqrt{A!}}\det\{\varphi_1, \varphi_2,..., \varphi_A\},
  \end{align}
where each single-particle wave packet is written as,
\begin{equation}
    \varphi_i(\bm r) = \phi_i(\bm r)\otimes \chi_i \otimes \tau_i.
\end{equation}
Here, $\phi_i(\bm r)$, $\chi_i$, and $\tau_i$ denote the spatial, spin, and isospin parts of the wave function, respectively. The spatial wave function is represented by a deformed Gaussian,
 \begin{align}\label{eq:spwp}
 \phi_i(\bm r)&=\prod_{\sigma=x,y,z}\left(\frac{2\nu_\sigma}{\pi}\right)^{1/4}\exp\left\{-\nu_{\sigma}\left(r_{\sigma}-\frac{Z_{i\sigma}}{\sqrt{\nu_{\sigma}}}\right)^2\right\}, 
 \end{align}
 where $Z_{i\sigma}$ are the complex-valued Gaussian centroids, and $\nu_{\sigma}$ are the width parameters. The spin function is given by,
 \begin{align}
 \chi_i &= a_i \chi_{\uparrow} + b_i \chi_{\downarrow},\hspace{2mm} \lvert a_i\rvert^2+\lvert b_i \rvert^2 = 1,
 \end{align}
 where, $a_i$ and $b_i$ are the spin direction parameters. The isospin wave function $\tau_i$ specifies whether a nucleon is a proton or a neutron. Because the intrinsic wave function does not have a definite parity, it is projected onto eigenstate of parity;  $\Phi^{\pi}=\hat{P}^{\pi}\Phi_{\rm int}$, where $\hat{P}^{\pi}= (1\pm \hat{P}_r)/2$ is the parity projection operator. The set of the variational parameters $\mathbf{Z}=\{Z_{i\sigma}, \nu_{\sigma}, a_i, b_i\}$ is determined by energy variation to minimize the energy of the system with constraints on the matter quadrupole deformation parameters, $\beta$ and $\gamma$. 
The optimized wave function $\Phi^{\pi}(\beta, \gamma)$ is obtained as a result of energy variation, which has a minimum energy for a given ($\beta$, $\gamma$).

 Then, the wave function $\Phi^{\pi}(\beta, \gamma)$ is projected to the eigenstate of the angular momentum,
 \begin{align}
 \Phi_{MK}^{J\pi}(\beta, \gamma) = \int d\Omega D^{J*}_{MK}(\Omega) \hat{R}(\Omega)\Phi^{\pi}(\beta, \gamma), 
 \end{align}
where $\Omega$ is the Euler angles. $D^{J*}_{MK}(\Omega)$ and $\hat{R}(\Omega)$ are Wigner's D-matrix and the rotation operator, respectively.
 
Finally, we construct the GCM wave function by superposing the projected wave functions employing $(\beta,\gamma)$ as generator coordinates,
 \begin{align}\label{eq:GCM}
 \Psi_{Mn}^{J\pi} = \sum_{i}\sum_{K}g^n_{iK} \Phi_{MK}^{J\pi}(\beta_i, \gamma_i),
 \end{align}
where, the quantum numbers other than $J$, $\pi$ and $M$ are denoted by $n$. The coefficients $g^n_{iK}$ and the eigen-energies are determined by solving Hill-Wheeler equation \cite{hw1953}. The electric transition probabilities are calculated from the GCM wave functions.
 
To investigate the nuclear shape associated with each state, the GCM overlap is evaluated as 
\begin{equation}\label{eq:ovlp}
     O^{J\pi}_{\alpha n}(\beta_i, \gamma_i) = \lvert \langle \widetilde{\Phi}^{J\pi}_{M} (\beta_i, \gamma_i)\lvert\Psi^{J\pi}_{Mn} \rangle\lvert^2,
 \end{equation}
 where 
 \begin{equation}\label{eq:14}
     \widetilde{\Phi}^{J\pi}_{M}(\beta_i, \gamma_i)  = \sum_K f_{iK}\Phi^{J\pi}_{MK}(\beta_i, \gamma_i),
 \end{equation}
 represents the $K$-mixed state which is the superposition of the wave functions with the same deformation parameters $\beta_i$ and $\gamma_i$ but different $K$ quantum numbers. The coefficient $f_{iK}$ is determined to minimize the energy. The GCM overlap quantifies the contribution of each intrinsic configuration to the corresponding wave function obtained by GCM, revealing the dominant deformation and nuclear shape.

 The overlap amplitudes for the neutron knockout reaction are calculated by taking the overlap between the wave functions for $^{43}$S and $^{44}$S,
\begin{align}\label{eq:ovlpf}
\mathcal{I}^{J\pi}_{M}(\bm r) &= C^{00}_{JMJ-M}I_{JL}(r)\mathcal{Y}_{JLM}(\hat r)\nonumber\\
 &= \sqrt{44}\langle\Psi^{J\pi}_{-Mn}( ^{43}\text{S})\lvert\Psi^{0+}(^{44}\text{S}) \rangle,
\end{align}
where $\Psi^{0+}(^{44}\text{S})$  and $\Psi^{J\pi}_{-Mn}(^{43}\text{S})$ denote the ground state of $^{44}{\rm S}$ and the ground and excited states of $^{43}{\rm S}$, respectively.
$\mathcal{Y}_{JLM}(\hat r)$ is the spinor spherical harmonic. 
Since AMD employs the Gaussian single-particle wave packets, $I_{JL}(r)$ do not always reproduce the correct asymptotic behavior.
Therefore, $I_{JL}(r)$ is smoothly connected to the exact asymptotic form at $r=a$,
\begin{align}
I_{JL}(a) &= A k_L(\kappa a),\\
\left.\frac{dI_{JL}(r)}{dr} \right|_{r=a} &= A\left.\frac{dk_L(\kappa r)}{dr}\right|_{r=a},
\end{align}
where $k_L(\kappa r)$ is the modified spherical Bessel function of second kind, $\kappa=\sqrt{2\mu S_n/\hbar^2}$ with $\mu$ and $S_n$ being the reduced mass and one-neutron separation energy. 
The asymptotic normalization constant $A$ and the matching radius $a$ are determined from the above equations.

The spectroscopic factor is the integral of $\mathcal{I}^{J\pi}_M(\bm r)$.
\begin{align}
    C^2S = \int d^3r\  \lvert \mathcal{I}^{J\pi}_{M}(\bm r)\rvert^2
    = \int dr \ r^2 \lvert I_{JL}(r)\rvert^2.
\end{align}

\subsection{Distorted Wave Impulse Approximation}\label{sec:DWIA}
In this section, we briefly describe the proton-induced neutron knockout ($p,pn$) reaction in inverse kinematics within the DWIA framework.
We label the incident proton, scattered proton, and knocked-out neutron as particles $0$, $1$, and $2$, respectively.
We refer to the reaction residue as nucleus $\mathrm{B}$.
The total energy and asymptotic momentum (in units of $\hbar$) of particle/nucleus $i \ (i = 0, 1, 2, \ \textrm{or} \ \mathrm{B})$ are denoted by $E_i$ and $\bm{K}_i$, respectively.
Quantities with superscript A are evaluated in the target-rest frame, whereas those without are evaluated in the center-of-mass (c.m.) frame of the reaction system.
The $z$-axis is set along the beam direction.

According to Ref.~\cite{KOgata24}, the momentum distribution of nucleus $\mathrm{B}$ is given by
\begin{align}
\frac{d^3 \sigma}{d \bm{K}_\mathrm{B}^\textrm{A}}
&=
\frac{(2 \pi)^4}{\hbar v_\alpha} \frac{1}{2 L + 1}
\int d E_1^\textrm{A} \int_0^{2 \pi} d \bar{\phi}_1^\textrm{A} \nonumber\\
&\quad \times
\mathcal{J}_\textrm{AG}
\frac{E_1^\textrm{A} E_2^\textrm{A}}{(\hbar c)^4 |\bm{Z}^\textrm{A}|}
\left( \frac{2 \pi \hbar}{\mathcal{M}_{pn}} \right)^2
\frac{d \sigma_{pn}}{d \Omega_{pn}}
\sum_M |\bar{T}_M|^2,
\label{eq:MomDist}
\end{align}
where $\bm{Z}^\textrm{A} \equiv \bm{K}_0^\textrm{A} - \bm{K}_\mathrm{B}^\textrm{A}$ and $\bar{\phi}_1^\textrm{A}$ is the azimuthal angle of $\bm{K}_1^\textrm{A}$ with respect to $\bm{Z}^\textrm{A}$.
$v_\alpha$ is the relative speed between particle $0$ and the target nucleus.
$\mathcal{J}_\textrm{AG}$ is the Jacobian for the transformation from the c.m. frame of the reaction system to the target-rest frame.
$\mathcal{M}_{pn}$ is the reduced energy of the $p$-$n$ system in the c.m.\ frame of the colliding two nucleons, and $d \sigma_{pn} / d \Omega_{pn}$ is the $p$-$n$ elastic cross section in free space.
The reduced transition matrix $\bar{T}_M$ is defined by
\begin{align}
\bar{T}_M
&=
\int d \bm{r}
\chi_{1, \bm{K}_1}^{(-) *} (\bm{r})
\chi_{2, \bm{K}_2}^{(-) *} (\bm{r})
\chi_{0, \bm{K}_0}^{(+) *} (\bm{r}) \nonumber\\
&\quad \times
e^{- i \bm{K}_0 \cdot \bm{r} / A}
I_{J L} (r) Y_{L, -M} (\hat{r}).
\label{eq:RedTmat}
\end{align}
Here, $\chi_{i, \bm{K}_i}^{(\pm)}$ is the distorted wave of particle $i$, which satisfies the outgoing $(+)$ and incoming $(-)$ boundary conditions.
$A$ is the mass number of the target.

In the cylindrical representation, $\bm{K}_\mathrm{B}^\textrm{A}$ is written as a linear combination of the $z$-axis component $K_{\mathrm{B} z}^\textrm{A}$ and its perpendicular one $K_{\mathrm{B} b}^\textrm{A}$.
Using Eq.~\eqref{eq:MomDist}, we obtain the longitudinal momentum distribution (LGMD) as
\begin{align}
\frac{d \sigma}{d K_{\mathrm{B} z}^\textrm{A}} =
\int_0^{2 \pi} d \phi_\mathrm{B}^\textrm{A}
\int_0^\infty K_{\mathrm{B} b}^\textrm{A} d K_{\mathrm{B} b}^\textrm{A}
\frac{d^3 \sigma}{d \bm{K}_\mathrm{B}^\textrm{A}},
\label{eq:LGMD}
\end{align}
where $\phi_\mathrm{B}^\textrm{A}$ is the azimuthal angle of $\bm{K}_\mathrm{B}^\textrm{A}$.
Integrating the LGMD with respect to $K_{\mathrm{B} z}^\textrm{A}$, $\sigma^\textrm{AMD}$ in Table~\ref{tab:D1S_D1M} is obtained.

\section{Results and discussion}\label{results}
\subsection{Large shape fluctuation of $^{44}$S and its interaction dependence}

Large-amplitude collective motion (LACM), or shape fluctuation, in the low-lying states of ${}^{44}{\rm S}$ has been pointed out in previous studies~\cite{suzuki_kimura_2021, suzuki_2022}. 
Here, we revisit this feature and investigate how it depends on the choice of effective interaction. 
For this purpose, we analyze the low-lying $0^+$ and $2^+$ states of ${}^{44}{\rm S}$ using two different Gogny interactions.

\begin{figure}[htbp]
    \centering
    \includegraphics[width=\linewidth]{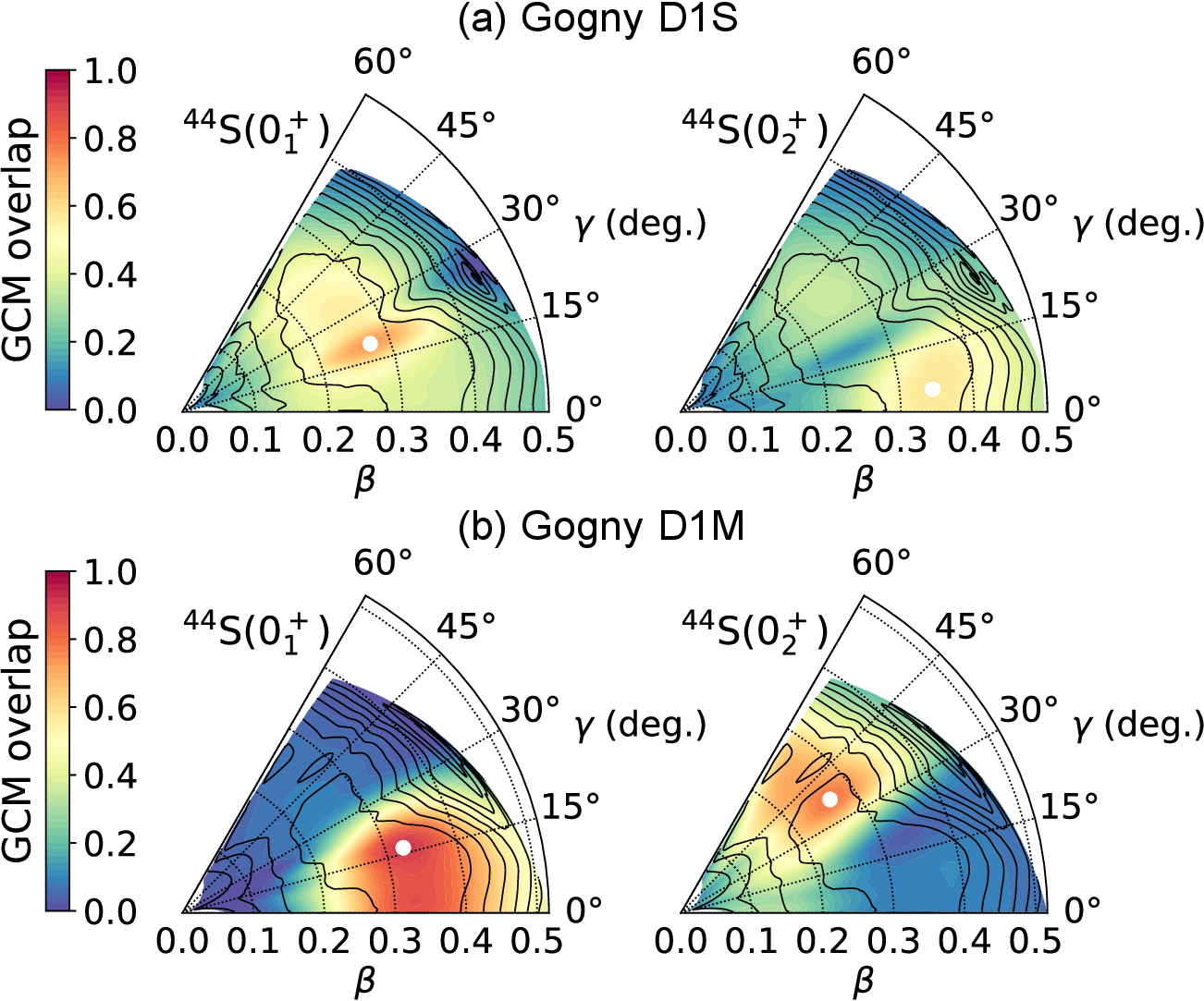}
    \caption{The energy surfaces and GCM overlaps for the $0^+$ states of $^{44}$S obtained with (a) the Gogny D1S and (b) the Gogny D1M parameter sets. The contour lines represent the energies of the $0^+$ states at intervals of 1 MeV relative to their respective minima. The color plots show the values of the GCM overlap defined in Eq.~\eqref{eq:ovlp}. The filled white circles indicate the positions of the maximum GCM overlap.}
    \label{fig:44S.J0+_ovlp}
\end{figure}

Figure~\ref{fig:44S.J0+_ovlp} shows the energy surfaces and GCM overlaps of the $0^+$ states of ${}^{44}{\rm S}$ calculated with the Gogny D1S and D1M parameter sets. The results obtained with the D1S interaction exhibit pronounced LACM, whereas those with D1M show localized configurations.

The contour lines in Fig.~\ref{fig:44S.J0+_ovlp}~(a) show the energy surface projected to $J^\pi=0^+$ obtained from D1S. 
It is almost flat along the $\gamma$ deformation. 
In particular, around $\beta \sim 0.3$, the energy varies by less than about $1.2$ MeV over a wide range of $\gamma$. Consistent with this feature, as already pointed out in previous studies, the GCM overlaps of the $0^+_1$ and $0^+_2$ states exhibit a broad distribution extending along the $\gamma$ direction at $\beta \sim 0.3$. Thus, the $0^+$ states calculated by D1S parameter set exhibit large shape fluctuations, particularly in $\gamma$ deformation.

In contrast, for the Gogny D1M interaction, we found that the $0^+$ states do not exhibit sizable shape fluctuations, but are characterized by relatively rigid intrinsic shapes. 
As shown in Fig.~\ref{fig:44S.J0+_ovlp}~(b), the GCM overlaps of the $0^+_1$ and $0^+_2$ states are localized in the prolate ($\gamma < 30^\circ$) and oblate ($\gamma > 30^\circ$) regions, with maximum overlaps of $86\%$ and $80\%$, respectively. 
Although the energy surfaces obtained with the D1S and D1M interactions differ only modestly, the resulting GCM overlaps are qualitatively different. This suggests that the manifestation of LACM is sensitively influenced by the non-diagonal coupling between different deformations.

\begin{figure}[htbp]
    \centering
    \includegraphics[width=\linewidth]{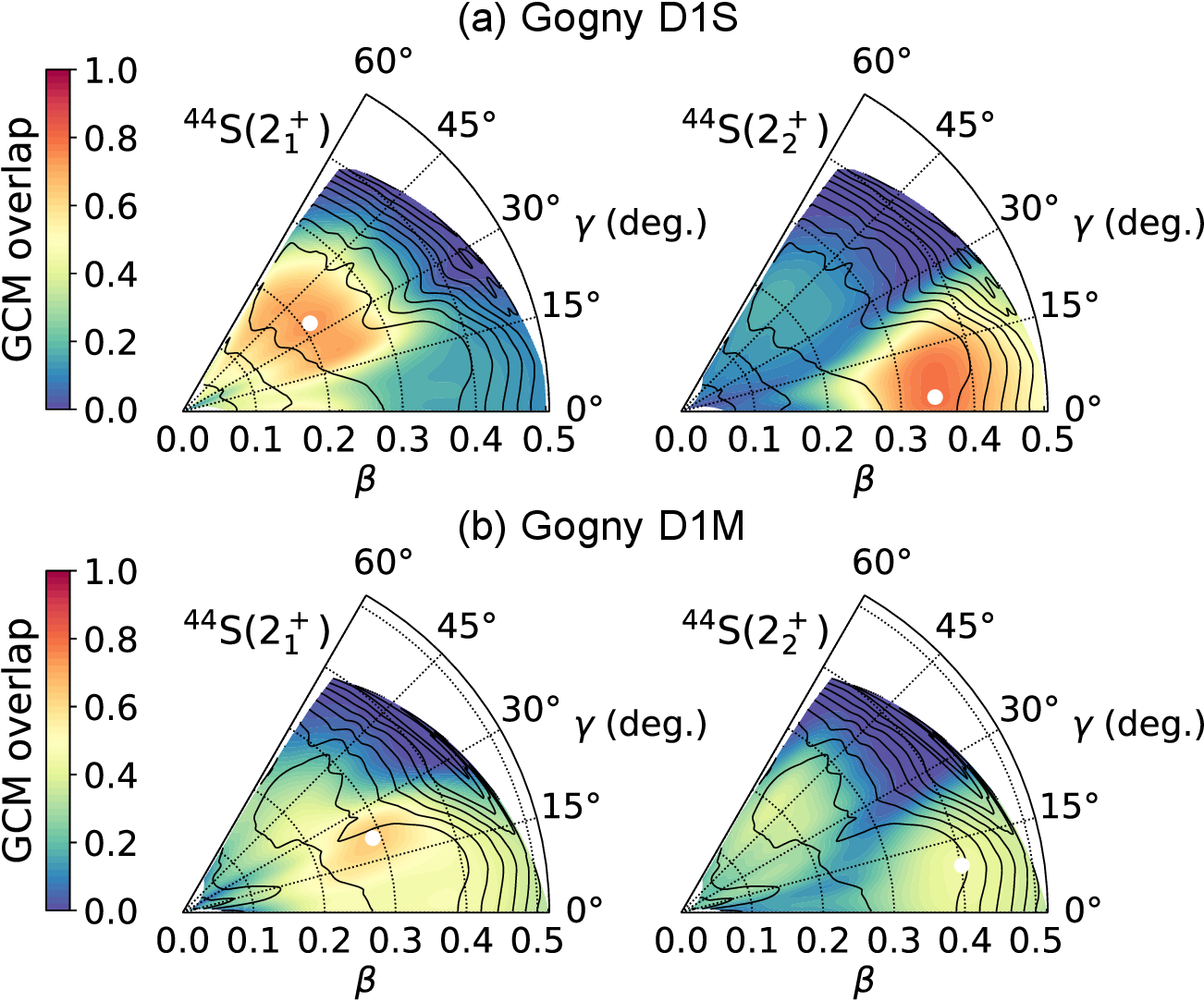}
    \caption{Same as Fig.~\ref{fig:44S.J0+_ovlp}, but for the $2^+$ states. 
    The contour lines represent the energies obtained after the  $K$ mixing, 
    shown at an interval of 1 MeV with respect to their minimum energies.}
    \label{fig:44S_J4+.ovlp-D1S+D1M}
\end{figure}
\begin{figure}[htbp]
    \centering   \includegraphics[width=\linewidth]{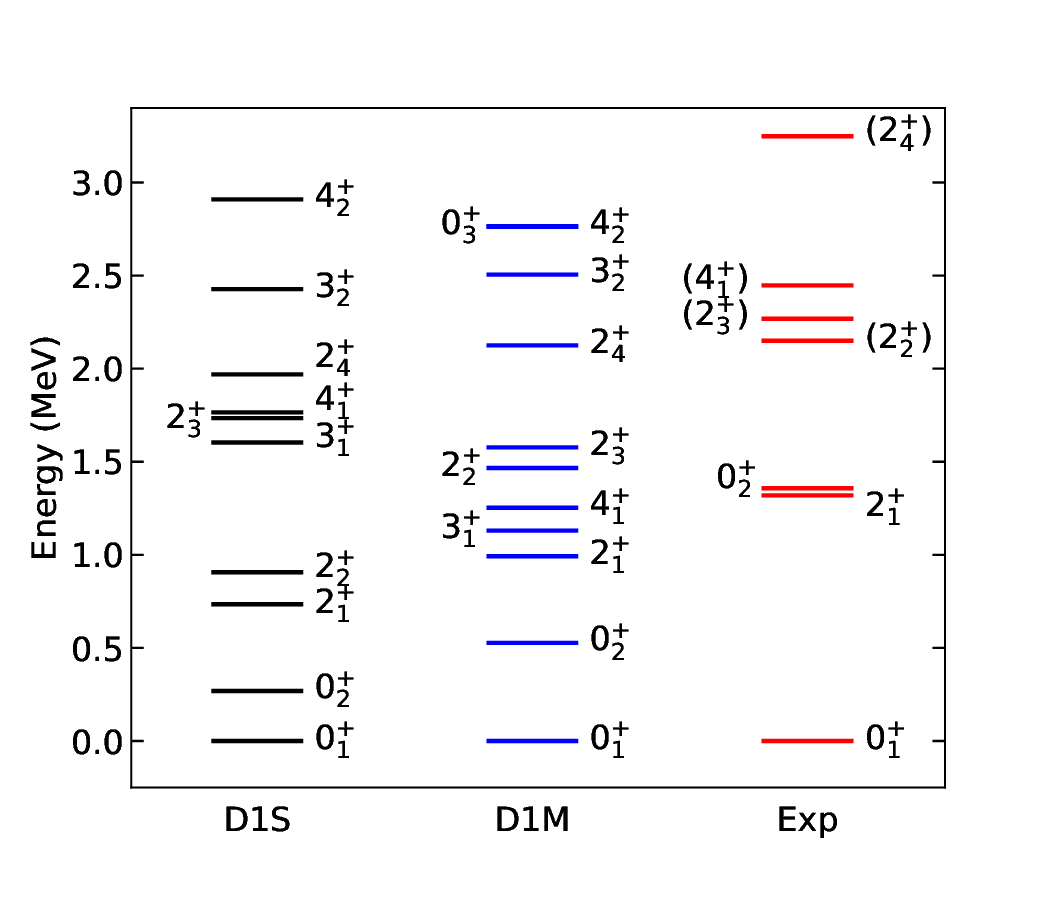}
    \caption{Excitation spectra of $^{44}$S. Experimental data is taken from~\cite{Santiago-Gonzalez_PhysRevC.83.061305}.}
    \label{fig:ex_44S}
\end{figure}

In contrast to the $0^+$ states, the $2^+_1$ and $2^+_2$ states exhibit an opposite pattern.
While the $2^+$ states obtained with the Gogny D1S interaction are characterized by relatively rigid intrinsic shapes, those from the D1M interaction show pronounced shape fluctuations. 
This is clearly visible in the GCM overlaps shown in Fig.~\ref{fig:44S_J4+.ovlp-D1S+D1M}. 
For the D1S interaction, the overlaps are localized in either prolate or oblate deformations, whereas for D1M they have broad distributions.
As in the $0^+$ case, the diagonal energy surfaces show no qualitative difference between the two interactions, again indicating a strong dependence on the non-diagonal coupling.

Figure~\ref{fig:ex_44S} compares the excitation spectra obtained with the two parameter sets.
Both calculations yield multiple low-lying $2^+$ states as observed experimentally; however, the excitation energy of the $0^+_2$ state is  underestimated in both D1S and D1M calculations, resulting in a reduced $0^+_2$ -- $0^+_1$ energy spacing compared to experiment. 
This deficiency, which has already been pointed out in previous studies, is likely related to an insufficient treatment of pairing correlations in the present framework.

In summary, we found that the realization of LACM in $^{44}$S is strongly dependent on the effective interaction. 
In the following sections, we show that such contrasting theoretical predictions can be tested experimentally through observables such as electromagnetic transitions and spectroscopic factors.

\subsection{Signature of large-amplitude collective motion in electromagnetic transitions}
The differences in the shape fluctuation described by the Gogny D1S and D1M interactions are reflected in electromagnetic transition strengths. 
Table~\ref{tab:BE2} summarizes the monopole and quadrupole transition strengths calculated for 
$^{44}$S, together with available experimental data.

\begin{table}[htbp]
    \centering
    \begin{ruledtabular}
    \begin{tabular}{lccc}
     Observable & D1S & D1M & Exp. \\
     \hline
     $\rho^2(E0, 0_1^+ \rightarrow 0_2^+)$ & $13.3\times10^{-3}$ & $4.4\times10^{-3}$ & $8.7(7)\times10^{-3}$\textsuperscript{a}\\
     $\rho^2(E0, 2_1^+ \rightarrow 2_2^+)$ & $3.2\times10^{-3}$ & $10.7\times10^{-3}$ &-\\
     $B(E2, 0_1^+ \rightarrow 2_1^+)$ & 318 & 405 & $314(88)$\textsuperscript{b}\\
     &&&$221(28)$\textsuperscript{c} \\
     $B(E2, 0_1^+ \rightarrow 2_2^+)$ & 105 & 155 & - \\
     $B(E2, 0_2^+ \rightarrow 2_1^+)$ & 30 & 25 & 42(13)\textsuperscript{a} \\
     $B(E2, 0_2^+ \rightarrow 2_2^+)$ & 390 & 160 & - \\
     $B(E2, 2_1^+ \rightarrow 2_2^+)$ & 1 & 30 & - \\
    \end{tabular}
    \end{ruledtabular}
    
    \caption{
    The monopole transition strengths, and quadrupole transition strengths ($\rm e^2fm^4$) for $^{44}$S. The experimental data are taken from \textsuperscript{a}Ref.~\cite{Force_2010}, \textsuperscript{b}Ref.~\cite{GLASMACHER1997163}, and \textsuperscript{c}Ref.~\cite{Longfellow2021_PRC}.
    \label{tab:BE2}
    }
\end{table}

A prominent difference between the two interactions appears in the monopole transitions connecting the $0_1^+$-- $0_2^+$ states and 
$2_1^+$-- $2_2^+$ states, as well as in the quadrupole transition between the $2_1^+$ and $2_2^+$ states. 
The D1S interaction predicts a significantly larger value of $\rho^2(E0; 0_1^+ \rightarrow 0_2^+)$ than D1M.
In addition, the transitions between the $2^+$ states are much more enhanced in D1M than in D1S, 
revealing an opposite pattern to that observed in the $0^+$ sector.
By contrast, the $E2$ transition strengths connecting the $0^+$ and $2^+$ states show only moderate differences between the two interactions.

It is well known that the enhancement of the monopole transition  is roughly explained by the two-configuration mixing model~\cite{Wood1992}. 
Here, we revisit it to provide a unified and qualitative interpretation of the above-mentioned features.
Let us approximate the $0^+$ and $2^+$ states as admixtures of prolate and oblate configurations,
\begin{align}
  \ket{0_1^{+}} &=  \cos{\theta_0}\ket{0^{+}_\mathrm{pr}}+\sin{\theta_0}\ket{0^{+}_\mathrm{ob}},\\
  \ket{0_2^{+}} &= -\sin{\theta_0}\ket{0^{+}_\mathrm{pr}}+\cos{\theta_0}\ket{0^{+}_\mathrm{ob}},\\
  \ket{2_1^{+}} &=  \cos{\theta_2}\ket{2^{+}_\mathrm{pr}}+\sin{\theta_2}\ket{2^{+}_\mathrm{ob}},\\
  \ket{2_2^{+}} &= -\sin{\theta_2}\ket{2^{+}_\mathrm{pr}}+\cos{\theta_2}\ket{2^{+}_\mathrm{ob}},
\end{align}
where $\theta_0$ and $\theta_2$ control the mixing between the prolate and oblate configurations. 
With these expressions, the monopole transition matrix element between the $0^+$ states is given by,
\begin{align}
\braket{0_2^+ |\hat{\mathcal{M}} | 0_1^+} &=\cos\theta_0 \sin\theta_0 \nonumber \\
&\times [\braket{0^+_\mathrm{ob}|\hat{\mathcal{M}}|0^+_\mathrm{ob}} 
       - \braket{0^+_\mathrm{pr}|\hat{\mathcal{M}}|0^+_\mathrm{pr}}],
\end{align}
where $\hat{\mathcal{M}}$ denotes the transition operator and $\braket{0^+_\mathrm{ob}|\hat{\mathcal{M}}|0^+_\mathrm{pr}}$ is assumed to be negligible due to the structural mismatch. From this result, we observe if there is no mixing (D1M case; $\sin\theta_0=0$), the transition matrix element vanishes, whereas it can be finite if there is significant mixing (D1S case). 
The same argument also explains the enhanced transitions between the $2^+$ states in the D1M case.

For the quadrupole transitions between the $0^+$ and $2^+$ states, as a representative case, the transition between the $0^+_1$ and $2^+_1$ states reads,
\begin{align}
\braket{2_1^+|\hat{\mathcal{M}}|0_1^+} &= \cos\theta_0 \cos\theta_2 \braket{2^+_\mathrm{pr}|\hat{\mathcal{M}}|0^+_\mathrm{pr}} \nonumber \\
&+ \sin\theta_0 \sin\theta_2 \braket{2^+_\mathrm{ob}|\hat{\mathcal{M}}|0^+_\mathrm{ob}}.
\end{align}
This shows that if there is shape mixing in either $0^+$ or $2^+$ states, the transition matrix element is always non-zero. 
Similar formulae apply for the transitions between other combinations of the $0^+$ and $2^+$ states, and hence, the transitions between $0^+$ and $2^+$ states are less sensitive to the degree of shape mixing, although their strengths still depend quantitatively on the mixing angles.

From this analysis, we conclude that monopole transitions between $0^+$ and $2^+$ states, and the quadrupole transition between $2^+$ states, provide particularly sensitive probes of shape fluctuations.
In contrast, the transitions between $0^+$ and $2^+$ states are less suited for discriminating between D1S and D1M.

\subsection{Low-lying structures of $^{43}$S relevant to $^{44}\text{S}(p,pn)^{43}\text{S}$ reaction}
As another selective probe of LACM, spectroscopic factors measured in nucleon knockout reactions provide complementary information. 
In the following, we discuss the $^{44}\mathrm{S}(p,pn)^{43}\mathrm{S}$ reaction. 
To this end, we briefly summarize the low-lying states of $^{43}$S that participate in this reaction.

\begin{figure}[htbp]
    \centering  \includegraphics[width=\linewidth]{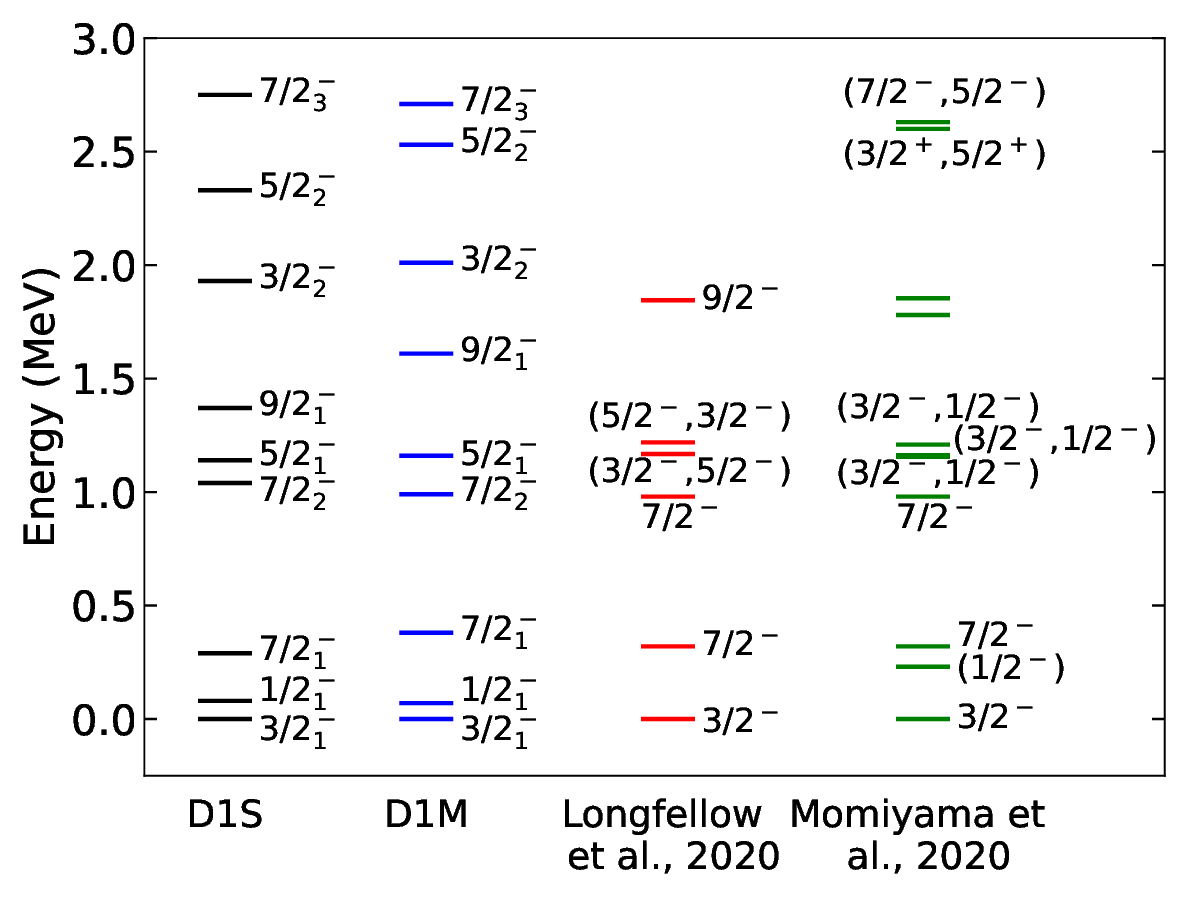}
    \caption{Excitation spectra of $^{43}$S. Experimental data are taken from Refs.~\cite{Longfellow2021_PRC, Momiyama_43S}.}
    \label{fig:spectra_43S}
\end{figure}
\begin{figure*}[htbp]
    \centering
    \includegraphics[width=\textwidth]{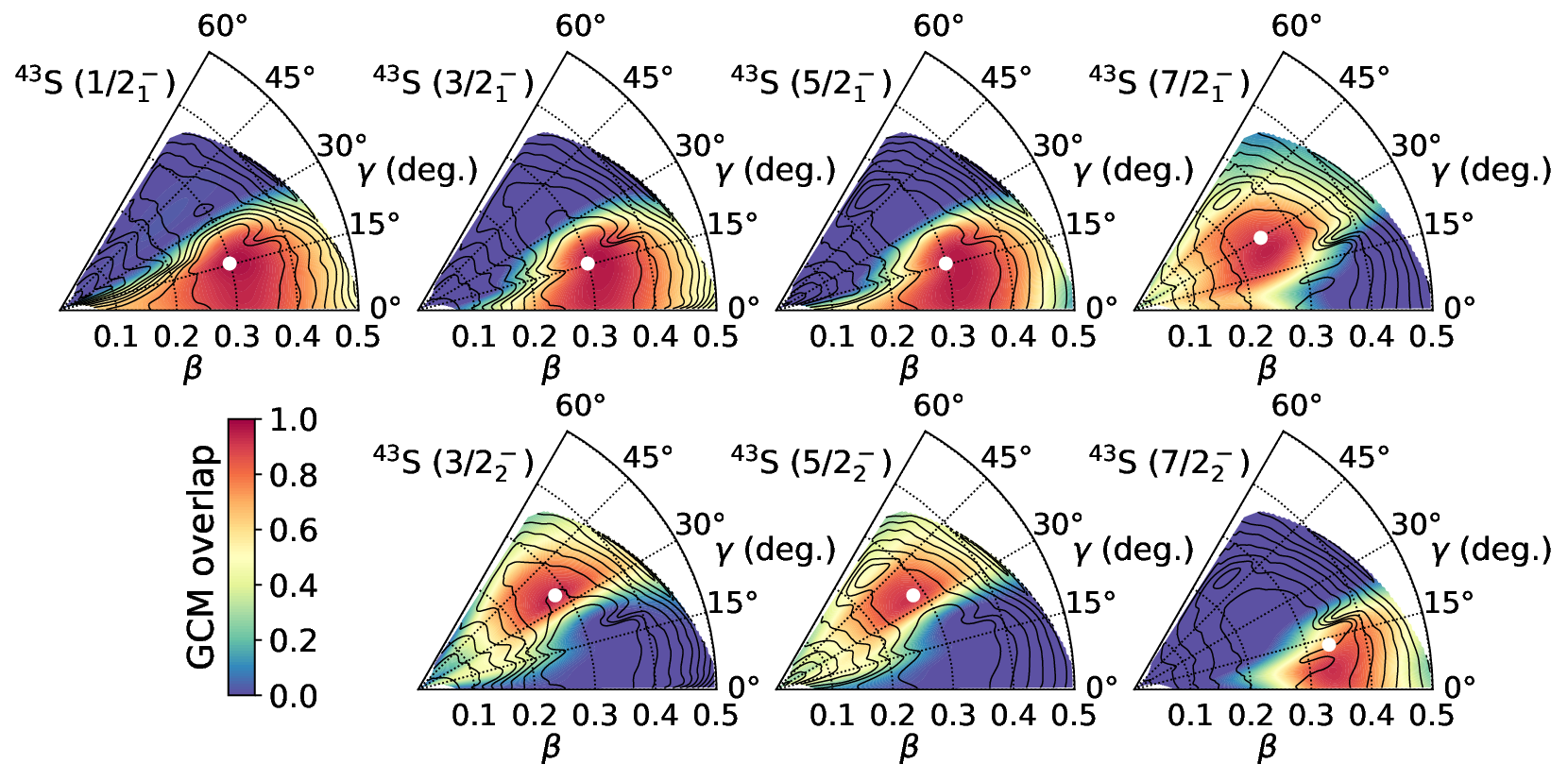}
    \caption{The energy surfaces and GCM overlaps for the $1/2^-$, $3/2^-$, $5/2^-$ and $7/2^-$ states of $^{43}$S calculated with Gogny D1S effective interaction. The contour lines represent the energies of the $K$-mixed states at an interval of 1 MeV from their minimum energies. The filled white circles represent the maximum value of the GCM overlap.}
    \label{fig:ovlp_43S}
\end{figure*}

As shown in Fig.~\ref{fig:spectra_43S}, the excitation spectra calculated with the Gogny D1S and D1M interactions are nearly identical for the ground and low-lying states.
They can be classified into three groups based on their intrinsic configurations, as illustrated in Fig.~\ref{fig:ovlp_43S}. 

The $1/2^-_1$, $3/2^-_1$, $5/2^-_1$, and $7/2^-_2$ states form a prolate rotational band built on the $K^\pi=1/2^-$ configuration. 
Their GCM overlaps (Fig.~\ref{fig:ovlp_43S}) are concentrated in the prolate region of the $\beta$ -- $\gamma$ plane, with nearly identical distributions. 
In contrast, the $3/2^-_2$ and $5/2^-_2$ states belonging to the $K^\pi=3/2^-$ band exhibit oblate intrinsic shapes, with GCM overlaps localized in the $\gamma>30^\circ$ region. 
Finally, the isomeric $7/2^-_1$ state shows a different character. 
Its GCM overlap spreads from the oblate region toward $\gamma\sim30^\circ$ and toward spherical region, indicating a significant admixture of triaxial configurations, which was discussed as a triaxially deformed state~\cite{Kimura2016}.

These characteristics of the low-lying states described above are common to both the D1S and D1M interactions, indicating that the intrinsic shapes of the low-lying states are rather rigid and only weakly dependent on the effective interaction. 
Moreover, many of these states were theoretically predicted and have been confirmed experimentally. 
This suggests $^{43}$S as a reliable reference for interpreting the spectroscopic factors discussed in the next section.

\subsection{Overlap amplitudes and spectroscopic factors for the $^{44}\text{S}(p,pn)^{43}\text{S}$ reaction}
We now investigate how the shape fluctuation in $^{44}$S can be probed through the $^{44}$S$(p,pn)^{43}$S knockout reaction. 
In nucleon-knockout reactions, the yield of a given final state is closely related to the overlap between the
initial and final states. Therefore, the strong shape mixing in $^{44}{\rm S}$ (D1S case) may lead to population
of  $^{43}$S with various nuclear shapes, whereas the rigid prolate-deformed $^{44}{\rm S}$ (D1M case) should result in more selective population.
In the following, we discuss the overlap functions and spectroscopic factors ($C^2S$) for the $^{44}\text{S}(p,pn)^{43}\text{S}$ reaction. The knockout cross sections are discussed in the next section.

\begin{figure}[htbp]
    \centering   \includegraphics[width=\linewidth]{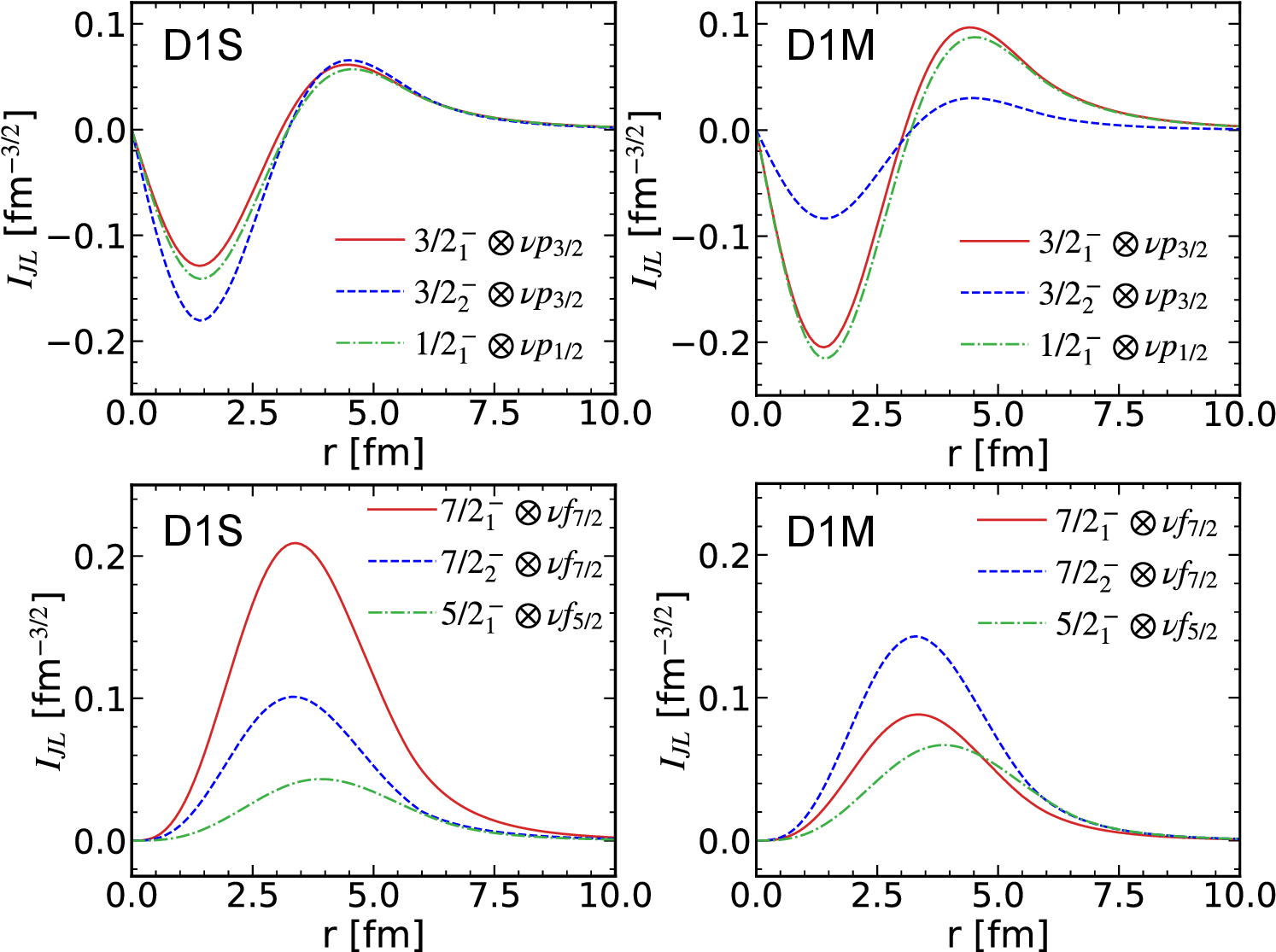}
    \caption{The overlap amplitudes corresponding to the $^{43}\mathrm{S}(J^{\pi})\otimes\nu L_J$ configurations for D1S and D1M case.}
    \label{fig:ovamp_43S-44S}
\end{figure}

We begin with the results obtained using the D1S interaction. 
As shown in Fig.~\ref{fig:ovamp_43S-44S} and Table~\ref{tab:D1S_D1M}, the overlap amplitudes and corresponding $C^2S$ values for the $1/2_1^-$, $3/2_1^-$, and $7/2_2^-$ states of $^{43}$S, which belong to the prolate band, are substantial. Although the $5/2_1^-$ state belongs to the same prolate band, the $f_{5/2}$ neutron orbit lies well above the Fermi level, leading to negligible overlap. 
In addition to the prolate band, the $3/2_2^-$ state belonging to the oblate band exhibits a sizable overlap and $C^2S$. For the same reason as for the $5/2^-_1$ state, the $5/2_2^-$ state is also negligible and is not shown in the table. 
Furthermore, the transition to the $7/2_1^-$ (triaxial) state shows the largest overlap and $C^2S$, reflecting the large GCM overlap  and the larger degeneracy factor.
These results clearly demonstrate strong mixing of prolate, oblate, and triaxial shapes in the ground state of $^{44}$S.

\renewcommand{\arraystretch}{1.2}
\begin{table*}[htbp]
\centering
\begin{ruledtabular}
\begin{tabular}{c c cccc cccc}
\multicolumn{2}{c}{} &
\multicolumn{4}{c}{Gogny D1S} &
\multicolumn{4}{c}{Gogny D1M} \\
\cline{3-6} \cline{7-10}

$^{43}$S$(J^\pi)$ & Shape
&$E_x$ (MeV)& $C^2S$ & $\sigma^{\text{AMD}}$ (mb)
& $C^2S \times \sigma_{\text{sp}}^{\text{WS}}$ (mb)
&$E_x$ (MeV)& $C^2S$ & $\sigma^{\text{AMD}}$ (mb)
& $C^2S \times \sigma_{\text{sp}}^{\text{WS}}$ (mb) \\
\hline

$1/2_1^-$ & Prolate
&0.08& 0.25 & 1.53 & 1.78
&0.07& 0.54 & 3.39 & 3.85 \\

$3/2_1^-$ & Prolate
&0.0& 0.24 & 1.60 & 1.80
&0.0& 0.57 & 3.74 & 4.27 \\

$5/2_1^-$ & Prolate
&1.14& 0.09 & 0.61 & 0.36
&1.19& 0.21 & 1.37 & 0.84 \\

$7/2_2^-$ & Prolate
&1.04& 0.32 & 1.31 & 1.58
&0.99& 0.60 & 2.38 & 2.97 \\

$3/2_2^-$ & Oblate
&1.93& 0.32 & 1.70 & 2.07
&2.01& 0.06 & 0.34 & 0.39 \\

$7/2_1^-$ & Triaxial
&0.29& 1.45 & 6.37 & 7.45
&0.38& 0.25 & 1.06 & 1.28 \\

\end{tabular}
\end{ruledtabular}
\caption{Comparison of spectroscopic factors and integrated cross sections
for $^{44}\mathrm{S}(p,pn)^{43}\mathrm{S}$ reaction calculated using Gogny D1S and D1M interactions. The intrinsic shapes of $^{43}\rm{S}$ are common to both interactions.}
\label{tab:D1S_D1M}
\end{table*}

In contrast, the D1M interaction predicts a predominantly prolate ground state of $^{44}$S. 
Consequently, the overlap amplitudes and corresponding $C^2S$ values for the prolate-band states of $^{43}$S are enhanced compared with the D1S case. 
Meanwhile, the $3/2_2^-$ (oblate) and $7/2_1^-$ (triaxial) states show suppressed $C^2S$ values because of their shape mismatch with the prolate-dominated ground state of $^{44}$S.
This demonstrates that spectroscopic factors, especially for states with identical spin-parity but different intrinsic configurations, offer a probe of LACM in this nucleus. 
Particularly, the $C^2S$ values for the $3/2_1^-$ (prolate) and $3/2_2^-$ (oblate) states, as well as the $7/2_1^-$ (triaxial) and $7/2_2^-$ (prolate) states, are sensitive to the shape mixing in $^{44}$S.

At present, no experimental data are available for the $^{44}\mathrm{S}(p,pn)^{43}\mathrm{S}$ reaction. 
However, the one-neutron knockout experiment with a $^{9}\mathrm{Be}$ target reported in Ref.~\cite{Momiyama_43S} provides $C^2S$ values for several low-lying states of $^{43}\mathrm{S}$. 
The observed significant $C^2S$ strengths among the $3/2^-$ and $7/2^-$ states are qualitatively consistent with the shape-mixing picture predicted by the D1S interaction, although a direct correspondence with the oblate $3/2^-_2$ state obtained in our AMD calculation cannot be established.
Quantitatively, however, the calculated $C^2S$ values are smaller than the experimental ones by roughly a factor of two. This motivates a direct measurement of the $^{44}\mathrm{S}(p,pn)^{43}\mathrm{S}$ reaction to clarify the role of shape mixing in $^{44}$S.

\subsection{$^{44}\text{S}(p,pn)^{43}\text{S}$ at 250~MeV/nucleon}\label{sec:44Sppn43S}


\begin{figure}[htbp]
    \centering
    \includegraphics[width=\linewidth,bb=0 0 255 226]{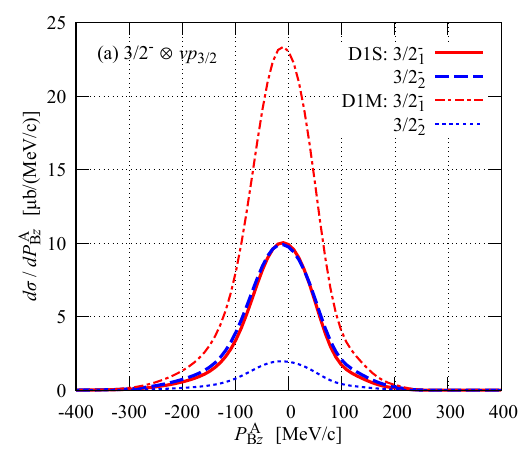}
    \includegraphics[width=\linewidth,bb=0 0 255 226]{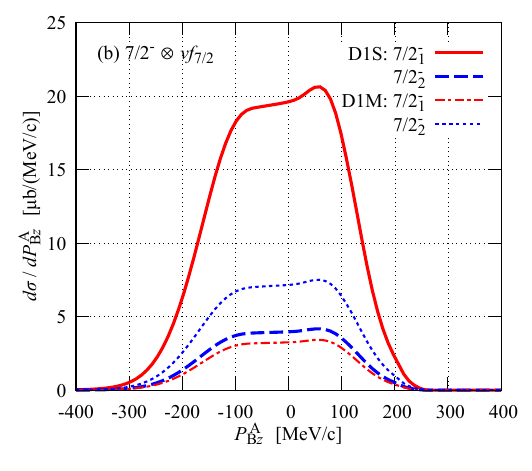}
    \caption{Longitudinal momentum distribution of the $^{44}$S($p,pn$)$^{43}$S reaction at $250$~MeV/nucleon.
    Thick (thin) lines represent the results calculated with the Gogny D1S (D1M) overlap amplitude.
    (a) The neutrons are knocked out from the $p_{3/2}$ orbits, which correspond to the $3/2_1^-$ (solid and dashed) and $3/2_2^-$ (dashed and dotted) states of $^{43}$S.
    (b) Same as (a) but the neutrons are knocked out from the $f_{7/2}$ orbits.
    \label{fig:LGMD_44Sppn43S_250AMeV}}
\end{figure}

As mentioned in the previous sections, we calculate the $^{44}$S($p,pn$)$^{43}$S reaction with the DWIA code \textsc{pikoe}~\cite{KOgata24}.
The incident energy is $250$~MeV/nucleon, which is a typical energy of the RIKEN accelerator facility.
We consider the transition from the ground state of $^{44}$S to the $3/2_1^-$, $3/2_2^-$, $7/2_1^-$, and $7/2_2^-$ states of $^{43}$S.
$d \sigma_{pn} / d \Omega_{pn}$ in Eq.~\eqref{eq:MomDist} is evaluated using the Franey-Love nucleon-nucleon effective interaction~\cite{MAFraney85}.
The optical potentials from the Dirac phenomenology with the EDAD1 parameter set~\cite{SHama90, EDCooper93, EDCooper09} are used to calculate the distorted waves in Eq.~\eqref{eq:RedTmat}.
The excitation energy of $^{43}$S is included in the neutron separation energy.

Figure~\ref{fig:LGMD_44Sppn43S_250AMeV} shows the LGMDs of the ($p,pn$) reactions from the (a) $p_{3/2}$ and (b) $f_{7/2}$ neutron orbits, as a function of $P_{\mathrm{B} z}^\textrm{A} \equiv \hbar K_{\mathrm{B} z}^\textrm{A}$.
The solid (dashed) line shows the results of the transition to the first (second) $J^\pi$ state of $^{43}$S with the Gogny D1S overlap amplitude, whereas the dot-dashed (dotted) one is the result of the D1M case.
Even if the spin-parity assignment is difficult from a comparison between an experimental $^{43}$S spectrum and theoretical ones, the $3/2^-$ and $7/2^-$ states can be distinguished by their LGMDs, as they have $l$ dependence~\cite{SChen19, YLSun20, BDLinh21, FBrowne21, MEnciu22, HWang23}.
Because of phase-volume effects associated with energy and momentum conservation, the LGMDs are suppressed on the high momentum side, $P_{\mathrm{B} z}^\textrm{A} > 0$, and become asymmetric~\cite{KOgata15}.

In Fig.~\ref{fig:LGMD_44Sppn43S_250AMeV}(a), the solid and dashed lines are almost the same, while the dot-dashed line is about $12$ times larger than the dotted one around the peak.
These features are consistent with the corresponding $C^2 S$ values and $\sigma^\textrm{AMD}$ in Table~\ref{tab:D1S_D1M}, indicating that the difference among the overlap amplitudes can be reflected quantitatively in the corresponding LGMDs.
A similar trend is seen in Fig.~\ref{fig:LGMD_44Sppn43S_250AMeV}(b).
The large magnitude of the solid line reflects that using the D1S interaction, the ground state of $^{44}$S has a large GCM overlap in the triaxial region.
On the other hand, the fact that the dotted line is larger than the dot-dashed one indicates that the ground state of $^{44}$S calculated with the D1M interaction favors the prolate shape.
We obtain qualitatively similar results using single-particle amplitudes, which are generated from the Woods-Saxon shape potentials with the geometries by Bohr and Mottelson~\cite{bohr-mottelson} and normalized to unity.
In Table~\ref{tab:D1S_D1M}, the integrated cross sections multiplied by the $C^2 S$ values are shown.

\renewcommand{\arraystretch}{1.2}
\begin{table}[htbp]
\centering
\begin{ruledtabular}
\begin{tabular}{c cc cc}
{} &
\multicolumn{2}{c}{Gogny D1S} &
\multicolumn{2}{c}{Gogny D1M} \\
\cline{2-3} \cline{4-5}
$^{43}$S($J^\pi$)
& $R (C^2 S)$ & $R (\sigma^\textrm{AMD})$     
& $R (C^2 S)$ & $R (\sigma^\textrm{AMD})$ \\  
\hline
$3/2^-$
& $1.33$ & $0.99$     
& $0.11$ & $0.09$ \\  
$7/2^-$
& $0.22$ & $0.20$     
& $2.40$ & $2.22$ \\  
\end{tabular}
\end{ruledtabular}
\caption{Ratios of $C^2 S$ and $\sigma^\textrm{AMD}$ for the $3/2_2^-$ and $7/2_2^-$ states relative to the $3/2_1^-$ and $7/2_1^-$ states.
\label{tab:Ratio}}
\end{table}

For a more quantitative discussion, we consider the ratios of the $C^2 S$ values and $\sigma^\textrm{AMD}$, defined as $R (Q) \equiv Q (J_2^\pi) / Q (J_1^\pi)$, where $Q$ represents the above two quantities.
The results are shown in Table~\ref{tab:Ratio}.
One sees that $R (\sigma^\textrm{AMD})$ are slightly smaller than $R (C^2 S)$; the difference is about $15\%$ on average and $25\%$ at most.
This reduction is mainly due to the different spatial distribution of overlap functions and absorption by the optical potentials.
For example, in the top-left panel of Fig.~\ref{fig:ovamp_43S-44S}, the dashed line has larger amplitude than the solid one in the nuclear interior, which is reflected in the $C^2 S$ value of the $3/2_2^-$ state, whereas these are almost the same around the nuclear surface.
The absorption effects suppress contributions from the former region, resulting in the $25\%$ decrease of $R (\sigma^\textrm{AMD})$ relative to $R (C^2 S)$.
Although this reduction should be noted, the choice of the Gogny interactions is crucial in a comparison between theoretical results with experimental data.

In conclusion, the comparable LGMDs of the ($p,pn$) reaction for the transition to the prolate and oblate states of the residual nucleus will suggest the shape coexistence in the target.
In the present study, if the D1S case is realized, the strong shape mixing would occur in the ground state of $^{44}$S.
On the other hand, if the D1M case is true, $^{44}$S would have a rigid prolate shape in its ground state.
\section{Summary and conclusions}\label{summary}
In summary, the present study highlights the occurrence of large-amplitude collective motion in $^{44}$S, which strongly depends on the choice of effective interaction. The Gogny D1S interaction predicts strong shape mixing in the $0^+$ states and relatively weak mixing in the $2^+$ states, whereas the D1M interaction favours more rigid shapes in the $0^+$ states but stronger mixing in the $2^+$ states.  Electric transition strengths and knockout reactions can therefore serve as suitable probes of the shape fluctuations in this nucleus. We find that the monopole and quadrupole transition strengths for the $0_1^+\rightarrow0_2^+$ and $2_1^+\rightarrow2_2^+$ transitions are sensitive to the associated shape differences between the two interactions and are effective observables to test these predictions.

The analysis of overlap amplitudes and spectroscopic factors for the $^{44}$S$(p,pn)^{43}$S reaction reflects the ground-state structure of $^{44}$S. For the D1S interaction, sizable spectroscopic factors are obtained for states associated with different intrinsic shapes of $^{43}$S, indicating strong shape mixing in the $^{44}$S ground state. On the other hand, the D1M interaction mainly populates prolate-band states in $^{43}$S, reflecting the weakly mixed, dominant prolate character of $^{44}$S. The spectroscopic factors for the $3/2^-$ and $7/2^-$ states are particularly sensitive to these structural differences, suggesting that measurements of these states could provide valuable information on the shape of $^{44}$S. The calculated longitudinal momentum distributions for these states also show clear interaction-dependent differences, demonstrating that reaction observables are sensitive to the underlying shape mixing. Therefore, experimental studies of the $^{44}$S$(p,pn)^{43}$S reaction could provide a direct test of the predicted LACM-driven shape mixing in $^{44}$S.

\begin{acknowledgments}
R.B. acknowledges RIKEN for International Program Associate (IPA) fellowship. 
The authors acknowledge Grant-in-Aid for Scientific Research (Nos.\ JP25K17393,\ JP25K17400,\ JP23K22485) from Japan Society for the Promotion of Science (JSPS).

\end{acknowledgments}



\bibliography{references}

\providecommand{\noopsort}[1]{}\providecommand{\singleletter}[1]{#1}%
\begin{thebibliography}{55}%
\makeatletter
\providecommand \@ifxundefined [1]{%
 \@ifx{#1\undefined}
}%
\providecommand \@ifnum [1]{%
 \ifnum #1\expandafter \@firstoftwo
 \else \expandafter \@secondoftwo
 \fi
}%
\providecommand \@ifx [1]{%
 \ifx #1\expandafter \@firstoftwo
 \else \expandafter \@secondoftwo
 \fi
}%
\providecommand \natexlab [1]{#1}%
\providecommand \enquote  [1]{``#1''}%
\providecommand \bibnamefont  [1]{#1}%
\providecommand \bibfnamefont [1]{#1}%
\providecommand \citenamefont [1]{#1}%
\providecommand \href@noop [0]{\@secondoftwo}%
\providecommand \href [0]{\begingroup \@sanitize@url \@href}%
\providecommand \@href[1]{\@@startlink{#1}\@@href}%
\providecommand \@@href[1]{\endgroup#1\@@endlink}%
\providecommand \@sanitize@url [0]{\catcode `\\12\catcode `\$12\catcode `\&12\catcode `\#12\catcode `\^12\catcode `\_12\catcode `\%12\relax}%
\providecommand \@@startlink[1]{}%
\providecommand \@@endlink[0]{}%
\providecommand \url  [0]{\begingroup\@sanitize@url \@url }%
\providecommand \@url [1]{\endgroup\@href {#1}{\urlprefix }}%
\providecommand \urlprefix  [0]{URL }%
\providecommand \Eprint [0]{\href }%
\providecommand \doibase [0]{https://doi.org/}%
\providecommand \selectlanguage [0]{\@gobble}%
\providecommand \bibinfo  [0]{\@secondoftwo}%
\providecommand \bibfield  [0]{\@secondoftwo}%
\providecommand \translation [1]{[#1]}%
\providecommand \BibitemOpen [0]{}%
\providecommand \bibitemStop [0]{}%
\providecommand \bibitemNoStop [0]{.\EOS\space}%
\providecommand \EOS [0]{\spacefactor3000\relax}%
\providecommand \BibitemShut  [1]{\csname bibitem#1\endcsname}%
\let\auto@bib@innerbib\@empty
\bibitem [{\citenamefont {Thibault}\ \emph {et~al.}(1975)\citenamefont {Thibault}, \citenamefont {Klapisch}, \citenamefont {Rigaud}, \citenamefont {Poskanzer}, \citenamefont {Prieels}, \citenamefont {Lessard},\ and\ \citenamefont {Reisdorf}}]{thibault_PhysRevC.12.644}%
  \BibitemOpen
  \bibfield  {author} {\bibinfo {author} {\bibfnamefont {C.}~\bibnamefont {Thibault}}, \bibinfo {author} {\bibfnamefont {R.}~\bibnamefont {Klapisch}}, \bibinfo {author} {\bibfnamefont {C.}~\bibnamefont {Rigaud}}, \bibinfo {author} {\bibfnamefont {A.~M.}\ \bibnamefont {Poskanzer}}, \bibinfo {author} {\bibfnamefont {R.}~\bibnamefont {Prieels}}, \bibinfo {author} {\bibfnamefont {L.}~\bibnamefont {Lessard}},\ and\ \bibinfo {author} {\bibfnamefont {W.}~\bibnamefont {Reisdorf}},\ }\bibfield  {title} {\bibinfo {title} {Direct measurement of the masses of $^{11}\mathrm{Li}$ and $^{26\ensuremath{-}32}\mathrm{Na}$ with an on-line mass spectrometer},\ }\href {https://doi.org/10.1103/PhysRevC.12.644} {\bibfield  {journal} {\bibinfo  {journal} {Phys. Rev. C}\ }\textbf {\bibinfo {volume} {12}},\ \bibinfo {pages} {644} (\bibinfo {year} {1975})}\BibitemShut {NoStop}%
\bibitem [{\citenamefont {Campi}\ \emph {et~al.}(1975)\citenamefont {Campi}, \citenamefont {Flocard}, \citenamefont {Kerman},\ and\ \citenamefont {Koonin}}]{CAMPI1975193}%
  \BibitemOpen
  \bibfield  {author} {\bibinfo {author} {\bibfnamefont {X.}~\bibnamefont {Campi}}, \bibinfo {author} {\bibfnamefont {H.}~\bibnamefont {Flocard}}, \bibinfo {author} {\bibfnamefont {A.}~\bibnamefont {Kerman}},\ and\ \bibinfo {author} {\bibfnamefont {S.}~\bibnamefont {Koonin}},\ }\bibfield  {title} {\bibinfo {title} {Shape transition in the neutron rich sodium isotopes},\ }\href {https://doi.org/https://doi.org/10.1016/0375-9474(75)90065-2} {\bibfield  {journal} {\bibinfo  {journal} {Nucl. Phys. A}\ }\textbf {\bibinfo {volume} {251}},\ \bibinfo {pages} {193} (\bibinfo {year} {1975})}\BibitemShut {NoStop}%
\bibitem [{\citenamefont {D\'etraz}\ \emph {et~al.}(1979)\citenamefont {D\'etraz}, \citenamefont {Guillemaud}, \citenamefont {Huber}, \citenamefont {Klapisch}, \citenamefont {Langevin}, \citenamefont {Naulin}, \citenamefont {Thibault}, \citenamefont {Carraz},\ and\ \citenamefont {Touchard}}]{detraz_PhysRevC.19.164}%
  \BibitemOpen
  \bibfield  {author} {\bibinfo {author} {\bibfnamefont {C.}~\bibnamefont {D\'etraz}}, \bibinfo {author} {\bibfnamefont {D.}~\bibnamefont {Guillemaud}}, \bibinfo {author} {\bibfnamefont {G.}~\bibnamefont {Huber}}, \bibinfo {author} {\bibfnamefont {R.}~\bibnamefont {Klapisch}}, \bibinfo {author} {\bibfnamefont {M.}~\bibnamefont {Langevin}}, \bibinfo {author} {\bibfnamefont {F.}~\bibnamefont {Naulin}}, \bibinfo {author} {\bibfnamefont {C.}~\bibnamefont {Thibault}}, \bibinfo {author} {\bibfnamefont {L.~C.}\ \bibnamefont {Carraz}},\ and\ \bibinfo {author} {\bibfnamefont {F.}~\bibnamefont {Touchard}},\ }\bibfield  {title} {\bibinfo {title} {Beta decay of $^{27\ensuremath{-}32}\mathrm{Na}$ and their descendants},\ }\href {https://doi.org/10.1103/PhysRevC.19.164} {\bibfield  {journal} {\bibinfo  {journal} {Phys. Rev. C}\ }\textbf {\bibinfo {volume} {19}},\ \bibinfo {pages} {164} (\bibinfo {year} {1979})}\BibitemShut {NoStop}%
\bibitem [{\citenamefont {Warburton}\ \emph {et~al.}(1990)\citenamefont {Warburton}, \citenamefont {Becker},\ and\ \citenamefont {Brown}}]{ioi_PhysRevC.41.1147}%
  \BibitemOpen
  \bibfield  {author} {\bibinfo {author} {\bibfnamefont {E.~K.}\ \bibnamefont {Warburton}}, \bibinfo {author} {\bibfnamefont {J.~A.}\ \bibnamefont {Becker}},\ and\ \bibinfo {author} {\bibfnamefont {B.~A.}\ \bibnamefont {Brown}},\ }\bibfield  {title} {\bibinfo {title} {Mass systematics for a=29--44 nuclei: The deformed a\ensuremath{\sim}32 region},\ }\href {https://doi.org/10.1103/PhysRevC.41.1147} {\bibfield  {journal} {\bibinfo  {journal} {Phys. Rev. C}\ }\textbf {\bibinfo {volume} {41}},\ \bibinfo {pages} {1147} (\bibinfo {year} {1990})}\BibitemShut {NoStop}%
\bibitem [{\citenamefont {Motobayashi}\ \emph {et~al.}(1995)\citenamefont {Motobayashi}, \citenamefont {Ikeda}, \citenamefont {Ieki}, \citenamefont {Inoue}, \citenamefont {Iwasa}, \citenamefont {Kikuchi}, \citenamefont {Kurokawa}, \citenamefont {Moriya}, \citenamefont {Ogawa}, \citenamefont {Murakami}, \citenamefont {Shimoura}, \citenamefont {Yanagisawa}, \citenamefont {Nakamura}, \citenamefont {Watanabe}, \citenamefont {Ishihara}, \citenamefont {Teranishi}, \citenamefont {Okuno},\ and\ \citenamefont {Casten}}]{MOTOBAYASHI19959}%
  \BibitemOpen
  \bibfield  {author} {\bibinfo {author} {\bibfnamefont {T.}~\bibnamefont {Motobayashi}}, \bibinfo {author} {\bibfnamefont {Y.}~\bibnamefont {Ikeda}}, \bibinfo {author} {\bibfnamefont {K.}~\bibnamefont {Ieki}}, \bibinfo {author} {\bibfnamefont {M.}~\bibnamefont {Inoue}}, \bibinfo {author} {\bibfnamefont {N.}~\bibnamefont {Iwasa}}, \bibinfo {author} {\bibfnamefont {T.}~\bibnamefont {Kikuchi}}, \bibinfo {author} {\bibfnamefont {M.}~\bibnamefont {Kurokawa}}, \bibinfo {author} {\bibfnamefont {S.}~\bibnamefont {Moriya}}, \bibinfo {author} {\bibfnamefont {S.}~\bibnamefont {Ogawa}}, \bibinfo {author} {\bibfnamefont {H.}~\bibnamefont {Murakami}}, \bibinfo {author} {\bibfnamefont {S.}~\bibnamefont {Shimoura}}, \bibinfo {author} {\bibfnamefont {Y.}~\bibnamefont {Yanagisawa}}, \bibinfo {author} {\bibfnamefont {T.}~\bibnamefont {Nakamura}}, \bibinfo {author} {\bibfnamefont {Y.}~\bibnamefont {Watanabe}}, \bibinfo {author} {\bibfnamefont {M.}~\bibnamefont {Ishihara}}, \bibinfo {author} {\bibfnamefont {T.}~\bibnamefont
  {Teranishi}}, \bibinfo {author} {\bibfnamefont {H.}~\bibnamefont {Okuno}},\ and\ \bibinfo {author} {\bibfnamefont {R.}~\bibnamefont {Casten}},\ }\bibfield  {title} {\bibinfo {title} {Large deformation of the very neutron-rich nucleus 32mg from intermediate-energy coulomb excitation},\ }\href {https://doi.org/https://doi.org/10.1016/0370-2693(95)00012-A} {\bibfield  {journal} {\bibinfo  {journal} {Phys. Lett. B}\ }\textbf {\bibinfo {volume} {346}},\ \bibinfo {pages} {9} (\bibinfo {year} {1995})}\BibitemShut {NoStop}%
\bibitem [{\citenamefont {Iwasaki}\ \emph {et~al.}(2001)\citenamefont {Iwasaki}, \citenamefont {Motobayashi}, \citenamefont {Sakurai}, \citenamefont {Yoneda}, \citenamefont {Gomi}, \citenamefont {Aoi}, \citenamefont {Fukuda}, \citenamefont {Fülöp}, \citenamefont {Futakami}, \citenamefont {Gacsi}, \citenamefont {Higurashi}, \citenamefont {Imai}, \citenamefont {Iwasa}, \citenamefont {Kubo}, \citenamefont {Kunibu}, \citenamefont {Kurokawa}, \citenamefont {Liu}, \citenamefont {Minemura}, \citenamefont {Saito}, \citenamefont {Serata}, \citenamefont {Shimoura}, \citenamefont {Takeuchi}, \citenamefont {Watanabe}, \citenamefont {Yamada}, \citenamefont {Yanagisawa},\ and\ \citenamefont {Ishihara}}]{IWASAKI2001227}%
  \BibitemOpen
  \bibfield  {author} {\bibinfo {author} {\bibfnamefont {H.}~\bibnamefont {Iwasaki}}, \bibinfo {author} {\bibfnamefont {T.}~\bibnamefont {Motobayashi}}, \bibinfo {author} {\bibfnamefont {H.}~\bibnamefont {Sakurai}}, \bibinfo {author} {\bibfnamefont {K.}~\bibnamefont {Yoneda}}, \bibinfo {author} {\bibfnamefont {T.}~\bibnamefont {Gomi}}, \bibinfo {author} {\bibfnamefont {N.}~\bibnamefont {Aoi}}, \bibinfo {author} {\bibfnamefont {N.}~\bibnamefont {Fukuda}}, \bibinfo {author} {\bibfnamefont {Z.}~\bibnamefont {Fülöp}}, \bibinfo {author} {\bibfnamefont {U.}~\bibnamefont {Futakami}}, \bibinfo {author} {\bibfnamefont {Z.}~\bibnamefont {Gacsi}}, \bibinfo {author} {\bibfnamefont {Y.}~\bibnamefont {Higurashi}}, \bibinfo {author} {\bibfnamefont {N.}~\bibnamefont {Imai}}, \bibinfo {author} {\bibfnamefont {N.}~\bibnamefont {Iwasa}}, \bibinfo {author} {\bibfnamefont {T.}~\bibnamefont {Kubo}}, \bibinfo {author} {\bibfnamefont {M.}~\bibnamefont {Kunibu}}, \bibinfo {author} {\bibfnamefont {M.}~\bibnamefont {Kurokawa}},
  \bibinfo {author} {\bibfnamefont {Z.}~\bibnamefont {Liu}}, \bibinfo {author} {\bibfnamefont {T.}~\bibnamefont {Minemura}}, \bibinfo {author} {\bibfnamefont {A.}~\bibnamefont {Saito}}, \bibinfo {author} {\bibfnamefont {M.}~\bibnamefont {Serata}}, \bibinfo {author} {\bibfnamefont {S.}~\bibnamefont {Shimoura}}, \bibinfo {author} {\bibfnamefont {S.}~\bibnamefont {Takeuchi}}, \bibinfo {author} {\bibfnamefont {Y.}~\bibnamefont {Watanabe}}, \bibinfo {author} {\bibfnamefont {K.}~\bibnamefont {Yamada}}, \bibinfo {author} {\bibfnamefont {Y.}~\bibnamefont {Yanagisawa}},\ and\ \bibinfo {author} {\bibfnamefont {M.}~\bibnamefont {Ishihara}},\ }\bibfield  {title} {\bibinfo {title} {Large collectivity of 34mg},\ }\href {https://doi.org/https://doi.org/10.1016/S0370-2693(01)01244-8} {\bibfield  {journal} {\bibinfo  {journal} {Phys. Lett. B}\ }\textbf {\bibinfo {volume} {522}},\ \bibinfo {pages} {227} (\bibinfo {year} {2001})}\BibitemShut {NoStop}%
\bibitem [{\citenamefont {Yanagisawa}\ \emph {et~al.}(2003)\citenamefont {Yanagisawa}, \citenamefont {Notani}, \citenamefont {Sakurai}, \citenamefont {Kunibu}, \citenamefont {Akiyoshi}, \citenamefont {Aoi}, \citenamefont {Baba}, \citenamefont {Demichi}, \citenamefont {Fukuda}, \citenamefont {Hasegawa}, \citenamefont {Higurashi}, \citenamefont {Ishihara}, \citenamefont {Iwasa}, \citenamefont {Iwasaki}, \citenamefont {Gomi}, \citenamefont {Kanno}, \citenamefont {Kurokawa}, \citenamefont {Matsuyama}, \citenamefont {Michimasa}, \citenamefont {Minemura}, \citenamefont {Mizoi}, \citenamefont {Nakamura}, \citenamefont {Saito}, \citenamefont {Serata}, \citenamefont {Shimoura}, \citenamefont {Sugimoto}, \citenamefont {Takeshita}, \citenamefont {Takeuchi}, \citenamefont {Ue}, \citenamefont {Yamada}, \citenamefont {Yoneda},\ and\ \citenamefont {Motobayashi}}]{YANAGISAWA200384}%
  \BibitemOpen
  \bibfield  {author} {\bibinfo {author} {\bibfnamefont {Y.}~\bibnamefont {Yanagisawa}}, \bibinfo {author} {\bibfnamefont {M.}~\bibnamefont {Notani}}, \bibinfo {author} {\bibfnamefont {H.}~\bibnamefont {Sakurai}}, \bibinfo {author} {\bibfnamefont {M.}~\bibnamefont {Kunibu}}, \bibinfo {author} {\bibfnamefont {H.}~\bibnamefont {Akiyoshi}}, \bibinfo {author} {\bibfnamefont {N.}~\bibnamefont {Aoi}}, \bibinfo {author} {\bibfnamefont {H.}~\bibnamefont {Baba}}, \bibinfo {author} {\bibfnamefont {K.}~\bibnamefont {Demichi}}, \bibinfo {author} {\bibfnamefont {N.}~\bibnamefont {Fukuda}}, \bibinfo {author} {\bibfnamefont {H.}~\bibnamefont {Hasegawa}}, \bibinfo {author} {\bibfnamefont {Y.}~\bibnamefont {Higurashi}}, \bibinfo {author} {\bibfnamefont {M.}~\bibnamefont {Ishihara}}, \bibinfo {author} {\bibfnamefont {N.}~\bibnamefont {Iwasa}}, \bibinfo {author} {\bibfnamefont {H.}~\bibnamefont {Iwasaki}}, \bibinfo {author} {\bibfnamefont {T.}~\bibnamefont {Gomi}}, \bibinfo {author} {\bibfnamefont {S.}~\bibnamefont {Kanno}},
  \bibinfo {author} {\bibfnamefont {M.}~\bibnamefont {Kurokawa}}, \bibinfo {author} {\bibfnamefont {Y.}~\bibnamefont {Matsuyama}}, \bibinfo {author} {\bibfnamefont {S.}~\bibnamefont {Michimasa}}, \bibinfo {author} {\bibfnamefont {T.}~\bibnamefont {Minemura}}, \bibinfo {author} {\bibfnamefont {T.}~\bibnamefont {Mizoi}}, \bibinfo {author} {\bibfnamefont {T.}~\bibnamefont {Nakamura}}, \bibinfo {author} {\bibfnamefont {A.}~\bibnamefont {Saito}}, \bibinfo {author} {\bibfnamefont {M.}~\bibnamefont {Serata}}, \bibinfo {author} {\bibfnamefont {S.}~\bibnamefont {Shimoura}}, \bibinfo {author} {\bibfnamefont {T.}~\bibnamefont {Sugimoto}}, \bibinfo {author} {\bibfnamefont {E.}~\bibnamefont {Takeshita}}, \bibinfo {author} {\bibfnamefont {S.}~\bibnamefont {Takeuchi}}, \bibinfo {author} {\bibfnamefont {K.}~\bibnamefont {Ue}}, \bibinfo {author} {\bibfnamefont {K.}~\bibnamefont {Yamada}}, \bibinfo {author} {\bibfnamefont {K.}~\bibnamefont {Yoneda}},\ and\ \bibinfo {author} {\bibfnamefont {T.}~\bibnamefont {Motobayashi}},\
  }\bibfield  {title} {\bibinfo {title} {The first excited state of 30ne studied by proton inelastic scattering in reversed kinematics},\ }\href {https://doi.org/https://doi.org/10.1016/S0370-2693(03)00802-5} {\bibfield  {journal} {\bibinfo  {journal} {Phys. Lett. B}\ }\textbf {\bibinfo {volume} {566}},\ \bibinfo {pages} {84} (\bibinfo {year} {2003})}\BibitemShut {NoStop}%
\bibitem [{\citenamefont {Rodr\'{\i}guez-Guzm\'an}\ \emph {et~al.}(2002)\citenamefont {Rodr\'{\i}guez-Guzm\'an}, \citenamefont {Egido},\ and\ \citenamefont {Robledo}}]{Rodriguez_2002}%
  \BibitemOpen
  \bibfield  {author} {\bibinfo {author} {\bibfnamefont {R.}~\bibnamefont {Rodr\'{\i}guez-Guzm\'an}}, \bibinfo {author} {\bibfnamefont {J.~L.}\ \bibnamefont {Egido}},\ and\ \bibinfo {author} {\bibfnamefont {L.~M.}\ \bibnamefont {Robledo}},\ }\bibfield  {title} {\bibinfo {title} {Quadrupole collectivity in $n\ensuremath{\approx}28$ nuclei with the angular momentum projected generator coordinate method},\ }\href {https://doi.org/10.1103/PhysRevC.65.024304} {\bibfield  {journal} {\bibinfo  {journal} {Phys. Rev. C}\ }\textbf {\bibinfo {volume} {65}},\ \bibinfo {pages} {024304} (\bibinfo {year} {2002})}\BibitemShut {NoStop}%
\bibitem [{\citenamefont {Otsuka}\ \emph {et~al.}(2005)\citenamefont {Otsuka}, \citenamefont {Suzuki}, \citenamefont {Fujimoto}, \citenamefont {Grawe},\ and\ \citenamefont {Akaishi}}]{Otsuka_2005}%
  \BibitemOpen
  \bibfield  {author} {\bibinfo {author} {\bibfnamefont {T.}~\bibnamefont {Otsuka}}, \bibinfo {author} {\bibfnamefont {T.}~\bibnamefont {Suzuki}}, \bibinfo {author} {\bibfnamefont {R.}~\bibnamefont {Fujimoto}}, \bibinfo {author} {\bibfnamefont {H.}~\bibnamefont {Grawe}},\ and\ \bibinfo {author} {\bibfnamefont {Y.}~\bibnamefont {Akaishi}},\ }\bibfield  {title} {\bibinfo {title} {Evolution of nuclear shells due to the tensor force},\ }\href {https://doi.org/10.1103/PhysRevLett.95.232502} {\bibfield  {journal} {\bibinfo  {journal} {Phys. Rev. Lett.}\ }\textbf {\bibinfo {volume} {95}},\ \bibinfo {pages} {232502} (\bibinfo {year} {2005})}\BibitemShut {NoStop}%
\bibitem [{\citenamefont {Sorlin}\ and\ \citenamefont {Porquet}(2008)}]{SORLIN2008602}%
  \BibitemOpen
  \bibfield  {author} {\bibinfo {author} {\bibfnamefont {O.}~\bibnamefont {Sorlin}}\ and\ \bibinfo {author} {\bibfnamefont {M.-G.}\ \bibnamefont {Porquet}},\ }\bibfield  {title} {\bibinfo {title} {Nuclear magic numbers: New features far from stability},\ }\href {https://doi.org/https://doi.org/10.1016/j.ppnp.2008.05.001} {\bibfield  {journal} {\bibinfo  {journal} {Progress in Particle and Nuclear Physics}\ }\textbf {\bibinfo {volume} {61}},\ \bibinfo {pages} {602} (\bibinfo {year} {2008})}\BibitemShut {NoStop}%
\bibitem [{\citenamefont {Glasmacher}\ \emph {et~al.}(1997)\citenamefont {Glasmacher}, \citenamefont {Brown}, \citenamefont {Chromik}, \citenamefont {Cottle}, \citenamefont {Fauerbach}, \citenamefont {Ibbotson}, \citenamefont {Kemper}, \citenamefont {Morrissey}, \citenamefont {Scheit}, \citenamefont {Sklenicka},\ and\ \citenamefont {Steiner}}]{GLASMACHER1997163}%
  \BibitemOpen
  \bibfield  {author} {\bibinfo {author} {\bibfnamefont {T.}~\bibnamefont {Glasmacher}}, \bibinfo {author} {\bibfnamefont {B.}~\bibnamefont {Brown}}, \bibinfo {author} {\bibfnamefont {M.}~\bibnamefont {Chromik}}, \bibinfo {author} {\bibfnamefont {P.}~\bibnamefont {Cottle}}, \bibinfo {author} {\bibfnamefont {M.}~\bibnamefont {Fauerbach}}, \bibinfo {author} {\bibfnamefont {R.}~\bibnamefont {Ibbotson}}, \bibinfo {author} {\bibfnamefont {K.}~\bibnamefont {Kemper}}, \bibinfo {author} {\bibfnamefont {D.}~\bibnamefont {Morrissey}}, \bibinfo {author} {\bibfnamefont {H.}~\bibnamefont {Scheit}}, \bibinfo {author} {\bibfnamefont {D.}~\bibnamefont {Sklenicka}},\ and\ \bibinfo {author} {\bibfnamefont {M.}~\bibnamefont {Steiner}},\ }\bibfield  {title} {\bibinfo {title} {{Collectivity in $^{44}$S}},\ }\href {https://doi.org/https://doi.org/10.1016/S0370-2693(97)00077-4} {\bibfield  {journal} {\bibinfo  {journal} {Physics Letters B}\ }\textbf {\bibinfo {volume} {395}},\ \bibinfo {pages} {163} (\bibinfo {year}
  {1997})}\BibitemShut {NoStop}%
\bibitem [{\citenamefont {Gaudefroy}\ \emph {et~al.}(2006)\citenamefont {Gaudefroy}, \citenamefont {Sorlin}, \citenamefont {Beaumel}, \citenamefont {Blumenfeld}, \citenamefont {Dombr\'adi}, \citenamefont {Fortier}, \citenamefont {Franchoo}, \citenamefont {G\'elin}, \citenamefont {Gibelin}, \citenamefont {Gr\'evy}, \citenamefont {Hammache}, \citenamefont {Ibrahim}, \citenamefont {Kemper}, \citenamefont {Kratz}, \citenamefont {Lukyanov}, \citenamefont {Monrozeau}, \citenamefont {Nalpas}, \citenamefont {Nowacki}, \citenamefont {Ostrowski}, \citenamefont {Otsuka}, \citenamefont {Penionzhkevich}, \citenamefont {Piekarewicz}, \citenamefont {Pollacco}, \citenamefont {Roussel-Chomaz}, \citenamefont {Rich}, \citenamefont {Scarpaci}, \citenamefont {St.~Laurent}, \citenamefont {Sohler}, \citenamefont {Stanoiu}, \citenamefont {Suzuki}, \citenamefont {Tryggestad},\ and\ \citenamefont {Verney}}]{Gaudefroy_2006}%
  \BibitemOpen
  \bibfield  {author} {\bibinfo {author} {\bibfnamefont {L.}~\bibnamefont {Gaudefroy}}, \bibinfo {author} {\bibfnamefont {O.}~\bibnamefont {Sorlin}}, \bibinfo {author} {\bibfnamefont {D.}~\bibnamefont {Beaumel}}, \bibinfo {author} {\bibfnamefont {Y.}~\bibnamefont {Blumenfeld}}, \bibinfo {author} {\bibfnamefont {Z.}~\bibnamefont {Dombr\'adi}}, \bibinfo {author} {\bibfnamefont {S.}~\bibnamefont {Fortier}}, \bibinfo {author} {\bibfnamefont {S.}~\bibnamefont {Franchoo}}, \bibinfo {author} {\bibfnamefont {M.}~\bibnamefont {G\'elin}}, \bibinfo {author} {\bibfnamefont {J.}~\bibnamefont {Gibelin}}, \bibinfo {author} {\bibfnamefont {S.}~\bibnamefont {Gr\'evy}}, \bibinfo {author} {\bibfnamefont {F.}~\bibnamefont {Hammache}}, \bibinfo {author} {\bibfnamefont {F.}~\bibnamefont {Ibrahim}}, \bibinfo {author} {\bibfnamefont {K.~W.}\ \bibnamefont {Kemper}}, \bibinfo {author} {\bibfnamefont {K.-L.}\ \bibnamefont {Kratz}}, \bibinfo {author} {\bibfnamefont {S.~M.}\ \bibnamefont {Lukyanov}}, \bibinfo {author} {\bibfnamefont
  {C.}~\bibnamefont {Monrozeau}}, \bibinfo {author} {\bibfnamefont {L.}~\bibnamefont {Nalpas}}, \bibinfo {author} {\bibfnamefont {F.}~\bibnamefont {Nowacki}}, \bibinfo {author} {\bibfnamefont {A.~N.}\ \bibnamefont {Ostrowski}}, \bibinfo {author} {\bibfnamefont {T.}~\bibnamefont {Otsuka}}, \bibinfo {author} {\bibfnamefont {Y.-E.}\ \bibnamefont {Penionzhkevich}}, \bibinfo {author} {\bibfnamefont {J.}~\bibnamefont {Piekarewicz}}, \bibinfo {author} {\bibfnamefont {E.~C.}\ \bibnamefont {Pollacco}}, \bibinfo {author} {\bibfnamefont {P.}~\bibnamefont {Roussel-Chomaz}}, \bibinfo {author} {\bibfnamefont {E.}~\bibnamefont {Rich}}, \bibinfo {author} {\bibfnamefont {J.~A.}\ \bibnamefont {Scarpaci}}, \bibinfo {author} {\bibfnamefont {M.~G.}\ \bibnamefont {St.~Laurent}}, \bibinfo {author} {\bibfnamefont {D.}~\bibnamefont {Sohler}}, \bibinfo {author} {\bibfnamefont {M.}~\bibnamefont {Stanoiu}}, \bibinfo {author} {\bibfnamefont {T.}~\bibnamefont {Suzuki}}, \bibinfo {author} {\bibfnamefont {E.}~\bibnamefont {Tryggestad}},\
  and\ \bibinfo {author} {\bibfnamefont {D.}~\bibnamefont {Verney}},\ }\bibfield  {title} {\bibinfo {title} {Reduction of the spin-orbit splittings at the $n=28$ shell closure},\ }\href {https://doi.org/10.1103/PhysRevLett.97.092501} {\bibfield  {journal} {\bibinfo  {journal} {Phys. Rev. Lett.}\ }\textbf {\bibinfo {volume} {97}},\ \bibinfo {pages} {092501} (\bibinfo {year} {2006})}\BibitemShut {NoStop}%
\bibitem [{\citenamefont {Bastin}\ \emph {et~al.}(2007)\citenamefont {Bastin}, \citenamefont {Gr\'evy}, \citenamefont {Sohler}, \citenamefont {Sorlin}, \citenamefont {Dombr\'adi}, \citenamefont {Achouri}, \citenamefont {Ang\'elique}, \citenamefont {Azaiez}, \citenamefont {Baiborodin}, \citenamefont {Borcea}, \citenamefont {Bourgeois}, \citenamefont {Buta}, \citenamefont {B\"urger}, \citenamefont {Chapman}, \citenamefont {Dalouzy}, \citenamefont {Dlouhy}, \citenamefont {Drouard}, \citenamefont {Elekes}, \citenamefont {Franchoo}, \citenamefont {Iacob}, \citenamefont {Laurent}, \citenamefont {Lazar}, \citenamefont {Liang}, \citenamefont {Li\'enard}, \citenamefont {Mrazek}, \citenamefont {Nalpas}, \citenamefont {Negoita}, \citenamefont {Orr}, \citenamefont {Penionzhkevich}, \citenamefont {Podoly\'ak}, \citenamefont {Pougheon}, \citenamefont {Roussel-Chomaz}, \citenamefont {Saint-Laurent}, \citenamefont {Stanoiu}, \citenamefont {Stefan}, \citenamefont {Nowacki},\ and\ \citenamefont {Poves}}]{Bastin_2007}%
  \BibitemOpen
  \bibfield  {author} {\bibinfo {author} {\bibfnamefont {B.}~\bibnamefont {Bastin}}, \bibinfo {author} {\bibfnamefont {S.}~\bibnamefont {Gr\'evy}}, \bibinfo {author} {\bibfnamefont {D.}~\bibnamefont {Sohler}}, \bibinfo {author} {\bibfnamefont {O.}~\bibnamefont {Sorlin}}, \bibinfo {author} {\bibfnamefont {Z.}~\bibnamefont {Dombr\'adi}}, \bibinfo {author} {\bibfnamefont {N.~L.}\ \bibnamefont {Achouri}}, \bibinfo {author} {\bibfnamefont {J.~C.}\ \bibnamefont {Ang\'elique}}, \bibinfo {author} {\bibfnamefont {F.}~\bibnamefont {Azaiez}}, \bibinfo {author} {\bibfnamefont {D.}~\bibnamefont {Baiborodin}}, \bibinfo {author} {\bibfnamefont {R.}~\bibnamefont {Borcea}}, \bibinfo {author} {\bibfnamefont {C.}~\bibnamefont {Bourgeois}}, \bibinfo {author} {\bibfnamefont {A.}~\bibnamefont {Buta}}, \bibinfo {author} {\bibfnamefont {A.}~\bibnamefont {B\"urger}}, \bibinfo {author} {\bibfnamefont {R.}~\bibnamefont {Chapman}}, \bibinfo {author} {\bibfnamefont {J.~C.}\ \bibnamefont {Dalouzy}}, \bibinfo {author} {\bibfnamefont
  {Z.}~\bibnamefont {Dlouhy}}, \bibinfo {author} {\bibfnamefont {A.}~\bibnamefont {Drouard}}, \bibinfo {author} {\bibfnamefont {Z.}~\bibnamefont {Elekes}}, \bibinfo {author} {\bibfnamefont {S.}~\bibnamefont {Franchoo}}, \bibinfo {author} {\bibfnamefont {S.}~\bibnamefont {Iacob}}, \bibinfo {author} {\bibfnamefont {B.}~\bibnamefont {Laurent}}, \bibinfo {author} {\bibfnamefont {M.}~\bibnamefont {Lazar}}, \bibinfo {author} {\bibfnamefont {X.}~\bibnamefont {Liang}}, \bibinfo {author} {\bibfnamefont {E.}~\bibnamefont {Li\'enard}}, \bibinfo {author} {\bibfnamefont {J.}~\bibnamefont {Mrazek}}, \bibinfo {author} {\bibfnamefont {L.}~\bibnamefont {Nalpas}}, \bibinfo {author} {\bibfnamefont {F.}~\bibnamefont {Negoita}}, \bibinfo {author} {\bibfnamefont {N.~A.}\ \bibnamefont {Orr}}, \bibinfo {author} {\bibfnamefont {Y.}~\bibnamefont {Penionzhkevich}}, \bibinfo {author} {\bibfnamefont {Z.}~\bibnamefont {Podoly\'ak}}, \bibinfo {author} {\bibfnamefont {F.}~\bibnamefont {Pougheon}}, \bibinfo {author} {\bibfnamefont
  {P.}~\bibnamefont {Roussel-Chomaz}}, \bibinfo {author} {\bibfnamefont {M.~G.}\ \bibnamefont {Saint-Laurent}}, \bibinfo {author} {\bibfnamefont {M.}~\bibnamefont {Stanoiu}}, \bibinfo {author} {\bibfnamefont {I.}~\bibnamefont {Stefan}}, \bibinfo {author} {\bibfnamefont {F.}~\bibnamefont {Nowacki}},\ and\ \bibinfo {author} {\bibfnamefont {A.}~\bibnamefont {Poves}},\ }\bibfield  {title} {\bibinfo {title} {Collapse of the $n=28$ shell closure in $^{42}\mathrm{S}\mathrm{i}$},\ }\href {https://doi.org/10.1103/PhysRevLett.99.022503} {\bibfield  {journal} {\bibinfo  {journal} {Phys. Rev. Lett.}\ }\textbf {\bibinfo {volume} {99}},\ \bibinfo {pages} {022503} (\bibinfo {year} {2007})}\BibitemShut {NoStop}%
\bibitem [{\citenamefont {Sorlin}\ and\ \citenamefont {Porquet}(2012)}]{Sorlin_2013}%
  \BibitemOpen
  \bibfield  {author} {\bibinfo {author} {\bibfnamefont {O.}~\bibnamefont {Sorlin}}\ and\ \bibinfo {author} {\bibfnamefont {M.-G.}\ \bibnamefont {Porquet}},\ }\bibfield  {title} {\bibinfo {title} {Evolution of the n = 28 shell closure: a test bench for nuclear forces},\ }\href {https://doi.org/10.1088/0031-8949/2013/T152/014003} {\bibfield  {journal} {\bibinfo  {journal} {Physica Scripta}\ }\textbf {\bibinfo {volume} {2013}},\ \bibinfo {pages} {014003} (\bibinfo {year} {2012})}\BibitemShut {NoStop}%
\bibitem [{\citenamefont {Mayer}(1949)}]{Goeppert_1949}%
  \BibitemOpen
  \bibfield  {author} {\bibinfo {author} {\bibfnamefont {M.~G.}\ \bibnamefont {Mayer}},\ }\bibfield  {title} {\bibinfo {title} {On closed shells in nuclei. ii},\ }\href {https://doi.org/10.1103/PhysRev.75.1969} {\bibfield  {journal} {\bibinfo  {journal} {Phys. Rev.}\ }\textbf {\bibinfo {volume} {75}},\ \bibinfo {pages} {1969} (\bibinfo {year} {1949})}\BibitemShut {NoStop}%
\bibitem [{\citenamefont {Haxel}\ \emph {et~al.}(1949)\citenamefont {Haxel}, \citenamefont {Jensen},\ and\ \citenamefont {Suess}}]{Haxel_1949}%
  \BibitemOpen
  \bibfield  {author} {\bibinfo {author} {\bibfnamefont {O.}~\bibnamefont {Haxel}}, \bibinfo {author} {\bibfnamefont {J.~H.~D.}\ \bibnamefont {Jensen}},\ and\ \bibinfo {author} {\bibfnamefont {H.~E.}\ \bibnamefont {Suess}},\ }\bibfield  {title} {\bibinfo {title} {On the "magic numbers" in nuclear structure},\ }\href {https://doi.org/10.1103/PhysRev.75.1766.2} {\bibfield  {journal} {\bibinfo  {journal} {Phys. Rev.}\ }\textbf {\bibinfo {volume} {75}},\ \bibinfo {pages} {1766} (\bibinfo {year} {1949})}\BibitemShut {NoStop}%
\bibitem [{\citenamefont {C\'aceres}\ \emph {et~al.}(2012)\citenamefont {C\'aceres}, \citenamefont {Sohler}, \citenamefont {Gr\'evy}, \citenamefont {Sorlin}, \citenamefont {Dombr\'adi}, \citenamefont {Bastin}, \citenamefont {Achouri}, \citenamefont {Ang\'elique}, \citenamefont {Azaiez}, \citenamefont {Baiborodin}, \citenamefont {Borcea}, \citenamefont {Bourgeois}, \citenamefont {Buta}, \citenamefont {B\"urger}, \citenamefont {Chapman}, \citenamefont {Dalouzy}, \citenamefont {Dlouhy}, \citenamefont {Drouard}, \citenamefont {Elekes}, \citenamefont {Franchoo}, \citenamefont {Gaudefroy}, \citenamefont {Iacob}, \citenamefont {Laurent}, \citenamefont {Lazar}, \citenamefont {Liang}, \citenamefont {Li\'enard}, \citenamefont {Mrazek}, \citenamefont {Nalpas}, \citenamefont {Negoita}, \citenamefont {Nowacki}, \citenamefont {Orr}, \citenamefont {Penionzhkevich}, \citenamefont {Podoly\'ak}, \citenamefont {Pougheon}, \citenamefont {Poves}, \citenamefont {Roussel-Chomaz}, \citenamefont {Saint-Laurent}, \citenamefont
  {Stanoiu},\ and\ \citenamefont {Stefan}}]{Caceres_PhysRevC.85.024311}%
  \BibitemOpen
  \bibfield  {author} {\bibinfo {author} {\bibfnamefont {L.}~\bibnamefont {C\'aceres}}, \bibinfo {author} {\bibfnamefont {D.}~\bibnamefont {Sohler}}, \bibinfo {author} {\bibfnamefont {S.}~\bibnamefont {Gr\'evy}}, \bibinfo {author} {\bibfnamefont {O.}~\bibnamefont {Sorlin}}, \bibinfo {author} {\bibfnamefont {Z.}~\bibnamefont {Dombr\'adi}}, \bibinfo {author} {\bibfnamefont {B.}~\bibnamefont {Bastin}}, \bibinfo {author} {\bibfnamefont {N.~L.}\ \bibnamefont {Achouri}}, \bibinfo {author} {\bibfnamefont {J.~C.}\ \bibnamefont {Ang\'elique}}, \bibinfo {author} {\bibfnamefont {F.}~\bibnamefont {Azaiez}}, \bibinfo {author} {\bibfnamefont {D.}~\bibnamefont {Baiborodin}}, \bibinfo {author} {\bibfnamefont {R.}~\bibnamefont {Borcea}}, \bibinfo {author} {\bibfnamefont {C.}~\bibnamefont {Bourgeois}}, \bibinfo {author} {\bibfnamefont {A.}~\bibnamefont {Buta}}, \bibinfo {author} {\bibfnamefont {A.}~\bibnamefont {B\"urger}}, \bibinfo {author} {\bibfnamefont {R.}~\bibnamefont {Chapman}}, \bibinfo {author} {\bibfnamefont {J.~C.}\
  \bibnamefont {Dalouzy}}, \bibinfo {author} {\bibfnamefont {Z.}~\bibnamefont {Dlouhy}}, \bibinfo {author} {\bibfnamefont {A.}~\bibnamefont {Drouard}}, \bibinfo {author} {\bibfnamefont {Z.}~\bibnamefont {Elekes}}, \bibinfo {author} {\bibfnamefont {S.}~\bibnamefont {Franchoo}}, \bibinfo {author} {\bibfnamefont {L.}~\bibnamefont {Gaudefroy}}, \bibinfo {author} {\bibfnamefont {S.}~\bibnamefont {Iacob}}, \bibinfo {author} {\bibfnamefont {B.}~\bibnamefont {Laurent}}, \bibinfo {author} {\bibfnamefont {M.}~\bibnamefont {Lazar}}, \bibinfo {author} {\bibfnamefont {X.}~\bibnamefont {Liang}}, \bibinfo {author} {\bibfnamefont {E.}~\bibnamefont {Li\'enard}}, \bibinfo {author} {\bibfnamefont {J.}~\bibnamefont {Mrazek}}, \bibinfo {author} {\bibfnamefont {L.}~\bibnamefont {Nalpas}}, \bibinfo {author} {\bibfnamefont {F.}~\bibnamefont {Negoita}}, \bibinfo {author} {\bibfnamefont {F.}~\bibnamefont {Nowacki}}, \bibinfo {author} {\bibfnamefont {N.~A.}\ \bibnamefont {Orr}}, \bibinfo {author} {\bibfnamefont {Y.}~\bibnamefont
  {Penionzhkevich}}, \bibinfo {author} {\bibfnamefont {Z.}~\bibnamefont {Podoly\'ak}}, \bibinfo {author} {\bibfnamefont {F.}~\bibnamefont {Pougheon}}, \bibinfo {author} {\bibfnamefont {A.}~\bibnamefont {Poves}}, \bibinfo {author} {\bibfnamefont {P.}~\bibnamefont {Roussel-Chomaz}}, \bibinfo {author} {\bibfnamefont {M.~G.}\ \bibnamefont {Saint-Laurent}}, \bibinfo {author} {\bibfnamefont {M.}~\bibnamefont {Stanoiu}},\ and\ \bibinfo {author} {\bibfnamefont {I.}~\bibnamefont {Stefan}},\ }\bibfield  {title} {\bibinfo {title} {In-beam spectroscopic studies of the ${}^{44}$s nucleus},\ }\href {https://doi.org/10.1103/PhysRevC.85.024311} {\bibfield  {journal} {\bibinfo  {journal} {Phys. Rev. C}\ }\textbf {\bibinfo {volume} {85}},\ \bibinfo {pages} {024311} (\bibinfo {year} {2012})}\BibitemShut {NoStop}%
\bibitem [{\citenamefont {Force}\ \emph {et~al.}(2010)\citenamefont {Force}, \citenamefont {Gr\'evy}, \citenamefont {Gaudefroy}, \citenamefont {Sorlin}, \citenamefont {C\'aceres}, \citenamefont {Rotaru}, \citenamefont {Mrazek}, \citenamefont {Achouri}, \citenamefont {Ang\'elique}, \citenamefont {Azaiez}, \citenamefont {Bastin}, \citenamefont {Borcea}, \citenamefont {Buta}, \citenamefont {Daugas}, \citenamefont {Dlouhy}, \citenamefont {Dombr\'adi}, \citenamefont {De~Oliveira}, \citenamefont {Negoita}, \citenamefont {Penionzhkevich}, \citenamefont {Saint-Laurent}, \citenamefont {Sohler}, \citenamefont {Stanoiu}, \citenamefont {Stefan}, \citenamefont {Stodel},\ and\ \citenamefont {Nowacki}}]{Force_2010}%
  \BibitemOpen
  \bibfield  {author} {\bibinfo {author} {\bibfnamefont {C.}~\bibnamefont {Force}}, \bibinfo {author} {\bibfnamefont {S.}~\bibnamefont {Gr\'evy}}, \bibinfo {author} {\bibfnamefont {L.}~\bibnamefont {Gaudefroy}}, \bibinfo {author} {\bibfnamefont {O.}~\bibnamefont {Sorlin}}, \bibinfo {author} {\bibfnamefont {L.}~\bibnamefont {C\'aceres}}, \bibinfo {author} {\bibfnamefont {F.}~\bibnamefont {Rotaru}}, \bibinfo {author} {\bibfnamefont {J.}~\bibnamefont {Mrazek}}, \bibinfo {author} {\bibfnamefont {N.~L.}\ \bibnamefont {Achouri}}, \bibinfo {author} {\bibfnamefont {J.~C.}\ \bibnamefont {Ang\'elique}}, \bibinfo {author} {\bibfnamefont {F.}~\bibnamefont {Azaiez}}, \bibinfo {author} {\bibfnamefont {B.}~\bibnamefont {Bastin}}, \bibinfo {author} {\bibfnamefont {R.}~\bibnamefont {Borcea}}, \bibinfo {author} {\bibfnamefont {A.}~\bibnamefont {Buta}}, \bibinfo {author} {\bibfnamefont {J.~M.}\ \bibnamefont {Daugas}}, \bibinfo {author} {\bibfnamefont {Z.}~\bibnamefont {Dlouhy}}, \bibinfo {author} {\bibfnamefont
  {Z.}~\bibnamefont {Dombr\'adi}}, \bibinfo {author} {\bibfnamefont {F.}~\bibnamefont {De~Oliveira}}, \bibinfo {author} {\bibfnamefont {F.}~\bibnamefont {Negoita}}, \bibinfo {author} {\bibfnamefont {Y.}~\bibnamefont {Penionzhkevich}}, \bibinfo {author} {\bibfnamefont {M.~G.}\ \bibnamefont {Saint-Laurent}}, \bibinfo {author} {\bibfnamefont {D.}~\bibnamefont {Sohler}}, \bibinfo {author} {\bibfnamefont {M.}~\bibnamefont {Stanoiu}}, \bibinfo {author} {\bibfnamefont {I.}~\bibnamefont {Stefan}}, \bibinfo {author} {\bibfnamefont {C.}~\bibnamefont {Stodel}},\ and\ \bibinfo {author} {\bibfnamefont {F.}~\bibnamefont {Nowacki}},\ }\bibfield  {title} {\bibinfo {title} {{Prolate-Spherical Shape Coexistence at $N=28$ in $^{44}\mathrm{S}$}},\ }\href {https://doi.org/10.1103/PhysRevLett.105.102501} {\bibfield  {journal} {\bibinfo  {journal} {Phys. Rev. Lett.}\ }\textbf {\bibinfo {volume} {105}},\ \bibinfo {pages} {102501} (\bibinfo {year} {2010})}\BibitemShut {NoStop}%
\bibitem [{\citenamefont {Riley}\ \emph {et~al.}(2025)\citenamefont {Riley}, \citenamefont {Conroy}, \citenamefont {Himmelreich}, \citenamefont {Heinze}, \citenamefont {Kosa}, \citenamefont {McNulty}, \citenamefont {Cottle}, \citenamefont {Spieker}, \citenamefont {Volya}, \citenamefont {Conley}, \citenamefont {Houlihan}, \citenamefont {Kelly}, \citenamefont {Kemper}, \citenamefont {Ali}, \citenamefont {Beck}, \citenamefont {Gillespie}, \citenamefont {Hausmann}, \citenamefont {Noji}, \citenamefont {Pereira}, \citenamefont {Weisshaar}, \citenamefont {Chung-Jung}, \citenamefont {Farris}, \citenamefont {Gade}, \citenamefont {Grauvogel}, \citenamefont {Hill}, \citenamefont {Rahman}, \citenamefont {Zegers}, \citenamefont {Longfellow},\ and\ \citenamefont {Pathirana}}]{Riley_2025}%
  \BibitemOpen
  \bibfield  {author} {\bibinfo {author} {\bibfnamefont {L.~A.}\ \bibnamefont {Riley}}, \bibinfo {author} {\bibfnamefont {I.}~\bibnamefont {Conroy}}, \bibinfo {author} {\bibfnamefont {A.~M.}\ \bibnamefont {Himmelreich}}, \bibinfo {author} {\bibfnamefont {M.}~\bibnamefont {Heinze}}, \bibinfo {author} {\bibfnamefont {J.}~\bibnamefont {Kosa}}, \bibinfo {author} {\bibfnamefont {B.}~\bibnamefont {McNulty}}, \bibinfo {author} {\bibfnamefont {P.~D.}\ \bibnamefont {Cottle}}, \bibinfo {author} {\bibfnamefont {M.}~\bibnamefont {Spieker}}, \bibinfo {author} {\bibfnamefont {A.}~\bibnamefont {Volya}}, \bibinfo {author} {\bibfnamefont {A.~L.}\ \bibnamefont {Conley}}, \bibinfo {author} {\bibfnamefont {D.}~\bibnamefont {Houlihan}}, \bibinfo {author} {\bibfnamefont {B.}~\bibnamefont {Kelly}}, \bibinfo {author} {\bibfnamefont {K.~W.}\ \bibnamefont {Kemper}}, \bibinfo {author} {\bibfnamefont {S.~M.}\ \bibnamefont {Ali}}, \bibinfo {author} {\bibfnamefont {T.}~\bibnamefont {Beck}}, \bibinfo {author} {\bibfnamefont {S.~A.}\
  \bibnamefont {Gillespie}}, \bibinfo {author} {\bibfnamefont {M.}~\bibnamefont {Hausmann}}, \bibinfo {author} {\bibfnamefont {S.}~\bibnamefont {Noji}}, \bibinfo {author} {\bibfnamefont {J.}~\bibnamefont {Pereira}}, \bibinfo {author} {\bibfnamefont {D.}~\bibnamefont {Weisshaar}}, \bibinfo {author} {\bibfnamefont {J.}~\bibnamefont {Chung-Jung}}, \bibinfo {author} {\bibfnamefont {P.}~\bibnamefont {Farris}}, \bibinfo {author} {\bibfnamefont {A.}~\bibnamefont {Gade}}, \bibinfo {author} {\bibfnamefont {G.}~\bibnamefont {Grauvogel}}, \bibinfo {author} {\bibfnamefont {A.~M.}\ \bibnamefont {Hill}}, \bibinfo {author} {\bibfnamefont {Z.}~\bibnamefont {Rahman}}, \bibinfo {author} {\bibfnamefont {R.~G.~T.}\ \bibnamefont {Zegers}}, \bibinfo {author} {\bibfnamefont {B.}~\bibnamefont {Longfellow}},\ and\ \bibinfo {author} {\bibfnamefont {N.~D.}\ \bibnamefont {Pathirana}},\ }\bibfield  {title} {\bibinfo {title} {Determination of proton and neutron contributions to the
  ${0}_{\text{g.s.}}^{+}\ensuremath{\rightarrow}{2}_{1}^{+}$ excitations in $^{42}\mathrm{Si}$ and $^{44}\mathrm{S}$ using inelastic proton scattering in inverse kinematics and intermediate-energy coulomb excitation},\ }\href {https://doi.org/10.1103/b8xj-ycqk} {\bibfield  {journal} {\bibinfo  {journal} {Phys. Rev. C}\ }\textbf {\bibinfo {volume} {112}},\ \bibinfo {pages} {014331} (\bibinfo {year} {2025})}\BibitemShut {NoStop}%
\bibitem [{\citenamefont {Santiago-Gonzalez}\ \emph {et~al.}(2011)\citenamefont {Santiago-Gonzalez}, \citenamefont {Wiedenh\"over}, \citenamefont {Abramkina}, \citenamefont {Avila}, \citenamefont {Baugher}, \citenamefont {Bazin}, \citenamefont {Brown}, \citenamefont {Cottle}, \citenamefont {Gade}, \citenamefont {Glasmacher}, \citenamefont {Kemper}, \citenamefont {McDaniel}, \citenamefont {Rojas}, \citenamefont {Ratkiewicz}, \citenamefont {Meharchand}, \citenamefont {Simpson}, \citenamefont {Tostevin}, \citenamefont {Volya},\ and\ \citenamefont {Weisshaar}}]{Santiago-Gonzalez_PhysRevC.83.061305}%
  \BibitemOpen
  \bibfield  {author} {\bibinfo {author} {\bibfnamefont {D.}~\bibnamefont {Santiago-Gonzalez}}, \bibinfo {author} {\bibfnamefont {I.}~\bibnamefont {Wiedenh\"over}}, \bibinfo {author} {\bibfnamefont {V.}~\bibnamefont {Abramkina}}, \bibinfo {author} {\bibfnamefont {M.~L.}\ \bibnamefont {Avila}}, \bibinfo {author} {\bibfnamefont {T.}~\bibnamefont {Baugher}}, \bibinfo {author} {\bibfnamefont {D.}~\bibnamefont {Bazin}}, \bibinfo {author} {\bibfnamefont {B.~A.}\ \bibnamefont {Brown}}, \bibinfo {author} {\bibfnamefont {P.~D.}\ \bibnamefont {Cottle}}, \bibinfo {author} {\bibfnamefont {A.}~\bibnamefont {Gade}}, \bibinfo {author} {\bibfnamefont {T.}~\bibnamefont {Glasmacher}}, \bibinfo {author} {\bibfnamefont {K.~W.}\ \bibnamefont {Kemper}}, \bibinfo {author} {\bibfnamefont {S.}~\bibnamefont {McDaniel}}, \bibinfo {author} {\bibfnamefont {A.}~\bibnamefont {Rojas}}, \bibinfo {author} {\bibfnamefont {A.}~\bibnamefont {Ratkiewicz}}, \bibinfo {author} {\bibfnamefont {R.}~\bibnamefont {Meharchand}}, \bibinfo {author}
  {\bibfnamefont {E.~C.}\ \bibnamefont {Simpson}}, \bibinfo {author} {\bibfnamefont {J.~A.}\ \bibnamefont {Tostevin}}, \bibinfo {author} {\bibfnamefont {A.}~\bibnamefont {Volya}},\ and\ \bibinfo {author} {\bibfnamefont {D.}~\bibnamefont {Weisshaar}},\ }\bibfield  {title} {\bibinfo {title} {{Triple configuration coexistence in $^{44}\mathrm{S}$}},\ }\href {https://doi.org/10.1103/PhysRevC.83.061305} {\bibfield  {journal} {\bibinfo  {journal} {Phys. Rev. C}\ }\textbf {\bibinfo {volume} {83}},\ \bibinfo {pages} {061305} (\bibinfo {year} {2011})}\BibitemShut {NoStop}%
\bibitem [{\citenamefont {P{\'e}ru}\ \emph {et~al.}(2000)\citenamefont {P{\'e}ru}, \citenamefont {Girod},\ and\ \citenamefont {Berger}}]{Peru2000}%
  \BibitemOpen
  \bibfield  {author} {\bibinfo {author} {\bibfnamefont {S.}~\bibnamefont {P{\'e}ru}}, \bibinfo {author} {\bibfnamefont {M.}~\bibnamefont {Girod}},\ and\ \bibinfo {author} {\bibfnamefont {J.~F.}\ \bibnamefont {Berger}},\ }\bibfield  {title} {\bibinfo {title} {{Evolution of the N = 20 and N = 28 shell closures in neutron-rich nuclei}},\ }\href {https://doi.org/10.1007/s100500070053} {\bibfield  {journal} {\bibinfo  {journal} {The European Physical Journal A - Hadrons and Nuclei}\ }\textbf {\bibinfo {volume} {9}},\ \bibinfo {pages} {35} (\bibinfo {year} {2000})}\BibitemShut {NoStop}%
\bibitem [{\citenamefont {Li}\ \emph {et~al.}(2011)\citenamefont {Li}, \citenamefont {Yao}, \citenamefont {Vretenar}, \citenamefont {Nik\ifmmode \check{s}\else \v{s}\fi{}i\ifmmode~\acute{c}\else \'{c}\fi{}}, \citenamefont {Chen},\ and\ \citenamefont {Meng}}]{Li_PhysRevC.84.054304}%
  \BibitemOpen
  \bibfield  {author} {\bibinfo {author} {\bibfnamefont {Z.~P.}\ \bibnamefont {Li}}, \bibinfo {author} {\bibfnamefont {J.~M.}\ \bibnamefont {Yao}}, \bibinfo {author} {\bibfnamefont {D.}~\bibnamefont {Vretenar}}, \bibinfo {author} {\bibfnamefont {T.}~\bibnamefont {Nik\ifmmode \check{s}\else \v{s}\fi{}i\ifmmode~\acute{c}\else \'{c}\fi{}}}, \bibinfo {author} {\bibfnamefont {H.}~\bibnamefont {Chen}},\ and\ \bibinfo {author} {\bibfnamefont {J.}~\bibnamefont {Meng}},\ }\bibfield  {title} {\bibinfo {title} {{Energy density functional analysis of shape evolution in $N=28$ isotones}},\ }\href {https://doi.org/10.1103/PhysRevC.84.054304} {\bibfield  {journal} {\bibinfo  {journal} {Phys. Rev. C}\ }\textbf {\bibinfo {volume} {84}},\ \bibinfo {pages} {054304} (\bibinfo {year} {2011})}\BibitemShut {NoStop}%
\bibitem [{\citenamefont {Chevrier}\ and\ \citenamefont {Gaudefroy}(2014)}]{Chevrier_2014}%
  \BibitemOpen
  \bibfield  {author} {\bibinfo {author} {\bibfnamefont {R.}~\bibnamefont {Chevrier}}\ and\ \bibinfo {author} {\bibfnamefont {L.}~\bibnamefont {Gaudefroy}},\ }\bibfield  {title} {\bibinfo {title} {Shell model structure of ${}^{43}s$ and ${}^{44}s$ re-examined},\ }\href {https://doi.org/10.1103/PhysRevC.89.051301} {\bibfield  {journal} {\bibinfo  {journal} {Phys. Rev. C}\ }\textbf {\bibinfo {volume} {89}},\ \bibinfo {pages} {051301} (\bibinfo {year} {2014})}\BibitemShut {NoStop}%
\bibitem [{\citenamefont {Rodr\'{\i}guez}\ and\ \citenamefont {Egido}(2011)}]{Rodriguez_PhysRevC.84.051307}%
  \BibitemOpen
  \bibfield  {author} {\bibinfo {author} {\bibfnamefont {T.~R.}\ \bibnamefont {Rodr\'{\i}guez}}\ and\ \bibinfo {author} {\bibfnamefont {J.~L.}\ \bibnamefont {Egido}},\ }\bibfield  {title} {\bibinfo {title} {{Configuration mixing description of the nucleus ${}^{44}$S}},\ }\href {https://doi.org/10.1103/PhysRevC.84.051307} {\bibfield  {journal} {\bibinfo  {journal} {Phys. Rev. C}\ }\textbf {\bibinfo {volume} {84}},\ \bibinfo {pages} {051307} (\bibinfo {year} {2011})}\BibitemShut {NoStop}%
\bibitem [{\citenamefont {Suzuki}\ and\ \citenamefont {Kimura}(2021)}]{suzuki_kimura_2021}%
  \BibitemOpen
  \bibfield  {author} {\bibinfo {author} {\bibfnamefont {Y.}~\bibnamefont {Suzuki}}\ and\ \bibinfo {author} {\bibfnamefont {M.}~\bibnamefont {Kimura}},\ }\href {https://doi.org/10.1103/PhysRevC.104.024327} {\bibfield  {journal} {\bibinfo  {journal} {Phys. Rev. C}\ }\textbf {\bibinfo {volume} {104}},\ \bibinfo {pages} {024327} (\bibinfo {year} {2021})}\BibitemShut {NoStop}%
\bibitem [{\citenamefont {Suzuki}\ \emph {et~al.}(2022)\citenamefont {Suzuki}, \citenamefont {Horiuchi},\ and\ \citenamefont {Kimura}}]{suzuki_2022}%
  \BibitemOpen
  \bibfield  {author} {\bibinfo {author} {\bibfnamefont {Y.}~\bibnamefont {Suzuki}}, \bibinfo {author} {\bibfnamefont {W.}~\bibnamefont {Horiuchi}},\ and\ \bibinfo {author} {\bibfnamefont {M.}~\bibnamefont {Kimura}},\ }\bibfield  {title} {\bibinfo {title} {{Erosion of N = 28 shell closure: Shape coexistence and monopole transition}},\ }\href {https://doi.org/10.1093/ptep/ptac071} {\bibfield  {journal} {\bibinfo  {journal} {Progress of Theoretical and Experimental Physics}\ }\textbf {\bibinfo {volume} {2022}},\ \bibinfo {pages} {063D02} (\bibinfo {year} {2022})},\ \Eprint {https://arxiv.org/abs/https://academic.oup.com/ptep/article-pdf/2022/6/063D02/44062687/ptac071.pdf} {https://academic.oup.com/ptep/article-pdf/2022/6/063D02/44062687/ptac071.pdf} \BibitemShut {NoStop}%
\bibitem [{\citenamefont {Heyde}\ and\ \citenamefont {Meyer}(1988)}]{Heyde_1988}%
  \BibitemOpen
  \bibfield  {author} {\bibinfo {author} {\bibfnamefont {K.}~\bibnamefont {Heyde}}\ and\ \bibinfo {author} {\bibfnamefont {R.~A.}\ \bibnamefont {Meyer}},\ }\bibfield  {title} {\bibinfo {title} {Monopole strength as a measure of nuclear shape mixing},\ }\href {https://doi.org/10.1103/PhysRevC.37.2170} {\bibfield  {journal} {\bibinfo  {journal} {Phys. Rev. C}\ }\textbf {\bibinfo {volume} {37}},\ \bibinfo {pages} {2170} (\bibinfo {year} {1988})}\BibitemShut {NoStop}%
\bibitem [{\citenamefont {Wood}\ \emph {et~al.}(1992{\natexlab{a}})\citenamefont {Wood}, \citenamefont {Heyde}, \citenamefont {Nazarewicz}, \citenamefont {Huyse},\ and\ \citenamefont {{van Duppen}}}]{WOOD1992101}%
  \BibitemOpen
  \bibfield  {author} {\bibinfo {author} {\bibfnamefont {J.}~\bibnamefont {Wood}}, \bibinfo {author} {\bibfnamefont {K.}~\bibnamefont {Heyde}}, \bibinfo {author} {\bibfnamefont {W.}~\bibnamefont {Nazarewicz}}, \bibinfo {author} {\bibfnamefont {M.}~\bibnamefont {Huyse}},\ and\ \bibinfo {author} {\bibfnamefont {P.}~\bibnamefont {{van Duppen}}},\ }\bibfield  {title} {\bibinfo {title} {Coexistence in even-mass nuclei},\ }\href {https://doi.org/https://doi.org/10.1016/0370-1573(92)90095-H} {\bibfield  {journal} {\bibinfo  {journal} {Physics Reports}\ }\textbf {\bibinfo {volume} {215}},\ \bibinfo {pages} {101} (\bibinfo {year} {1992}{\natexlab{a}})}\BibitemShut {NoStop}%
\bibitem [{\citenamefont {Heyde}\ and\ \citenamefont {Wood}(2011)}]{Heyde_2011}%
  \BibitemOpen
  \bibfield  {author} {\bibinfo {author} {\bibfnamefont {K.}~\bibnamefont {Heyde}}\ and\ \bibinfo {author} {\bibfnamefont {J.~L.}\ \bibnamefont {Wood}},\ }\bibfield  {title} {\bibinfo {title} {Shape coexistence in atomic nuclei},\ }\href {https://doi.org/10.1103/RevModPhys.83.1467} {\bibfield  {journal} {\bibinfo  {journal} {Rev. Mod. Phys.}\ }\textbf {\bibinfo {volume} {83}},\ \bibinfo {pages} {1467} (\bibinfo {year} {2011})}\BibitemShut {NoStop}%
\bibitem [{\citenamefont {Otsuka}\ \emph {et~al.}(2020)\citenamefont {Otsuka}, \citenamefont {Gade}, \citenamefont {Sorlin}, \citenamefont {Suzuki},\ and\ \citenamefont {Utsuno}}]{Otsuka_2020}%
  \BibitemOpen
  \bibfield  {author} {\bibinfo {author} {\bibfnamefont {T.}~\bibnamefont {Otsuka}}, \bibinfo {author} {\bibfnamefont {A.}~\bibnamefont {Gade}}, \bibinfo {author} {\bibfnamefont {O.}~\bibnamefont {Sorlin}}, \bibinfo {author} {\bibfnamefont {T.}~\bibnamefont {Suzuki}},\ and\ \bibinfo {author} {\bibfnamefont {Y.}~\bibnamefont {Utsuno}},\ }\bibfield  {title} {\bibinfo {title} {Evolution of shell structure in exotic nuclei},\ }\href {https://doi.org/10.1103/RevModPhys.92.015002} {\bibfield  {journal} {\bibinfo  {journal} {Rev. Mod. Phys.}\ }\textbf {\bibinfo {volume} {92}},\ \bibinfo {pages} {015002} (\bibinfo {year} {2020})}\BibitemShut {NoStop}%
\bibitem [{\citenamefont {Garrett}\ \emph {et~al.}(2022)\citenamefont {Garrett}, \citenamefont {Zielińska},\ and\ \citenamefont {Clément}}]{GARRETT2022103931}%
  \BibitemOpen
  \bibfield  {author} {\bibinfo {author} {\bibfnamefont {P.~E.}\ \bibnamefont {Garrett}}, \bibinfo {author} {\bibfnamefont {M.}~\bibnamefont {Zielińska}},\ and\ \bibinfo {author} {\bibfnamefont {E.}~\bibnamefont {Clément}},\ }\bibfield  {title} {\bibinfo {title} {An experimental view on shape coexistence in nuclei},\ }\href {https://doi.org/https://doi.org/10.1016/j.ppnp.2021.103931} {\bibfield  {journal} {\bibinfo  {journal} {Progress in Particle and Nuclear Physics}\ }\textbf {\bibinfo {volume} {124}},\ \bibinfo {pages} {103931} (\bibinfo {year} {2022})}\BibitemShut {NoStop}%
\bibitem [{\citenamefont {Chant}\ and\ \citenamefont {Roos}(1977)}]{NSChant77}%
  \BibitemOpen
  \bibfield  {author} {\bibinfo {author} {\bibfnamefont {N.~S.}\ \bibnamefont {Chant}}\ and\ \bibinfo {author} {\bibfnamefont {P.~G.}\ \bibnamefont {Roos}},\ }\bibfield  {title} {\bibinfo {title} {Distorted-wave impulse-approximation calculations for quasifree cluster knockout reactions},\ }\href {https://doi.org/10.1103/PhysRevC.15.57} {\bibfield  {journal} {\bibinfo  {journal} {Phys. Rev. C}\ }\textbf {\bibinfo {volume} {15}},\ \bibinfo {pages} {57} (\bibinfo {year} {1977})}\BibitemShut {NoStop}%
\bibitem [{\citenamefont {Chant}\ and\ \citenamefont {Roos}(1983)}]{NSChant83}%
  \BibitemOpen
  \bibfield  {author} {\bibinfo {author} {\bibfnamefont {N.~S.}\ \bibnamefont {Chant}}\ and\ \bibinfo {author} {\bibfnamefont {P.~G.}\ \bibnamefont {Roos}},\ }\bibfield  {title} {\bibinfo {title} {Spin orbit effects in quasifree knockout reactions},\ }\href {https://doi.org/10.1103/PhysRevC.27.1060} {\bibfield  {journal} {\bibinfo  {journal} {Phys. Rev. C}\ }\textbf {\bibinfo {volume} {27}},\ \bibinfo {pages} {1060} (\bibinfo {year} {1983})}\BibitemShut {NoStop}%
\bibitem [{\citenamefont {Wakasa}\ \emph {et~al.}(2017)\citenamefont {Wakasa}, \citenamefont {Ogata},\ and\ \citenamefont {Noro}}]{TWakasa17}%
  \BibitemOpen
  \bibfield  {author} {\bibinfo {author} {\bibfnamefont {T.}~\bibnamefont {Wakasa}}, \bibinfo {author} {\bibfnamefont {K.}~\bibnamefont {Ogata}},\ and\ \bibinfo {author} {\bibfnamefont {T.}~\bibnamefont {Noro}},\ }\bibfield  {title} {\bibinfo {title} {Proton-induced knockout reactions with polarized and unpolarized beams},\ }\href {https://doi.org/10.1016/j.ppnp.2017.06.002} {\bibfield  {journal} {\bibinfo  {journal} {Prog. Part. Nucl. Phys.}\ }\textbf {\bibinfo {volume} {96}},\ \bibinfo {pages} {32} (\bibinfo {year} {2017})}\BibitemShut {NoStop}%
\bibitem [{\citenamefont {Ogata}\ \emph {et~al.}(2024)\citenamefont {Ogata}, \citenamefont {Yoshida},\ and\ \citenamefont {Chazono}}]{KOgata24}%
  \BibitemOpen
  \bibfield  {author} {\bibinfo {author} {\bibfnamefont {K.}~\bibnamefont {Ogata}}, \bibinfo {author} {\bibfnamefont {K.}~\bibnamefont {Yoshida}},\ and\ \bibinfo {author} {\bibfnamefont {Y.}~\bibnamefont {Chazono}},\ }\bibfield  {title} {\bibinfo {title} {\textsc{pikoe}: A computer program for distorted-wave impulse approximation calculation for proton induced nucleon knockout reactions},\ }\href {https://doi.org/10.1016/j.cpc.2023.109058} {\bibfield  {journal} {\bibinfo  {journal} {Comp. Phys. Comm.}\ }\textbf {\bibinfo {volume} {297}},\ \bibinfo {pages} {109058} (\bibinfo {year} {2024})}\BibitemShut {NoStop}%
\bibitem [{\citenamefont {Barman}\ \emph {et~al.}(2025)\citenamefont {Barman}, \citenamefont {Horiuchi}, \citenamefont {Kimura},\ and\ \citenamefont {Chatterjee}}]{Barman_PhysRevC.111.064305}%
  \BibitemOpen
  \bibfield  {author} {\bibinfo {author} {\bibfnamefont {R.}~\bibnamefont {Barman}}, \bibinfo {author} {\bibfnamefont {W.}~\bibnamefont {Horiuchi}}, \bibinfo {author} {\bibfnamefont {M.}~\bibnamefont {Kimura}},\ and\ \bibinfo {author} {\bibfnamefont {R.}~\bibnamefont {Chatterjee}},\ }\bibfield  {title} {\bibinfo {title} {Investigating nuclear density profiles to reveal particle-hole configurations in the island of inversion},\ }\href {https://doi.org/10.1103/PhysRevC.111.064305} {\bibfield  {journal} {\bibinfo  {journal} {Phys. Rev. C}\ }\textbf {\bibinfo {volume} {111}},\ \bibinfo {pages} {064305} (\bibinfo {year} {2025})}\BibitemShut {NoStop}%
\bibitem [{\citenamefont {Berger}\ \emph {et~al.}(1991)\citenamefont {Berger}, \citenamefont {Girod},\ and\ \citenamefont {Gogny}}]{BERGER1991365}%
  \BibitemOpen
  \bibfield  {author} {\bibinfo {author} {\bibfnamefont {J.}~\bibnamefont {Berger}}, \bibinfo {author} {\bibfnamefont {M.}~\bibnamefont {Girod}},\ and\ \bibinfo {author} {\bibfnamefont {D.}~\bibnamefont {Gogny}},\ }\href {https://doi.org/https://doi.org/10.1016/0010-4655(91)90263-K} {\bibfield  {journal} {\bibinfo  {journal} {Computer Physics Communications}\ }\textbf {\bibinfo {volume} {63}},\ \bibinfo {pages} {365} (\bibinfo {year} {1991})}\BibitemShut {NoStop}%
\bibitem [{\citenamefont {Goriely}\ \emph {et~al.}(2009)\citenamefont {Goriely}, \citenamefont {Hilaire}, \citenamefont {Girod},\ and\ \citenamefont {P\'eru}}]{Goriely_2009}%
  \BibitemOpen
  \bibfield  {author} {\bibinfo {author} {\bibfnamefont {S.}~\bibnamefont {Goriely}}, \bibinfo {author} {\bibfnamefont {S.}~\bibnamefont {Hilaire}}, \bibinfo {author} {\bibfnamefont {M.}~\bibnamefont {Girod}},\ and\ \bibinfo {author} {\bibfnamefont {S.}~\bibnamefont {P\'eru}},\ }\bibfield  {title} {\bibinfo {title} {{First Gogny-Hartree-Fock-Bogoliubov Nuclear Mass Model}},\ }\href {https://doi.org/10.1103/PhysRevLett.102.242501} {\bibfield  {journal} {\bibinfo  {journal} {Phys. Rev. Lett.}\ }\textbf {\bibinfo {volume} {102}},\ \bibinfo {pages} {242501} (\bibinfo {year} {2009})}\BibitemShut {NoStop}%
\bibitem [{\citenamefont {Hill}\ and\ \citenamefont {Wheeler}(1953)}]{hw1953}%
  \BibitemOpen
  \bibfield  {author} {\bibinfo {author} {\bibfnamefont {D.~L.}\ \bibnamefont {Hill}}\ and\ \bibinfo {author} {\bibfnamefont {J.~A.}\ \bibnamefont {Wheeler}},\ }\href {https://doi.org/10.1103/PhysRev.89.1102} {\bibfield  {journal} {\bibinfo  {journal} {Phys. Rev.}\ }\textbf {\bibinfo {volume} {89}},\ \bibinfo {pages} {1102} (\bibinfo {year} {1953})}\BibitemShut {NoStop}%
\bibitem [{\citenamefont {Longfellow}\ \emph {et~al.}(2021)\citenamefont {Longfellow}, \citenamefont {Weisshaar}, \citenamefont {Gade}, \citenamefont {Brown}, \citenamefont {Bazin}, \citenamefont {Brown}, \citenamefont {Elman}, \citenamefont {Pereira}, \citenamefont {Rhodes},\ and\ \citenamefont {Spieker}}]{Longfellow2021_PRC}%
  \BibitemOpen
  \bibfield  {author} {\bibinfo {author} {\bibfnamefont {B.}~\bibnamefont {Longfellow}}, \bibinfo {author} {\bibfnamefont {D.}~\bibnamefont {Weisshaar}}, \bibinfo {author} {\bibfnamefont {A.}~\bibnamefont {Gade}}, \bibinfo {author} {\bibfnamefont {B.~A.}\ \bibnamefont {Brown}}, \bibinfo {author} {\bibfnamefont {D.}~\bibnamefont {Bazin}}, \bibinfo {author} {\bibfnamefont {K.~W.}\ \bibnamefont {Brown}}, \bibinfo {author} {\bibfnamefont {B.}~\bibnamefont {Elman}}, \bibinfo {author} {\bibfnamefont {J.}~\bibnamefont {Pereira}}, \bibinfo {author} {\bibfnamefont {D.}~\bibnamefont {Rhodes}},\ and\ \bibinfo {author} {\bibfnamefont {M.}~\bibnamefont {Spieker}},\ }\bibfield  {title} {\bibinfo {title} {Quadrupole collectivity in the neutron-rich sulfur isotopes},\ }\href {https://doi.org/10.1103/PhysRevC.103.054309} {\bibfield  {journal} {\bibinfo  {journal} {Phys. Rev. C}\ }\textbf {\bibinfo {volume} {103}},\ \bibinfo {pages} {054309} (\bibinfo {year} {2021})}\BibitemShut {NoStop}%
\bibitem [{\citenamefont {Wood}\ \emph {et~al.}(1992{\natexlab{b}})\citenamefont {Wood}, \citenamefont {Heyde}, \citenamefont {Nazarewicz}, \citenamefont {Huyse},\ and\ \citenamefont {{van Duppen}}}]{Wood1992}%
  \BibitemOpen
  \bibfield  {author} {\bibinfo {author} {\bibfnamefont {J.}~\bibnamefont {Wood}}, \bibinfo {author} {\bibfnamefont {K.}~\bibnamefont {Heyde}}, \bibinfo {author} {\bibfnamefont {W.}~\bibnamefont {Nazarewicz}}, \bibinfo {author} {\bibfnamefont {M.}~\bibnamefont {Huyse}},\ and\ \bibinfo {author} {\bibfnamefont {P.}~\bibnamefont {{van Duppen}}},\ }\bibfield  {title} {\bibinfo {title} {Coexistence in even-mass nuclei},\ }\href {https://doi.org/https://doi.org/10.1016/0370-1573(92)90095-H} {\bibfield  {journal} {\bibinfo  {journal} {Physics Reports}\ }\textbf {\bibinfo {volume} {215}},\ \bibinfo {pages} {101} (\bibinfo {year} {1992}{\natexlab{b}})}\BibitemShut {NoStop}%
\bibitem [{\citenamefont {Momiyama}\ \emph {et~al.}(2020)\citenamefont {Momiyama}, \citenamefont {Wimmer}, \citenamefont {Bazin}, \citenamefont {Belarge}, \citenamefont {Bender}, \citenamefont {Elman}, \citenamefont {Gade}, \citenamefont {Kemper}, \citenamefont {Kitamura}, \citenamefont {Longfellow}, \citenamefont {Lunderberg}, \citenamefont {Niikura}, \citenamefont {Ota}, \citenamefont {Schrock}, \citenamefont {Tostevin},\ and\ \citenamefont {Weisshaar}}]{Momiyama_43S}%
  \BibitemOpen
  \bibfield  {author} {\bibinfo {author} {\bibfnamefont {S.}~\bibnamefont {Momiyama}}, \bibinfo {author} {\bibfnamefont {K.}~\bibnamefont {Wimmer}}, \bibinfo {author} {\bibfnamefont {D.}~\bibnamefont {Bazin}}, \bibinfo {author} {\bibfnamefont {J.}~\bibnamefont {Belarge}}, \bibinfo {author} {\bibfnamefont {P.}~\bibnamefont {Bender}}, \bibinfo {author} {\bibfnamefont {B.}~\bibnamefont {Elman}}, \bibinfo {author} {\bibfnamefont {A.}~\bibnamefont {Gade}}, \bibinfo {author} {\bibfnamefont {K.~W.}\ \bibnamefont {Kemper}}, \bibinfo {author} {\bibfnamefont {N.}~\bibnamefont {Kitamura}}, \bibinfo {author} {\bibfnamefont {B.}~\bibnamefont {Longfellow}}, \bibinfo {author} {\bibfnamefont {E.}~\bibnamefont {Lunderberg}}, \bibinfo {author} {\bibfnamefont {M.}~\bibnamefont {Niikura}}, \bibinfo {author} {\bibfnamefont {S.}~\bibnamefont {Ota}}, \bibinfo {author} {\bibfnamefont {P.}~\bibnamefont {Schrock}}, \bibinfo {author} {\bibfnamefont {J.~A.}\ \bibnamefont {Tostevin}},\ and\ \bibinfo {author} {\bibfnamefont
  {D.}~\bibnamefont {Weisshaar}},\ }\bibfield  {title} {\bibinfo {title} {{Shell structure of $^{43}\mathrm{S}$ and collapse of the $N=28$ shell closure}},\ }\href {https://doi.org/10.1103/PhysRevC.102.034325} {\bibfield  {journal} {\bibinfo  {journal} {Phys. Rev. C}\ }\textbf {\bibinfo {volume} {102}},\ \bibinfo {pages} {034325} (\bibinfo {year} {2020})}\BibitemShut {NoStop}%
\bibitem [{\citenamefont {Kimura}\ \emph {et~al.}(2016)\citenamefont {Kimura}, \citenamefont {Suhara}, \citenamefont {Horiuchi},\ and\ \citenamefont {Kanada-En’yo}}]{Kimura2016}%
  \BibitemOpen
  \bibfield  {author} {\bibinfo {author} {\bibfnamefont {M.}~\bibnamefont {Kimura}}, \bibinfo {author} {\bibfnamefont {T.}~\bibnamefont {Suhara}}, \bibinfo {author} {\bibfnamefont {W.}~\bibnamefont {Horiuchi}},\ and\ \bibinfo {author} {\bibfnamefont {Y.}~\bibnamefont {Kanada-En’yo}},\ }\href {https://doi.org/10.1140/epja/i2016-16373-9} {\bibfield  {journal} {\bibinfo  {journal} {Eur. Phys. J. A}\ }\textbf {\bibinfo {volume} {52}},\ \bibinfo {pages} {373} (\bibinfo {year} {2016})}\BibitemShut {NoStop}%
\bibitem [{\citenamefont {Franey}\ and\ \citenamefont {Love}(1985)}]{MAFraney85}%
  \BibitemOpen
  \bibfield  {author} {\bibinfo {author} {\bibfnamefont {M.~A.}\ \bibnamefont {Franey}}\ and\ \bibinfo {author} {\bibfnamefont {W.~G.}\ \bibnamefont {Love}},\ }\bibfield  {title} {\bibinfo {title} {Nucleon-nucleon t-matrix interaction for scattering at intermediate energies},\ }\href {https://doi.org/10.1103/PhysRevC.31.488} {\bibfield  {journal} {\bibinfo  {journal} {Phys. Rev. C}\ }\textbf {\bibinfo {volume} {31}},\ \bibinfo {pages} {488} (\bibinfo {year} {1985})}\BibitemShut {NoStop}%
\bibitem [{\citenamefont {Hama}\ \emph {et~al.}(1990)\citenamefont {Hama}, \citenamefont {Clark}, \citenamefont {Cooper}, \citenamefont {Sherif},\ and\ \citenamefont {Mercer}}]{SHama90}%
  \BibitemOpen
  \bibfield  {author} {\bibinfo {author} {\bibfnamefont {S.}~\bibnamefont {Hama}}, \bibinfo {author} {\bibfnamefont {B.~C.}\ \bibnamefont {Clark}}, \bibinfo {author} {\bibfnamefont {E.~D.}\ \bibnamefont {Cooper}}, \bibinfo {author} {\bibfnamefont {H.~S.}\ \bibnamefont {Sherif}},\ and\ \bibinfo {author} {\bibfnamefont {R.~L.}\ \bibnamefont {Mercer}},\ }\bibfield  {title} {\bibinfo {title} {Global dirac optical potentials for elastic proton scattering from heavy nuclei},\ }\href {https://doi.org/10.1103/PhysRevC.41.2737} {\bibfield  {journal} {\bibinfo  {journal} {Phys. Rev. C}\ }\textbf {\bibinfo {volume} {41}},\ \bibinfo {pages} {2737} (\bibinfo {year} {1990})}\BibitemShut {NoStop}%
\bibitem [{\citenamefont {Cooper}\ \emph {et~al.}(1993)\citenamefont {Cooper}, \citenamefont {Hama}, \citenamefont {Clark},\ and\ \citenamefont {Mercer}}]{EDCooper93}%
  \BibitemOpen
  \bibfield  {author} {\bibinfo {author} {\bibfnamefont {E.~D.}\ \bibnamefont {Cooper}}, \bibinfo {author} {\bibfnamefont {S.}~\bibnamefont {Hama}}, \bibinfo {author} {\bibfnamefont {B.~C.}\ \bibnamefont {Clark}},\ and\ \bibinfo {author} {\bibfnamefont {R.~L.}\ \bibnamefont {Mercer}},\ }\bibfield  {title} {\bibinfo {title} {Global dirac phenomenology for proton-nucleus elastic scattering},\ }\href {https://doi.org/10.1103/PhysRevC.47.297} {\bibfield  {journal} {\bibinfo  {journal} {Phys. Rev. C}\ }\textbf {\bibinfo {volume} {47}},\ \bibinfo {pages} {297} (\bibinfo {year} {1993})}\BibitemShut {NoStop}%
\bibitem [{\citenamefont {Cooper}\ \emph {et~al.}(2009)\citenamefont {Cooper}, \citenamefont {Hama},\ and\ \citenamefont {Clark}}]{EDCooper09}%
  \BibitemOpen
  \bibfield  {author} {\bibinfo {author} {\bibfnamefont {E.~D.}\ \bibnamefont {Cooper}}, \bibinfo {author} {\bibfnamefont {S.}~\bibnamefont {Hama}},\ and\ \bibinfo {author} {\bibfnamefont {B.~C.}\ \bibnamefont {Clark}},\ }\bibfield  {title} {\bibinfo {title} {Global dirac optical potential from helium to lead},\ }\href {https://doi.org/10.1103/PhysRevC.80.034605} {\bibfield  {journal} {\bibinfo  {journal} {Phys. Rev. C}\ }\textbf {\bibinfo {volume} {80}},\ \bibinfo {pages} {034605} (\bibinfo {year} {2009})}\BibitemShut {NoStop}%
\bibitem [{\citenamefont {Chen}\ \emph {et~al.}(2019)\citenamefont {Chen}, \citenamefont {Lee}, \citenamefont {Doornenbal}, \citenamefont {Obertelli}, \citenamefont {Barbieri}, \citenamefont {Chazono}, \citenamefont {Navr\'atil}, \citenamefont {Ogata}, \citenamefont {Otsuka}, \citenamefont {Raimondi}, \citenamefont {Som\`a}, \citenamefont {Utsuno}, \citenamefont {Yoshida}, \citenamefont {Baba}, \citenamefont {Browne}, \citenamefont {Calvet}, \citenamefont {Ch\^ateau}, \citenamefont {Chiga}, \citenamefont {Corsi}, \citenamefont {Cort\'es}, \citenamefont {Delbart}, \citenamefont {Gheller}, \citenamefont {Giganon}, \citenamefont {Gillibert}, \citenamefont {Hilaire}, \citenamefont {Isobe}, \citenamefont {Kahlbow}, \citenamefont {Kobayashi}, \citenamefont {Kubota}, \citenamefont {Lapoux}, \citenamefont {Liu}, \citenamefont {Motobayashi}, \citenamefont {Murray}, \citenamefont {Otsu}, \citenamefont {Panin}, \citenamefont {Paul}, \citenamefont {Rodriguez}, \citenamefont {Sakurai}, \citenamefont {Sasano},
  \citenamefont {Steppenbeck}, \citenamefont {Stuhl}, \citenamefont {Sun}, \citenamefont {Togano}, \citenamefont {Uesaka}, \citenamefont {Wimmer}, \citenamefont {Yoneda}, \citenamefont {Achouri}, \citenamefont {Aktas}, \citenamefont {Aumann}, \citenamefont {Chung}, \citenamefont {Flavigny}, \citenamefont {Franchoo}, \citenamefont {Ga\ifmmode \check{s}\else \v{s}\fi{}pari\ifmmode~\acute{c}\else \'{c}\fi{}}, \citenamefont {Gerst}, \citenamefont {Gibelin}, \citenamefont {Hahn}, \citenamefont {Kim}, \citenamefont {Koiwai}, \citenamefont {Kondo}, \citenamefont {Koseoglou}, \citenamefont {Lehr}, \citenamefont {Linh}, \citenamefont {Lokotko}, \citenamefont {MacCormick}, \citenamefont {Moschner}, \citenamefont {Nakamura}, \citenamefont {Park}, \citenamefont {Rossi}, \citenamefont {Sahin}, \citenamefont {Sohler}, \citenamefont {S\"oderstr\"om}, \citenamefont {Takeuchi}, \citenamefont {T\"ornqvist}, \citenamefont {Vaquero}, \citenamefont {Wagner}, \citenamefont {Wang}, \citenamefont {Werner}, \citenamefont {Xu},
  \citenamefont {Yamada}, \citenamefont {Yan}, \citenamefont {Yang}, \citenamefont {Yasuda},\ and\ \citenamefont {Zanetti}}]{SChen19}%
  \BibitemOpen
  \bibfield  {author} {\bibinfo {author} {\bibfnamefont {S.}~\bibnamefont {Chen}}, \bibinfo {author} {\bibfnamefont {J.}~\bibnamefont {Lee}}, \bibinfo {author} {\bibfnamefont {P.}~\bibnamefont {Doornenbal}}, \bibinfo {author} {\bibfnamefont {A.}~\bibnamefont {Obertelli}}, \bibinfo {author} {\bibfnamefont {C.}~\bibnamefont {Barbieri}}, \bibinfo {author} {\bibfnamefont {Y.}~\bibnamefont {Chazono}}, \bibinfo {author} {\bibfnamefont {P.}~\bibnamefont {Navr\'atil}}, \bibinfo {author} {\bibfnamefont {K.}~\bibnamefont {Ogata}}, \bibinfo {author} {\bibfnamefont {T.}~\bibnamefont {Otsuka}}, \bibinfo {author} {\bibfnamefont {F.}~\bibnamefont {Raimondi}}, \bibinfo {author} {\bibfnamefont {V.}~\bibnamefont {Som\`a}}, \bibinfo {author} {\bibfnamefont {Y.}~\bibnamefont {Utsuno}}, \bibinfo {author} {\bibfnamefont {K.}~\bibnamefont {Yoshida}}, \bibinfo {author} {\bibfnamefont {H.}~\bibnamefont {Baba}}, \bibinfo {author} {\bibfnamefont {F.}~\bibnamefont {Browne}}, \bibinfo {author} {\bibfnamefont {D.}~\bibnamefont {Calvet}},
  \bibinfo {author} {\bibfnamefont {F.}~\bibnamefont {Ch\^ateau}}, \bibinfo {author} {\bibfnamefont {N.}~\bibnamefont {Chiga}}, \bibinfo {author} {\bibfnamefont {A.}~\bibnamefont {Corsi}}, \bibinfo {author} {\bibfnamefont {M.~L.}\ \bibnamefont {Cort\'es}}, \bibinfo {author} {\bibfnamefont {A.}~\bibnamefont {Delbart}}, \bibinfo {author} {\bibfnamefont {J.-M.}\ \bibnamefont {Gheller}}, \bibinfo {author} {\bibfnamefont {A.}~\bibnamefont {Giganon}}, \bibinfo {author} {\bibfnamefont {A.}~\bibnamefont {Gillibert}}, \bibinfo {author} {\bibfnamefont {C.}~\bibnamefont {Hilaire}}, \bibinfo {author} {\bibfnamefont {T.}~\bibnamefont {Isobe}}, \bibinfo {author} {\bibfnamefont {J.}~\bibnamefont {Kahlbow}}, \bibinfo {author} {\bibfnamefont {T.}~\bibnamefont {Kobayashi}}, \bibinfo {author} {\bibfnamefont {Y.}~\bibnamefont {Kubota}}, \bibinfo {author} {\bibfnamefont {V.}~\bibnamefont {Lapoux}}, \bibinfo {author} {\bibfnamefont {H.~N.}\ \bibnamefont {Liu}}, \bibinfo {author} {\bibfnamefont {T.}~\bibnamefont {Motobayashi}},
  \bibinfo {author} {\bibfnamefont {I.}~\bibnamefont {Murray}}, \bibinfo {author} {\bibfnamefont {H.}~\bibnamefont {Otsu}}, \bibinfo {author} {\bibfnamefont {V.}~\bibnamefont {Panin}}, \bibinfo {author} {\bibfnamefont {N.}~\bibnamefont {Paul}}, \bibinfo {author} {\bibfnamefont {W.}~\bibnamefont {Rodriguez}}, \bibinfo {author} {\bibfnamefont {H.}~\bibnamefont {Sakurai}}, \bibinfo {author} {\bibfnamefont {M.}~\bibnamefont {Sasano}}, \bibinfo {author} {\bibfnamefont {D.}~\bibnamefont {Steppenbeck}}, \bibinfo {author} {\bibfnamefont {L.}~\bibnamefont {Stuhl}}, \bibinfo {author} {\bibfnamefont {Y.~L.}\ \bibnamefont {Sun}}, \bibinfo {author} {\bibfnamefont {Y.}~\bibnamefont {Togano}}, \bibinfo {author} {\bibfnamefont {T.}~\bibnamefont {Uesaka}}, \bibinfo {author} {\bibfnamefont {K.}~\bibnamefont {Wimmer}}, \bibinfo {author} {\bibfnamefont {K.}~\bibnamefont {Yoneda}}, \bibinfo {author} {\bibfnamefont {N.}~\bibnamefont {Achouri}}, \bibinfo {author} {\bibfnamefont {O.}~\bibnamefont {Aktas}}, \bibinfo {author}
  {\bibfnamefont {T.}~\bibnamefont {Aumann}}, \bibinfo {author} {\bibfnamefont {L.~X.}\ \bibnamefont {Chung}}, \bibinfo {author} {\bibfnamefont {F.}~\bibnamefont {Flavigny}}, \bibinfo {author} {\bibfnamefont {S.}~\bibnamefont {Franchoo}}, \bibinfo {author} {\bibfnamefont {I.}~\bibnamefont {Ga\ifmmode \check{s}\else \v{s}\fi{}pari\ifmmode~\acute{c}\else \'{c}\fi{}}}, \bibinfo {author} {\bibfnamefont {R.-B.}\ \bibnamefont {Gerst}}, \bibinfo {author} {\bibfnamefont {J.}~\bibnamefont {Gibelin}}, \bibinfo {author} {\bibfnamefont {K.~I.}\ \bibnamefont {Hahn}}, \bibinfo {author} {\bibfnamefont {D.}~\bibnamefont {Kim}}, \bibinfo {author} {\bibfnamefont {T.}~\bibnamefont {Koiwai}}, \bibinfo {author} {\bibfnamefont {Y.}~\bibnamefont {Kondo}}, \bibinfo {author} {\bibfnamefont {P.}~\bibnamefont {Koseoglou}}, \bibinfo {author} {\bibfnamefont {C.}~\bibnamefont {Lehr}}, \bibinfo {author} {\bibfnamefont {B.~D.}\ \bibnamefont {Linh}}, \bibinfo {author} {\bibfnamefont {T.}~\bibnamefont {Lokotko}}, \bibinfo {author}
  {\bibfnamefont {M.}~\bibnamefont {MacCormick}}, \bibinfo {author} {\bibfnamefont {K.}~\bibnamefont {Moschner}}, \bibinfo {author} {\bibfnamefont {T.}~\bibnamefont {Nakamura}}, \bibinfo {author} {\bibfnamefont {S.~Y.}\ \bibnamefont {Park}}, \bibinfo {author} {\bibfnamefont {D.}~\bibnamefont {Rossi}}, \bibinfo {author} {\bibfnamefont {E.}~\bibnamefont {Sahin}}, \bibinfo {author} {\bibfnamefont {D.}~\bibnamefont {Sohler}}, \bibinfo {author} {\bibfnamefont {P.-A.}\ \bibnamefont {S\"oderstr\"om}}, \bibinfo {author} {\bibfnamefont {S.}~\bibnamefont {Takeuchi}}, \bibinfo {author} {\bibfnamefont {H.}~\bibnamefont {T\"ornqvist}}, \bibinfo {author} {\bibfnamefont {V.}~\bibnamefont {Vaquero}}, \bibinfo {author} {\bibfnamefont {V.}~\bibnamefont {Wagner}}, \bibinfo {author} {\bibfnamefont {S.}~\bibnamefont {Wang}}, \bibinfo {author} {\bibfnamefont {V.}~\bibnamefont {Werner}}, \bibinfo {author} {\bibfnamefont {X.}~\bibnamefont {Xu}}, \bibinfo {author} {\bibfnamefont {H.}~\bibnamefont {Yamada}}, \bibinfo {author}
  {\bibfnamefont {D.}~\bibnamefont {Yan}}, \bibinfo {author} {\bibfnamefont {Z.}~\bibnamefont {Yang}}, \bibinfo {author} {\bibfnamefont {M.}~\bibnamefont {Yasuda}},\ and\ \bibinfo {author} {\bibfnamefont {L.}~\bibnamefont {Zanetti}},\ }\bibfield  {title} {\bibinfo {title} {Quasifree neutron knockout from $^{54}\mathrm{Ca}$ corroborates arising ${N} = 34$ neutron magic number},\ }\href {https://doi.org/10.1103/PhysRevLett.123.142501} {\bibfield  {journal} {\bibinfo  {journal} {Phys. Rev. Lett.}\ }\textbf {\bibinfo {volume} {123}},\ \bibinfo {pages} {142501} (\bibinfo {year} {2019})}\BibitemShut {NoStop}%
\bibitem [{\citenamefont {Sun}\ \emph {et~al.}(2020)\citenamefont {Sun}, \citenamefont {Obertelli}, \citenamefont {Doornenbal}, \citenamefont {Barbieri}, \citenamefont {Chazono}, \citenamefont {Duguet}, \citenamefont {Liu}, \citenamefont {Navr\'atil}, \citenamefont {Nowacki}, \citenamefont {Ogata}, \citenamefont {Otsuka}, \citenamefont {Raimondi}, \citenamefont {Som\`a}, \citenamefont {Utsuno}, \citenamefont {Yoshida}, \citenamefont {Achouri}, \citenamefont {Baba}, \citenamefont {Browne}, \citenamefont {Calvet}, \citenamefont {Ch\^ateau}, \citenamefont {Chen}, \citenamefont {Chiga}, \citenamefont {Corsi}, \citenamefont {Cort\'es}, \citenamefont {Delbart}, \citenamefont {Gheller}, \citenamefont {Giganon}, \citenamefont {Gillibert}, \citenamefont {Hilaire}, \citenamefont {Isobe}, \citenamefont {Kobayashi}, \citenamefont {Kubota}, \citenamefont {Lapoux}, \citenamefont {Motobayashi}, \citenamefont {Murray}, \citenamefont {Otsu}, \citenamefont {Panin}, \citenamefont {Paul}, \citenamefont {Rodriguez}, \citenamefont
  {Sakurai}, \citenamefont {Sasano}, \citenamefont {Steppenbeck}, \citenamefont {Stuhl}, \citenamefont {Togano}, \citenamefont {Uesaka}, \citenamefont {Wimmer}, \citenamefont {Yoneda}, \citenamefont {Aktas}, \citenamefont {Aumann}, \citenamefont {Chung}, \citenamefont {Flavigny}, \citenamefont {Franchoo}, \citenamefont {Ga\ifmmode \check{s}\else \v{s}\fi{}pari\ifmmode~\acute{c}\else \'{c}\fi{}}, \citenamefont {Gerst}, \citenamefont {Gibelin}, \citenamefont {Hahn}, \citenamefont {Kim}, \citenamefont {Koiwai}, \citenamefont {Kondo}, \citenamefont {Koseoglou}, \citenamefont {Lee}, \citenamefont {Lehr}, \citenamefont {Linh}, \citenamefont {Lokotko}, \citenamefont {MacCormick}, \citenamefont {Moschner}, \citenamefont {Nakamura}, \citenamefont {Park}, \citenamefont {Rossi}, \citenamefont {Sahin}, \citenamefont {Sohler}, \citenamefont {S\"oderstr\"om}, \citenamefont {Takeuchi}, \citenamefont {T\"ornqvist}, \citenamefont {Vaquero}, \citenamefont {Wagner}, \citenamefont {Wang}, \citenamefont {Werner}, \citenamefont
  {Xu}, \citenamefont {Yamada}, \citenamefont {Yan}, \citenamefont {Yang}, \citenamefont {Yasuda},\ and\ \citenamefont {Zanetti}}]{YLSun20}%
  \BibitemOpen
  \bibfield  {author} {\bibinfo {author} {\bibfnamefont {Y.~L.}\ \bibnamefont {Sun}}, \bibinfo {author} {\bibfnamefont {A.}~\bibnamefont {Obertelli}}, \bibinfo {author} {\bibfnamefont {P.}~\bibnamefont {Doornenbal}}, \bibinfo {author} {\bibfnamefont {C.}~\bibnamefont {Barbieri}}, \bibinfo {author} {\bibfnamefont {Y.}~\bibnamefont {Chazono}}, \bibinfo {author} {\bibfnamefont {T.}~\bibnamefont {Duguet}}, \bibinfo {author} {\bibfnamefont {H.~N.}\ \bibnamefont {Liu}}, \bibinfo {author} {\bibfnamefont {P.}~\bibnamefont {Navr\'atil}}, \bibinfo {author} {\bibfnamefont {F.}~\bibnamefont {Nowacki}}, \bibinfo {author} {\bibfnamefont {K.}~\bibnamefont {Ogata}}, \bibinfo {author} {\bibfnamefont {T.}~\bibnamefont {Otsuka}}, \bibinfo {author} {\bibfnamefont {F.}~\bibnamefont {Raimondi}}, \bibinfo {author} {\bibfnamefont {V.}~\bibnamefont {Som\`a}}, \bibinfo {author} {\bibfnamefont {Y.}~\bibnamefont {Utsuno}}, \bibinfo {author} {\bibfnamefont {K.}~\bibnamefont {Yoshida}}, \bibinfo {author} {\bibfnamefont {N.}~\bibnamefont
  {Achouri}}, \bibinfo {author} {\bibfnamefont {H.}~\bibnamefont {Baba}}, \bibinfo {author} {\bibfnamefont {F.}~\bibnamefont {Browne}}, \bibinfo {author} {\bibfnamefont {D.}~\bibnamefont {Calvet}}, \bibinfo {author} {\bibfnamefont {F.}~\bibnamefont {Ch\^ateau}}, \bibinfo {author} {\bibfnamefont {S.}~\bibnamefont {Chen}}, \bibinfo {author} {\bibfnamefont {N.}~\bibnamefont {Chiga}}, \bibinfo {author} {\bibfnamefont {A.}~\bibnamefont {Corsi}}, \bibinfo {author} {\bibfnamefont {M.~L.}\ \bibnamefont {Cort\'es}}, \bibinfo {author} {\bibfnamefont {A.}~\bibnamefont {Delbart}}, \bibinfo {author} {\bibfnamefont {J.-M.}\ \bibnamefont {Gheller}}, \bibinfo {author} {\bibfnamefont {A.}~\bibnamefont {Giganon}}, \bibinfo {author} {\bibfnamefont {A.}~\bibnamefont {Gillibert}}, \bibinfo {author} {\bibfnamefont {C.}~\bibnamefont {Hilaire}}, \bibinfo {author} {\bibfnamefont {T.}~\bibnamefont {Isobe}}, \bibinfo {author} {\bibfnamefont {T.}~\bibnamefont {Kobayashi}}, \bibinfo {author} {\bibfnamefont {Y.}~\bibnamefont {Kubota}},
  \bibinfo {author} {\bibfnamefont {V.}~\bibnamefont {Lapoux}}, \bibinfo {author} {\bibfnamefont {T.}~\bibnamefont {Motobayashi}}, \bibinfo {author} {\bibfnamefont {I.}~\bibnamefont {Murray}}, \bibinfo {author} {\bibfnamefont {H.}~\bibnamefont {Otsu}}, \bibinfo {author} {\bibfnamefont {V.}~\bibnamefont {Panin}}, \bibinfo {author} {\bibfnamefont {N.}~\bibnamefont {Paul}}, \bibinfo {author} {\bibfnamefont {W.}~\bibnamefont {Rodriguez}}, \bibinfo {author} {\bibfnamefont {H.}~\bibnamefont {Sakurai}}, \bibinfo {author} {\bibfnamefont {M.}~\bibnamefont {Sasano}}, \bibinfo {author} {\bibfnamefont {D.}~\bibnamefont {Steppenbeck}}, \bibinfo {author} {\bibfnamefont {L.}~\bibnamefont {Stuhl}}, \bibinfo {author} {\bibfnamefont {Y.}~\bibnamefont {Togano}}, \bibinfo {author} {\bibfnamefont {T.}~\bibnamefont {Uesaka}}, \bibinfo {author} {\bibfnamefont {K.}~\bibnamefont {Wimmer}}, \bibinfo {author} {\bibfnamefont {K.}~\bibnamefont {Yoneda}}, \bibinfo {author} {\bibfnamefont {O.}~\bibnamefont {Aktas}}, \bibinfo {author}
  {\bibfnamefont {T.}~\bibnamefont {Aumann}}, \bibinfo {author} {\bibfnamefont {L.~X.}\ \bibnamefont {Chung}}, \bibinfo {author} {\bibfnamefont {F.}~\bibnamefont {Flavigny}}, \bibinfo {author} {\bibfnamefont {S.}~\bibnamefont {Franchoo}}, \bibinfo {author} {\bibfnamefont {I.}~\bibnamefont {Ga\ifmmode \check{s}\else \v{s}\fi{}pari\ifmmode~\acute{c}\else \'{c}\fi{}}}, \bibinfo {author} {\bibfnamefont {R.-B.}\ \bibnamefont {Gerst}}, \bibinfo {author} {\bibfnamefont {J.}~\bibnamefont {Gibelin}}, \bibinfo {author} {\bibfnamefont {K.~I.}\ \bibnamefont {Hahn}}, \bibinfo {author} {\bibfnamefont {D.}~\bibnamefont {Kim}}, \bibinfo {author} {\bibfnamefont {T.}~\bibnamefont {Koiwai}}, \bibinfo {author} {\bibfnamefont {Y.}~\bibnamefont {Kondo}}, \bibinfo {author} {\bibfnamefont {P.}~\bibnamefont {Koseoglou}}, \bibinfo {author} {\bibfnamefont {J.}~\bibnamefont {Lee}}, \bibinfo {author} {\bibfnamefont {C.}~\bibnamefont {Lehr}}, \bibinfo {author} {\bibfnamefont {B.~D.}\ \bibnamefont {Linh}}, \bibinfo {author} {\bibfnamefont
  {T.}~\bibnamefont {Lokotko}}, \bibinfo {author} {\bibfnamefont {M.}~\bibnamefont {MacCormick}}, \bibinfo {author} {\bibfnamefont {K.}~\bibnamefont {Moschner}}, \bibinfo {author} {\bibfnamefont {T.}~\bibnamefont {Nakamura}}, \bibinfo {author} {\bibfnamefont {S.~Y.}\ \bibnamefont {Park}}, \bibinfo {author} {\bibfnamefont {D.}~\bibnamefont {Rossi}}, \bibinfo {author} {\bibfnamefont {E.}~\bibnamefont {Sahin}}, \bibinfo {author} {\bibfnamefont {D.}~\bibnamefont {Sohler}}, \bibinfo {author} {\bibfnamefont {P.-A.}\ \bibnamefont {S\"oderstr\"om}}, \bibinfo {author} {\bibfnamefont {S.}~\bibnamefont {Takeuchi}}, \bibinfo {author} {\bibfnamefont {H.}~\bibnamefont {T\"ornqvist}}, \bibinfo {author} {\bibfnamefont {V.}~\bibnamefont {Vaquero}}, \bibinfo {author} {\bibfnamefont {V.}~\bibnamefont {Wagner}}, \bibinfo {author} {\bibfnamefont {S.}~\bibnamefont {Wang}}, \bibinfo {author} {\bibfnamefont {V.}~\bibnamefont {Werner}}, \bibinfo {author} {\bibfnamefont {X.}~\bibnamefont {Xu}}, \bibinfo {author} {\bibfnamefont
  {H.}~\bibnamefont {Yamada}}, \bibinfo {author} {\bibfnamefont {D.}~\bibnamefont {Yan}}, \bibinfo {author} {\bibfnamefont {Z.}~\bibnamefont {Yang}}, \bibinfo {author} {\bibfnamefont {M.}~\bibnamefont {Yasuda}},\ and\ \bibinfo {author} {\bibfnamefont {L.}~\bibnamefont {Zanetti}},\ }\bibfield  {title} {\bibinfo {title} {Restoration of the natural ${E}(1/2_1^+)$-${E}(3/2_1^+)$ energy splitting in odd-k isotopes towards ${N} = 40$},\ }\href {https://doi.org/10.1016/j.physletb.2020.135215} {\bibfield  {journal} {\bibinfo  {journal} {Phys. Lett. B}\ }\textbf {\bibinfo {volume} {802}},\ \bibinfo {pages} {135215} (\bibinfo {year} {2020})}\BibitemShut {NoStop}%
\bibitem [{\citenamefont {Linh}\ \emph {et~al.}(2021)\citenamefont {Linh}, \citenamefont {Corsi}, \citenamefont {Gillibert}, \citenamefont {Obertelli}, \citenamefont {Doornenbal}, \citenamefont {Barbieri}, \citenamefont {Chen}, \citenamefont {Chung}, \citenamefont {Duguet}, \citenamefont {G\'omez-Ramos}, \citenamefont {Holt}, \citenamefont {Moro}, \citenamefont {Navr\'atil}, \citenamefont {Ogata}, \citenamefont {Phuc}, \citenamefont {Shimizu}, \citenamefont {Som\`a}, \citenamefont {Utsuno}, \citenamefont {Achouri}, \citenamefont {Baba}, \citenamefont {Browne}, \citenamefont {Calvet}, \citenamefont {Ch\^ateau}, \citenamefont {Chiga}, \citenamefont {Cort\'es}, \citenamefont {Delbart}, \citenamefont {Gheller}, \citenamefont {Giganon}, \citenamefont {Hilaire}, \citenamefont {Isobe}, \citenamefont {Kobayashi}, \citenamefont {Kubota}, \citenamefont {Lapoux}, \citenamefont {Liu}, \citenamefont {Motobayashi}, \citenamefont {Murray}, \citenamefont {Otsu}, \citenamefont {Panin}, \citenamefont {Paul}, \citenamefont
  {Rodriguez}, \citenamefont {Sakurai}, \citenamefont {Sasano}, \citenamefont {Steppenbeck}, \citenamefont {Stuhl}, \citenamefont {Sun}, \citenamefont {Togano}, \citenamefont {Uesaka}, \citenamefont {Wimmer}, \citenamefont {Yoneda}, \citenamefont {Aktas}, \citenamefont {Aumann}, \citenamefont {Flavigny}, \citenamefont {Franchoo}, \citenamefont {Ga\ifmmode \check{s}\else \v{s}\fi{}pari\ifmmode~\acute{c}\else \'{c}\fi{}}, \citenamefont {Gerst}, \citenamefont {Gibelin}, \citenamefont {Hahn}, \citenamefont {Khai}, \citenamefont {Kim}, \citenamefont {Koiwai}, \citenamefont {Kondo}, \citenamefont {Koseoglou}, \citenamefont {Lee}, \citenamefont {Lehr}, \citenamefont {Lokotko}, \citenamefont {MacCormick}, \citenamefont {Moschner}, \citenamefont {Nakamura}, \citenamefont {Park}, \citenamefont {Rossi}, \citenamefont {Sahin}, \citenamefont {Sohler}, \citenamefont {S\"oderstr\"om}, \citenamefont {Takeuchi}, \citenamefont {Ton}, \citenamefont {T\"ornqvist}, \citenamefont {Vaquero}, \citenamefont {Wagner}, \citenamefont
  {Wang}, \citenamefont {Werner}, \citenamefont {Xu}, \citenamefont {Yamada}, \citenamefont {Yan}, \citenamefont {Yang}, \citenamefont {Yasuda},\ and\ \citenamefont {Zanetti}}]{BDLinh21}%
  \BibitemOpen
  \bibfield  {author} {\bibinfo {author} {\bibfnamefont {B.~D.}\ \bibnamefont {Linh}}, \bibinfo {author} {\bibfnamefont {A.}~\bibnamefont {Corsi}}, \bibinfo {author} {\bibfnamefont {A.}~\bibnamefont {Gillibert}}, \bibinfo {author} {\bibfnamefont {A.}~\bibnamefont {Obertelli}}, \bibinfo {author} {\bibfnamefont {P.}~\bibnamefont {Doornenbal}}, \bibinfo {author} {\bibfnamefont {C.}~\bibnamefont {Barbieri}}, \bibinfo {author} {\bibfnamefont {S.}~\bibnamefont {Chen}}, \bibinfo {author} {\bibfnamefont {L.~X.}\ \bibnamefont {Chung}}, \bibinfo {author} {\bibfnamefont {T.}~\bibnamefont {Duguet}}, \bibinfo {author} {\bibfnamefont {M.}~\bibnamefont {G\'omez-Ramos}}, \bibinfo {author} {\bibfnamefont {J.~D.}\ \bibnamefont {Holt}}, \bibinfo {author} {\bibfnamefont {A.}~\bibnamefont {Moro}}, \bibinfo {author} {\bibfnamefont {P.}~\bibnamefont {Navr\'atil}}, \bibinfo {author} {\bibfnamefont {K.}~\bibnamefont {Ogata}}, \bibinfo {author} {\bibfnamefont {N.~T.~T.}\ \bibnamefont {Phuc}}, \bibinfo {author} {\bibfnamefont
  {N.}~\bibnamefont {Shimizu}}, \bibinfo {author} {\bibfnamefont {V.}~\bibnamefont {Som\`a}}, \bibinfo {author} {\bibfnamefont {Y.}~\bibnamefont {Utsuno}}, \bibinfo {author} {\bibfnamefont {N.~L.}\ \bibnamefont {Achouri}}, \bibinfo {author} {\bibfnamefont {H.}~\bibnamefont {Baba}}, \bibinfo {author} {\bibfnamefont {F.}~\bibnamefont {Browne}}, \bibinfo {author} {\bibfnamefont {D.}~\bibnamefont {Calvet}}, \bibinfo {author} {\bibfnamefont {F.}~\bibnamefont {Ch\^ateau}}, \bibinfo {author} {\bibfnamefont {N.}~\bibnamefont {Chiga}}, \bibinfo {author} {\bibfnamefont {M.~L.}\ \bibnamefont {Cort\'es}}, \bibinfo {author} {\bibfnamefont {A.}~\bibnamefont {Delbart}}, \bibinfo {author} {\bibfnamefont {J.-M.}\ \bibnamefont {Gheller}}, \bibinfo {author} {\bibfnamefont {A.}~\bibnamefont {Giganon}}, \bibinfo {author} {\bibfnamefont {C.}~\bibnamefont {Hilaire}}, \bibinfo {author} {\bibfnamefont {T.}~\bibnamefont {Isobe}}, \bibinfo {author} {\bibfnamefont {T.}~\bibnamefont {Kobayashi}}, \bibinfo {author} {\bibfnamefont
  {Y.}~\bibnamefont {Kubota}}, \bibinfo {author} {\bibfnamefont {V.}~\bibnamefont {Lapoux}}, \bibinfo {author} {\bibfnamefont {H.~N.}\ \bibnamefont {Liu}}, \bibinfo {author} {\bibfnamefont {T.}~\bibnamefont {Motobayashi}}, \bibinfo {author} {\bibfnamefont {I.}~\bibnamefont {Murray}}, \bibinfo {author} {\bibfnamefont {H.}~\bibnamefont {Otsu}}, \bibinfo {author} {\bibfnamefont {V.}~\bibnamefont {Panin}}, \bibinfo {author} {\bibfnamefont {N.}~\bibnamefont {Paul}}, \bibinfo {author} {\bibfnamefont {W.}~\bibnamefont {Rodriguez}}, \bibinfo {author} {\bibfnamefont {H.}~\bibnamefont {Sakurai}}, \bibinfo {author} {\bibfnamefont {M.}~\bibnamefont {Sasano}}, \bibinfo {author} {\bibfnamefont {D.}~\bibnamefont {Steppenbeck}}, \bibinfo {author} {\bibfnamefont {L.}~\bibnamefont {Stuhl}}, \bibinfo {author} {\bibfnamefont {Y.~L.}\ \bibnamefont {Sun}}, \bibinfo {author} {\bibfnamefont {Y.}~\bibnamefont {Togano}}, \bibinfo {author} {\bibfnamefont {T.}~\bibnamefont {Uesaka}}, \bibinfo {author} {\bibfnamefont {K.}~\bibnamefont
  {Wimmer}}, \bibinfo {author} {\bibfnamefont {K.}~\bibnamefont {Yoneda}}, \bibinfo {author} {\bibfnamefont {O.}~\bibnamefont {Aktas}}, \bibinfo {author} {\bibfnamefont {T.}~\bibnamefont {Aumann}}, \bibinfo {author} {\bibfnamefont {F.}~\bibnamefont {Flavigny}}, \bibinfo {author} {\bibfnamefont {S.}~\bibnamefont {Franchoo}}, \bibinfo {author} {\bibfnamefont {I.}~\bibnamefont {Ga\ifmmode \check{s}\else \v{s}\fi{}pari\ifmmode~\acute{c}\else \'{c}\fi{}}}, \bibinfo {author} {\bibfnamefont {R.-B.}\ \bibnamefont {Gerst}}, \bibinfo {author} {\bibfnamefont {J.}~\bibnamefont {Gibelin}}, \bibinfo {author} {\bibfnamefont {K.~I.}\ \bibnamefont {Hahn}}, \bibinfo {author} {\bibfnamefont {N.~T.}\ \bibnamefont {Khai}}, \bibinfo {author} {\bibfnamefont {D.}~\bibnamefont {Kim}}, \bibinfo {author} {\bibfnamefont {T.}~\bibnamefont {Koiwai}}, \bibinfo {author} {\bibfnamefont {Y.}~\bibnamefont {Kondo}}, \bibinfo {author} {\bibfnamefont {P.}~\bibnamefont {Koseoglou}}, \bibinfo {author} {\bibfnamefont {J.}~\bibnamefont {Lee}},
  \bibinfo {author} {\bibfnamefont {C.}~\bibnamefont {Lehr}}, \bibinfo {author} {\bibfnamefont {T.}~\bibnamefont {Lokotko}}, \bibinfo {author} {\bibfnamefont {M.}~\bibnamefont {MacCormick}}, \bibinfo {author} {\bibfnamefont {K.}~\bibnamefont {Moschner}}, \bibinfo {author} {\bibfnamefont {T.}~\bibnamefont {Nakamura}}, \bibinfo {author} {\bibfnamefont {S.~Y.}\ \bibnamefont {Park}}, \bibinfo {author} {\bibfnamefont {D.}~\bibnamefont {Rossi}}, \bibinfo {author} {\bibfnamefont {E.}~\bibnamefont {Sahin}}, \bibinfo {author} {\bibfnamefont {D.}~\bibnamefont {Sohler}}, \bibinfo {author} {\bibfnamefont {P.-A.}\ \bibnamefont {S\"oderstr\"om}}, \bibinfo {author} {\bibfnamefont {S.}~\bibnamefont {Takeuchi}}, \bibinfo {author} {\bibfnamefont {N.~D.}\ \bibnamefont {Ton}}, \bibinfo {author} {\bibfnamefont {H.}~\bibnamefont {T\"ornqvist}}, \bibinfo {author} {\bibfnamefont {V.}~\bibnamefont {Vaquero}}, \bibinfo {author} {\bibfnamefont {V.}~\bibnamefont {Wagner}}, \bibinfo {author} {\bibfnamefont {H.}~\bibnamefont {Wang}},
  \bibinfo {author} {\bibfnamefont {V.}~\bibnamefont {Werner}}, \bibinfo {author} {\bibfnamefont {X.}~\bibnamefont {Xu}}, \bibinfo {author} {\bibfnamefont {Y.}~\bibnamefont {Yamada}}, \bibinfo {author} {\bibfnamefont {D.}~\bibnamefont {Yan}}, \bibinfo {author} {\bibfnamefont {Z.}~\bibnamefont {Yang}}, \bibinfo {author} {\bibfnamefont {M.}~\bibnamefont {Yasuda}},\ and\ \bibinfo {author} {\bibfnamefont {L.}~\bibnamefont {Zanetti}},\ }\bibfield  {title} {\bibinfo {title} {Investigation of the ground-state spin inversion in the neutron-rich $^{47,49}\mathrm{Cl}$ isotopes},\ }\href {https://doi.org/10.1103/PhysRevC.104.044331} {\bibfield  {journal} {\bibinfo  {journal} {Phys. Rev. C}\ }\textbf {\bibinfo {volume} {104}},\ \bibinfo {pages} {044331} (\bibinfo {year} {2021})}\BibitemShut {NoStop}%
\bibitem [{\citenamefont {Browne}\ \emph {et~al.}(2021)\citenamefont {Browne}, \citenamefont {Chen}, \citenamefont {Doornenbal}, \citenamefont {Obertelli}, \citenamefont {Ogata}, \citenamefont {Utsuno}, \citenamefont {Yoshida}, \citenamefont {Achouri}, \citenamefont {Baba}, \citenamefont {Calvet}, \citenamefont {Ch\^ateau}, \citenamefont {Chiga}, \citenamefont {Corsi}, \citenamefont {Cort\'es}, \citenamefont {Delbart}, \citenamefont {Gheller}, \citenamefont {Giganon}, \citenamefont {Gillibert}, \citenamefont {Hilaire}, \citenamefont {Isobe}, \citenamefont {Kobayashi}, \citenamefont {Kubota}, \citenamefont {Lapoux}, \citenamefont {Liu}, \citenamefont {Motobayashi}, \citenamefont {Murray}, \citenamefont {Otsu}, \citenamefont {Panin}, \citenamefont {Paul}, \citenamefont {Rodriguez}, \citenamefont {Sakurai}, \citenamefont {Sasano}, \citenamefont {Steppenbeck}, \citenamefont {Stuhl}, \citenamefont {Sun}, \citenamefont {Togano}, \citenamefont {Uesaka}, \citenamefont {Wimmer}, \citenamefont {Yoneda}, \citenamefont
  {Aktas}, \citenamefont {Aumann}, \citenamefont {Boretzky}, \citenamefont {Caesar}, \citenamefont {Chung}, \citenamefont {Flavigny}, \citenamefont {Franchoo}, \citenamefont {Gasparic}, \citenamefont {Gerst}, \citenamefont {Gibelin}, \citenamefont {Hahn}, \citenamefont {Holl}, \citenamefont {Kahlbow}, \citenamefont {Kim}, \citenamefont {K\"orper}, \citenamefont {Koiwai}, \citenamefont {Kondo}, \citenamefont {Koseoglou}, \citenamefont {Lee}, \citenamefont {Lehr}, \citenamefont {Linh}, \citenamefont {Lokotko}, \citenamefont {MacCormick}, \citenamefont {Miki}, \citenamefont {Moschner}, \citenamefont {Nakamura}, \citenamefont {Park}, \citenamefont {Rossi}, \citenamefont {Sahin}, \citenamefont {Schindler}, \citenamefont {Simon}, \citenamefont {S\"oderstr\"om}, \citenamefont {Sohler}, \citenamefont {Takeuchi}, \citenamefont {T\"ornqvist}, \citenamefont {Tscheuschner}, \citenamefont {Vaquero}, \citenamefont {Wagner}, \citenamefont {Wang}, \citenamefont {Werner}, \citenamefont {Xu}, \citenamefont {Yamada},
  \citenamefont {Yan}, \citenamefont {Yang}, \citenamefont {Yasuda},\ and\ \citenamefont {Zanetti}}]{FBrowne21}%
  \BibitemOpen
  \bibfield  {author} {\bibinfo {author} {\bibfnamefont {F.}~\bibnamefont {Browne}}, \bibinfo {author} {\bibfnamefont {S.}~\bibnamefont {Chen}}, \bibinfo {author} {\bibfnamefont {P.}~\bibnamefont {Doornenbal}}, \bibinfo {author} {\bibfnamefont {A.}~\bibnamefont {Obertelli}}, \bibinfo {author} {\bibfnamefont {K.}~\bibnamefont {Ogata}}, \bibinfo {author} {\bibfnamefont {Y.}~\bibnamefont {Utsuno}}, \bibinfo {author} {\bibfnamefont {K.}~\bibnamefont {Yoshida}}, \bibinfo {author} {\bibfnamefont {N.~L.}\ \bibnamefont {Achouri}}, \bibinfo {author} {\bibfnamefont {H.}~\bibnamefont {Baba}}, \bibinfo {author} {\bibfnamefont {D.}~\bibnamefont {Calvet}}, \bibinfo {author} {\bibfnamefont {F.}~\bibnamefont {Ch\^ateau}}, \bibinfo {author} {\bibfnamefont {N.}~\bibnamefont {Chiga}}, \bibinfo {author} {\bibfnamefont {A.}~\bibnamefont {Corsi}}, \bibinfo {author} {\bibfnamefont {M.~L.}\ \bibnamefont {Cort\'es}}, \bibinfo {author} {\bibfnamefont {A.}~\bibnamefont {Delbart}}, \bibinfo {author} {\bibfnamefont {J.-M.}\ \bibnamefont
  {Gheller}}, \bibinfo {author} {\bibfnamefont {A.}~\bibnamefont {Giganon}}, \bibinfo {author} {\bibfnamefont {A.}~\bibnamefont {Gillibert}}, \bibinfo {author} {\bibfnamefont {C.}~\bibnamefont {Hilaire}}, \bibinfo {author} {\bibfnamefont {T.}~\bibnamefont {Isobe}}, \bibinfo {author} {\bibfnamefont {T.}~\bibnamefont {Kobayashi}}, \bibinfo {author} {\bibfnamefont {Y.}~\bibnamefont {Kubota}}, \bibinfo {author} {\bibfnamefont {V.}~\bibnamefont {Lapoux}}, \bibinfo {author} {\bibfnamefont {H.~N.}\ \bibnamefont {Liu}}, \bibinfo {author} {\bibfnamefont {T.}~\bibnamefont {Motobayashi}}, \bibinfo {author} {\bibfnamefont {I.}~\bibnamefont {Murray}}, \bibinfo {author} {\bibfnamefont {H.}~\bibnamefont {Otsu}}, \bibinfo {author} {\bibfnamefont {V.}~\bibnamefont {Panin}}, \bibinfo {author} {\bibfnamefont {N.}~\bibnamefont {Paul}}, \bibinfo {author} {\bibfnamefont {W.}~\bibnamefont {Rodriguez}}, \bibinfo {author} {\bibfnamefont {H.}~\bibnamefont {Sakurai}}, \bibinfo {author} {\bibfnamefont {M.}~\bibnamefont {Sasano}},
  \bibinfo {author} {\bibfnamefont {D.}~\bibnamefont {Steppenbeck}}, \bibinfo {author} {\bibfnamefont {L.}~\bibnamefont {Stuhl}}, \bibinfo {author} {\bibfnamefont {Y.~L.}\ \bibnamefont {Sun}}, \bibinfo {author} {\bibfnamefont {Y.}~\bibnamefont {Togano}}, \bibinfo {author} {\bibfnamefont {T.}~\bibnamefont {Uesaka}}, \bibinfo {author} {\bibfnamefont {K.}~\bibnamefont {Wimmer}}, \bibinfo {author} {\bibfnamefont {K.}~\bibnamefont {Yoneda}}, \bibinfo {author} {\bibfnamefont {O.}~\bibnamefont {Aktas}}, \bibinfo {author} {\bibfnamefont {T.}~\bibnamefont {Aumann}}, \bibinfo {author} {\bibfnamefont {K.}~\bibnamefont {Boretzky}}, \bibinfo {author} {\bibfnamefont {C.}~\bibnamefont {Caesar}}, \bibinfo {author} {\bibfnamefont {L.~X.}\ \bibnamefont {Chung}}, \bibinfo {author} {\bibfnamefont {F.}~\bibnamefont {Flavigny}}, \bibinfo {author} {\bibfnamefont {S.}~\bibnamefont {Franchoo}}, \bibinfo {author} {\bibfnamefont {I.}~\bibnamefont {Gasparic}}, \bibinfo {author} {\bibfnamefont {R.-B.}\ \bibnamefont {Gerst}}, \bibinfo
  {author} {\bibfnamefont {J.}~\bibnamefont {Gibelin}}, \bibinfo {author} {\bibfnamefont {K.~I.}\ \bibnamefont {Hahn}}, \bibinfo {author} {\bibfnamefont {M.}~\bibnamefont {Holl}}, \bibinfo {author} {\bibfnamefont {J.}~\bibnamefont {Kahlbow}}, \bibinfo {author} {\bibfnamefont {D.}~\bibnamefont {Kim}}, \bibinfo {author} {\bibfnamefont {D.}~\bibnamefont {K\"orper}}, \bibinfo {author} {\bibfnamefont {T.}~\bibnamefont {Koiwai}}, \bibinfo {author} {\bibfnamefont {Y.}~\bibnamefont {Kondo}}, \bibinfo {author} {\bibfnamefont {P.}~\bibnamefont {Koseoglou}}, \bibinfo {author} {\bibfnamefont {J.}~\bibnamefont {Lee}}, \bibinfo {author} {\bibfnamefont {C.}~\bibnamefont {Lehr}}, \bibinfo {author} {\bibfnamefont {B.~D.}\ \bibnamefont {Linh}}, \bibinfo {author} {\bibfnamefont {T.}~\bibnamefont {Lokotko}}, \bibinfo {author} {\bibfnamefont {M.}~\bibnamefont {MacCormick}}, \bibinfo {author} {\bibfnamefont {K.}~\bibnamefont {Miki}}, \bibinfo {author} {\bibfnamefont {K.}~\bibnamefont {Moschner}}, \bibinfo {author} {\bibfnamefont
  {T.}~\bibnamefont {Nakamura}}, \bibinfo {author} {\bibfnamefont {S.~Y.}\ \bibnamefont {Park}}, \bibinfo {author} {\bibfnamefont {D.}~\bibnamefont {Rossi}}, \bibinfo {author} {\bibfnamefont {E.}~\bibnamefont {Sahin}}, \bibinfo {author} {\bibfnamefont {F.}~\bibnamefont {Schindler}}, \bibinfo {author} {\bibfnamefont {H.}~\bibnamefont {Simon}}, \bibinfo {author} {\bibfnamefont {P.-A.}\ \bibnamefont {S\"oderstr\"om}}, \bibinfo {author} {\bibfnamefont {D.}~\bibnamefont {Sohler}}, \bibinfo {author} {\bibfnamefont {S.}~\bibnamefont {Takeuchi}}, \bibinfo {author} {\bibfnamefont {H.}~\bibnamefont {T\"ornqvist}}, \bibinfo {author} {\bibfnamefont {J.}~\bibnamefont {Tscheuschner}}, \bibinfo {author} {\bibfnamefont {V.}~\bibnamefont {Vaquero}}, \bibinfo {author} {\bibfnamefont {V.}~\bibnamefont {Wagner}}, \bibinfo {author} {\bibfnamefont {S.}~\bibnamefont {Wang}}, \bibinfo {author} {\bibfnamefont {V.}~\bibnamefont {Werner}}, \bibinfo {author} {\bibfnamefont {X.}~\bibnamefont {Xu}}, \bibinfo {author} {\bibfnamefont
  {H.}~\bibnamefont {Yamada}}, \bibinfo {author} {\bibfnamefont {D.}~\bibnamefont {Yan}}, \bibinfo {author} {\bibfnamefont {Z.}~\bibnamefont {Yang}}, \bibinfo {author} {\bibfnamefont {M.}~\bibnamefont {Yasuda}},\ and\ \bibinfo {author} {\bibfnamefont {L.}~\bibnamefont {Zanetti}},\ }\bibfield  {title} {\bibinfo {title} {Pairing forces govern population of doubly magic $^{54}\mathrm{Ca}$ from direct reactions},\ }\href {https://doi.org/10.1103/PhysRevLett.126.252501} {\bibfield  {journal} {\bibinfo  {journal} {Phys. Rev. Lett.}\ }\textbf {\bibinfo {volume} {126}},\ \bibinfo {pages} {252501} (\bibinfo {year} {2021})}\BibitemShut {NoStop}%
\bibitem [{\citenamefont {Enciu}\ \emph {et~al.}(2022)\citenamefont {Enciu}, \citenamefont {Liu}, \citenamefont {Obertelli}, \citenamefont {Doornenbal}, \citenamefont {Nowacki}, \citenamefont {Ogata}, \citenamefont {Poves}, \citenamefont {Yoshida}, \citenamefont {Achouri}, \citenamefont {Baba}, \citenamefont {Browne}, \citenamefont {Calvet}, \citenamefont {Ch\^ateau}, \citenamefont {Chen}, \citenamefont {Chiga}, \citenamefont {Corsi}, \citenamefont {Cort\'es}, \citenamefont {Delbart}, \citenamefont {Gheller}, \citenamefont {Giganon}, \citenamefont {Gillibert}, \citenamefont {Hilaire}, \citenamefont {Isobe}, \citenamefont {Kobayashi}, \citenamefont {Kubota}, \citenamefont {Lapoux}, \citenamefont {Motobayashi}, \citenamefont {Murray}, \citenamefont {Otsu}, \citenamefont {Panin}, \citenamefont {Paul}, \citenamefont {Rodriguez}, \citenamefont {Sakurai}, \citenamefont {Sasano}, \citenamefont {Steppenbeck}, \citenamefont {Stuhl}, \citenamefont {Sun}, \citenamefont {Togano}, \citenamefont {Uesaka}, \citenamefont
  {Wimmer}, \citenamefont {Yoneda}, \citenamefont {Aktas}, \citenamefont {Aumann}, \citenamefont {Chung}, \citenamefont {Flavigny}, \citenamefont {Franchoo}, \citenamefont {Gasparic}, \citenamefont {Gerst}, \citenamefont {Gibelin}, \citenamefont {Hahn}, \citenamefont {Kim}, \citenamefont {Kondo}, \citenamefont {Koseoglou}, \citenamefont {Lee}, \citenamefont {Lehr}, \citenamefont {Li}, \citenamefont {Linh}, \citenamefont {Lokotko}, \citenamefont {MacCormick}, \citenamefont {Moschner}, \citenamefont {Nakamura}, \citenamefont {Park}, \citenamefont {Rossi}, \citenamefont {Sahin}, \citenamefont {S\"oderstr\"om}, \citenamefont {Sohler}, \citenamefont {Takeuchi}, \citenamefont {Toernqvist}, \citenamefont {Vaquero}, \citenamefont {Wagner}, \citenamefont {Wang}, \citenamefont {Werner}, \citenamefont {Xu}, \citenamefont {Yamada}, \citenamefont {Yan}, \citenamefont {Yang}, \citenamefont {Yasuda},\ and\ \citenamefont {Zanetti}}]{MEnciu22}%
  \BibitemOpen
  \bibfield  {author} {\bibinfo {author} {\bibfnamefont {M.}~\bibnamefont {Enciu}}, \bibinfo {author} {\bibfnamefont {H.~N.}\ \bibnamefont {Liu}}, \bibinfo {author} {\bibfnamefont {A.}~\bibnamefont {Obertelli}}, \bibinfo {author} {\bibfnamefont {P.}~\bibnamefont {Doornenbal}}, \bibinfo {author} {\bibfnamefont {F.}~\bibnamefont {Nowacki}}, \bibinfo {author} {\bibfnamefont {K.}~\bibnamefont {Ogata}}, \bibinfo {author} {\bibfnamefont {A.}~\bibnamefont {Poves}}, \bibinfo {author} {\bibfnamefont {K.}~\bibnamefont {Yoshida}}, \bibinfo {author} {\bibfnamefont {N.~L.}\ \bibnamefont {Achouri}}, \bibinfo {author} {\bibfnamefont {H.}~\bibnamefont {Baba}}, \bibinfo {author} {\bibfnamefont {F.}~\bibnamefont {Browne}}, \bibinfo {author} {\bibfnamefont {D.}~\bibnamefont {Calvet}}, \bibinfo {author} {\bibfnamefont {F.}~\bibnamefont {Ch\^ateau}}, \bibinfo {author} {\bibfnamefont {S.}~\bibnamefont {Chen}}, \bibinfo {author} {\bibfnamefont {N.}~\bibnamefont {Chiga}}, \bibinfo {author} {\bibfnamefont {A.}~\bibnamefont {Corsi}},
  \bibinfo {author} {\bibfnamefont {M.~L.}\ \bibnamefont {Cort\'es}}, \bibinfo {author} {\bibfnamefont {A.}~\bibnamefont {Delbart}}, \bibinfo {author} {\bibfnamefont {J.-M.}\ \bibnamefont {Gheller}}, \bibinfo {author} {\bibfnamefont {A.}~\bibnamefont {Giganon}}, \bibinfo {author} {\bibfnamefont {A.}~\bibnamefont {Gillibert}}, \bibinfo {author} {\bibfnamefont {C.}~\bibnamefont {Hilaire}}, \bibinfo {author} {\bibfnamefont {T.}~\bibnamefont {Isobe}}, \bibinfo {author} {\bibfnamefont {T.}~\bibnamefont {Kobayashi}}, \bibinfo {author} {\bibfnamefont {Y.}~\bibnamefont {Kubota}}, \bibinfo {author} {\bibfnamefont {V.}~\bibnamefont {Lapoux}}, \bibinfo {author} {\bibfnamefont {T.}~\bibnamefont {Motobayashi}}, \bibinfo {author} {\bibfnamefont {I.}~\bibnamefont {Murray}}, \bibinfo {author} {\bibfnamefont {H.}~\bibnamefont {Otsu}}, \bibinfo {author} {\bibfnamefont {V.}~\bibnamefont {Panin}}, \bibinfo {author} {\bibfnamefont {N.}~\bibnamefont {Paul}}, \bibinfo {author} {\bibfnamefont {W.}~\bibnamefont {Rodriguez}}, \bibinfo
  {author} {\bibfnamefont {H.}~\bibnamefont {Sakurai}}, \bibinfo {author} {\bibfnamefont {M.}~\bibnamefont {Sasano}}, \bibinfo {author} {\bibfnamefont {D.}~\bibnamefont {Steppenbeck}}, \bibinfo {author} {\bibfnamefont {L.}~\bibnamefont {Stuhl}}, \bibinfo {author} {\bibfnamefont {Y.~L.}\ \bibnamefont {Sun}}, \bibinfo {author} {\bibfnamefont {Y.}~\bibnamefont {Togano}}, \bibinfo {author} {\bibfnamefont {T.}~\bibnamefont {Uesaka}}, \bibinfo {author} {\bibfnamefont {K.}~\bibnamefont {Wimmer}}, \bibinfo {author} {\bibfnamefont {K.}~\bibnamefont {Yoneda}}, \bibinfo {author} {\bibfnamefont {O.}~\bibnamefont {Aktas}}, \bibinfo {author} {\bibfnamefont {T.}~\bibnamefont {Aumann}}, \bibinfo {author} {\bibfnamefont {L.~X.}\ \bibnamefont {Chung}}, \bibinfo {author} {\bibfnamefont {F.}~\bibnamefont {Flavigny}}, \bibinfo {author} {\bibfnamefont {S.}~\bibnamefont {Franchoo}}, \bibinfo {author} {\bibfnamefont {I.}~\bibnamefont {Gasparic}}, \bibinfo {author} {\bibfnamefont {R.-B.}\ \bibnamefont {Gerst}}, \bibinfo {author}
  {\bibfnamefont {J.}~\bibnamefont {Gibelin}}, \bibinfo {author} {\bibfnamefont {K.~I.}\ \bibnamefont {Hahn}}, \bibinfo {author} {\bibfnamefont {D.}~\bibnamefont {Kim}}, \bibinfo {author} {\bibfnamefont {Y.}~\bibnamefont {Kondo}}, \bibinfo {author} {\bibfnamefont {P.}~\bibnamefont {Koseoglou}}, \bibinfo {author} {\bibfnamefont {J.}~\bibnamefont {Lee}}, \bibinfo {author} {\bibfnamefont {C.}~\bibnamefont {Lehr}}, \bibinfo {author} {\bibfnamefont {P.~J.}\ \bibnamefont {Li}}, \bibinfo {author} {\bibfnamefont {B.~D.}\ \bibnamefont {Linh}}, \bibinfo {author} {\bibfnamefont {T.}~\bibnamefont {Lokotko}}, \bibinfo {author} {\bibfnamefont {M.}~\bibnamefont {MacCormick}}, \bibinfo {author} {\bibfnamefont {K.}~\bibnamefont {Moschner}}, \bibinfo {author} {\bibfnamefont {T.}~\bibnamefont {Nakamura}}, \bibinfo {author} {\bibfnamefont {S.~Y.}\ \bibnamefont {Park}}, \bibinfo {author} {\bibfnamefont {D.}~\bibnamefont {Rossi}}, \bibinfo {author} {\bibfnamefont {E.}~\bibnamefont {Sahin}}, \bibinfo {author} {\bibfnamefont
  {P.-A.}\ \bibnamefont {S\"oderstr\"om}}, \bibinfo {author} {\bibfnamefont {D.}~\bibnamefont {Sohler}}, \bibinfo {author} {\bibfnamefont {S.}~\bibnamefont {Takeuchi}}, \bibinfo {author} {\bibfnamefont {H.}~\bibnamefont {Toernqvist}}, \bibinfo {author} {\bibfnamefont {V.}~\bibnamefont {Vaquero}}, \bibinfo {author} {\bibfnamefont {V.}~\bibnamefont {Wagner}}, \bibinfo {author} {\bibfnamefont {S.}~\bibnamefont {Wang}}, \bibinfo {author} {\bibfnamefont {V.}~\bibnamefont {Werner}}, \bibinfo {author} {\bibfnamefont {X.}~\bibnamefont {Xu}}, \bibinfo {author} {\bibfnamefont {H.}~\bibnamefont {Yamada}}, \bibinfo {author} {\bibfnamefont {D.}~\bibnamefont {Yan}}, \bibinfo {author} {\bibfnamefont {Z.}~\bibnamefont {Yang}}, \bibinfo {author} {\bibfnamefont {M.}~\bibnamefont {Yasuda}},\ and\ \bibinfo {author} {\bibfnamefont {L.}~\bibnamefont {Zanetti}},\ }\bibfield  {title} {\bibinfo {title} {{Extended ${p}_{3/2}$ Neutron Orbital and the $N=32$ Shell Closure in $^{52}\mathrm{Ca}$}},\ }\href
  {https://doi.org/10.1103/PhysRevLett.129.262501} {\bibfield  {journal} {\bibinfo  {journal} {Phys. Rev. Lett.}\ }\textbf {\bibinfo {volume} {129}},\ \bibinfo {pages} {262501} (\bibinfo {year} {2022})}\BibitemShut {NoStop}%
\bibitem [{\citenamefont {Wang}\ \emph {et~al.}(2023)\citenamefont {Wang}, \citenamefont {Yasuda}, \citenamefont {Kondo}, \citenamefont {Nakamura}, \citenamefont {Tostevin}, \citenamefont {Ogata}, \citenamefont {Otsuka}, \citenamefont {Poves}, \citenamefont {Shimizu}, \citenamefont {Yoshida}, \citenamefont {Achouri}, \citenamefont {{Al Falou}}, \citenamefont {Atar}, \citenamefont {Aumann}, \citenamefont {Baba}, \citenamefont {Boretzky}, \citenamefont {Caesar}, \citenamefont {Calvet}, \citenamefont {Chae}, \citenamefont {Chiga}, \citenamefont {Corsi}, \citenamefont {Crawford}, \citenamefont {Delaunay}, \citenamefont {Delbart}, \citenamefont {Deshayes}, \citenamefont {Dombr\'{a}di}, \citenamefont {Douma}, \citenamefont {Elekes}, \citenamefont {Fallon}, \citenamefont {Ga\v{s}pari\'{c}}, \citenamefont {Gheller}, \citenamefont {Gibelin}, \citenamefont {Gillibert}, \citenamefont {Harakeh}, \citenamefont {Hirayama}, \citenamefont {Hoffman}, \citenamefont {Holl}, \citenamefont {Horvat}, \citenamefont {Horv\'{a}th},
  \citenamefont {Hwang}, \citenamefont {Isobe}, \citenamefont {Kahlbow}, \citenamefont {Kalantar-Nayestanaki}, \citenamefont {Kawase}, \citenamefont {Kim}, \citenamefont {Kisamori}, \citenamefont {Kobayashi}, \citenamefont {K\"{o}rper}, \citenamefont {Koyama}, \citenamefont {Kuti}, \citenamefont {Lapoux}, \citenamefont {Lindberg}, \citenamefont {Marqu\'{e}s}, \citenamefont {Masuoka}, \citenamefont {Mayer}, \citenamefont {Miki}, \citenamefont {Murakami}, \citenamefont {Najafi}, \citenamefont {Nakano}, \citenamefont {Nakatsuka}, \citenamefont {Nilsson}, \citenamefont {Obertelli}, \citenamefont {Orr}, \citenamefont {Otsu}, \citenamefont {Ozaki}, \citenamefont {Panin}, \citenamefont {Paschalis}, \citenamefont {Revel}, \citenamefont {Rossi}, \citenamefont {Saito}, \citenamefont {Saito}, \citenamefont {Sasano}, \citenamefont {Sato}, \citenamefont {Satou}, \citenamefont {Scheit}, \citenamefont {Schindler}, \citenamefont {Schrock}, \citenamefont {Shikata}, \citenamefont {Shimizu}, \citenamefont {Simon}, \citenamefont
  {Sohler}, \citenamefont {Sorlin}, \citenamefont {Stuhl}, \citenamefont {Takeuchi}, \citenamefont {Tanaka}, \citenamefont {Thoennessen}, \citenamefont {T\"{o}rnqvist}, \citenamefont {Togano}, \citenamefont {Tomai}, \citenamefont {Tscheuschner}, \citenamefont {Tsubota}, \citenamefont {Uesaka}, \citenamefont {Yang},\ and\ \citenamefont {Yoneda}}]{HWang23}%
  \BibitemOpen
  \bibfield  {author} {\bibinfo {author} {\bibfnamefont {H.}~\bibnamefont {Wang}}, \bibinfo {author} {\bibfnamefont {M.}~\bibnamefont {Yasuda}}, \bibinfo {author} {\bibfnamefont {Y.}~\bibnamefont {Kondo}}, \bibinfo {author} {\bibfnamefont {T.}~\bibnamefont {Nakamura}}, \bibinfo {author} {\bibfnamefont {J.}~\bibnamefont {Tostevin}}, \bibinfo {author} {\bibfnamefont {K.}~\bibnamefont {Ogata}}, \bibinfo {author} {\bibfnamefont {T.}~\bibnamefont {Otsuka}}, \bibinfo {author} {\bibfnamefont {A.}~\bibnamefont {Poves}}, \bibinfo {author} {\bibfnamefont {N.}~\bibnamefont {Shimizu}}, \bibinfo {author} {\bibfnamefont {K.}~\bibnamefont {Yoshida}}, \bibinfo {author} {\bibfnamefont {N.}~\bibnamefont {Achouri}}, \bibinfo {author} {\bibfnamefont {H.}~\bibnamefont {{Al Falou}}}, \bibinfo {author} {\bibfnamefont {L.}~\bibnamefont {Atar}}, \bibinfo {author} {\bibfnamefont {T.}~\bibnamefont {Aumann}}, \bibinfo {author} {\bibfnamefont {H.}~\bibnamefont {Baba}}, \bibinfo {author} {\bibfnamefont {K.}~\bibnamefont {Boretzky}},
  \bibinfo {author} {\bibfnamefont {C.}~\bibnamefont {Caesar}}, \bibinfo {author} {\bibfnamefont {D.}~\bibnamefont {Calvet}}, \bibinfo {author} {\bibfnamefont {H.}~\bibnamefont {Chae}}, \bibinfo {author} {\bibfnamefont {N.}~\bibnamefont {Chiga}}, \bibinfo {author} {\bibfnamefont {A.}~\bibnamefont {Corsi}}, \bibinfo {author} {\bibfnamefont {H.}~\bibnamefont {Crawford}}, \bibinfo {author} {\bibfnamefont {F.}~\bibnamefont {Delaunay}}, \bibinfo {author} {\bibfnamefont {A.}~\bibnamefont {Delbart}}, \bibinfo {author} {\bibfnamefont {Q.}~\bibnamefont {Deshayes}}, \bibinfo {author} {\bibfnamefont {Z.}~\bibnamefont {Dombr\'{a}di}}, \bibinfo {author} {\bibfnamefont {C.}~\bibnamefont {Douma}}, \bibinfo {author} {\bibfnamefont {Z.}~\bibnamefont {Elekes}}, \bibinfo {author} {\bibfnamefont {P.}~\bibnamefont {Fallon}}, \bibinfo {author} {\bibfnamefont {I.}~\bibnamefont {Ga\v{s}pari\'{c}}}, \bibinfo {author} {\bibfnamefont {J.-M.}\ \bibnamefont {Gheller}}, \bibinfo {author} {\bibfnamefont {J.}~\bibnamefont {Gibelin}},
  \bibinfo {author} {\bibfnamefont {A.}~\bibnamefont {Gillibert}}, \bibinfo {author} {\bibfnamefont {M.}~\bibnamefont {Harakeh}}, \bibinfo {author} {\bibfnamefont {A.}~\bibnamefont {Hirayama}}, \bibinfo {author} {\bibfnamefont {C.}~\bibnamefont {Hoffman}}, \bibinfo {author} {\bibfnamefont {M.}~\bibnamefont {Holl}}, \bibinfo {author} {\bibfnamefont {A.}~\bibnamefont {Horvat}}, \bibinfo {author} {\bibfnamefont {A.}~\bibnamefont {Horv\'{a}th}}, \bibinfo {author} {\bibfnamefont {J.}~\bibnamefont {Hwang}}, \bibinfo {author} {\bibfnamefont {T.}~\bibnamefont {Isobe}}, \bibinfo {author} {\bibfnamefont {J.}~\bibnamefont {Kahlbow}}, \bibinfo {author} {\bibfnamefont {N.}~\bibnamefont {Kalantar-Nayestanaki}}, \bibinfo {author} {\bibfnamefont {S.}~\bibnamefont {Kawase}}, \bibinfo {author} {\bibfnamefont {S.}~\bibnamefont {Kim}}, \bibinfo {author} {\bibfnamefont {K.}~\bibnamefont {Kisamori}}, \bibinfo {author} {\bibfnamefont {T.}~\bibnamefont {Kobayashi}}, \bibinfo {author} {\bibfnamefont {D.}~\bibnamefont {K\"{o}rper}},
  \bibinfo {author} {\bibfnamefont {S.}~\bibnamefont {Koyama}}, \bibinfo {author} {\bibfnamefont {I.}~\bibnamefont {Kuti}}, \bibinfo {author} {\bibfnamefont {V.}~\bibnamefont {Lapoux}}, \bibinfo {author} {\bibfnamefont {S.}~\bibnamefont {Lindberg}}, \bibinfo {author} {\bibfnamefont {F.}~\bibnamefont {Marqu\'{e}s}}, \bibinfo {author} {\bibfnamefont {S.}~\bibnamefont {Masuoka}}, \bibinfo {author} {\bibfnamefont {J.}~\bibnamefont {Mayer}}, \bibinfo {author} {\bibfnamefont {K.}~\bibnamefont {Miki}}, \bibinfo {author} {\bibfnamefont {T.}~\bibnamefont {Murakami}}, \bibinfo {author} {\bibfnamefont {M.}~\bibnamefont {Najafi}}, \bibinfo {author} {\bibfnamefont {K.}~\bibnamefont {Nakano}}, \bibinfo {author} {\bibfnamefont {N.}~\bibnamefont {Nakatsuka}}, \bibinfo {author} {\bibfnamefont {T.}~\bibnamefont {Nilsson}}, \bibinfo {author} {\bibfnamefont {A.}~\bibnamefont {Obertelli}}, \bibinfo {author} {\bibfnamefont {N.}~\bibnamefont {Orr}}, \bibinfo {author} {\bibfnamefont {H.}~\bibnamefont {Otsu}}, \bibinfo {author}
  {\bibfnamefont {T.}~\bibnamefont {Ozaki}}, \bibinfo {author} {\bibfnamefont {V.}~\bibnamefont {Panin}}, \bibinfo {author} {\bibfnamefont {S.}~\bibnamefont {Paschalis}}, \bibinfo {author} {\bibfnamefont {A.}~\bibnamefont {Revel}}, \bibinfo {author} {\bibfnamefont {D.}~\bibnamefont {Rossi}}, \bibinfo {author} {\bibfnamefont {A.}~\bibnamefont {Saito}}, \bibinfo {author} {\bibfnamefont {T.}~\bibnamefont {Saito}}, \bibinfo {author} {\bibfnamefont {M.}~\bibnamefont {Sasano}}, \bibinfo {author} {\bibfnamefont {H.}~\bibnamefont {Sato}}, \bibinfo {author} {\bibfnamefont {Y.}~\bibnamefont {Satou}}, \bibinfo {author} {\bibfnamefont {H.}~\bibnamefont {Scheit}}, \bibinfo {author} {\bibfnamefont {F.}~\bibnamefont {Schindler}}, \bibinfo {author} {\bibfnamefont {P.}~\bibnamefont {Schrock}}, \bibinfo {author} {\bibfnamefont {M.}~\bibnamefont {Shikata}}, \bibinfo {author} {\bibfnamefont {Y.}~\bibnamefont {Shimizu}}, \bibinfo {author} {\bibfnamefont {H.}~\bibnamefont {Simon}}, \bibinfo {author} {\bibfnamefont
  {D.}~\bibnamefont {Sohler}}, \bibinfo {author} {\bibfnamefont {O.}~\bibnamefont {Sorlin}}, \bibinfo {author} {\bibfnamefont {L.}~\bibnamefont {Stuhl}}, \bibinfo {author} {\bibfnamefont {S.}~\bibnamefont {Takeuchi}}, \bibinfo {author} {\bibfnamefont {M.}~\bibnamefont {Tanaka}}, \bibinfo {author} {\bibfnamefont {M.}~\bibnamefont {Thoennessen}}, \bibinfo {author} {\bibfnamefont {H.}~\bibnamefont {T\"{o}rnqvist}}, \bibinfo {author} {\bibfnamefont {Y.}~\bibnamefont {Togano}}, \bibinfo {author} {\bibfnamefont {T.}~\bibnamefont {Tomai}}, \bibinfo {author} {\bibfnamefont {J.}~\bibnamefont {Tscheuschner}}, \bibinfo {author} {\bibfnamefont {J.}~\bibnamefont {Tsubota}}, \bibinfo {author} {\bibfnamefont {T.}~\bibnamefont {Uesaka}}, \bibinfo {author} {\bibfnamefont {Z.}~\bibnamefont {Yang}},\ and\ \bibinfo {author} {\bibfnamefont {K.}~\bibnamefont {Yoneda}},\ }\bibfield  {title} {\bibinfo {title} {Intruder configurations in $^{29}\mathrm{Ne}$ at the transition into the island of inversion: Detailed structure study of
  $^{28}\mathrm{Ne}$},\ }\href {https://doi.org/https://doi.org/10.1016/j.physletb.2023.138038} {\bibfield  {journal} {\bibinfo  {journal} {Physics Letters B}\ }\textbf {\bibinfo {volume} {843}},\ \bibinfo {pages} {138038} (\bibinfo {year} {2023})}\BibitemShut {NoStop}%
\bibitem [{\citenamefont {Ogata}\ \emph {et~al.}(2015)\citenamefont {Ogata}, \citenamefont {Yoshida},\ and\ \citenamefont {Minomo}}]{KOgata15}%
  \BibitemOpen
  \bibfield  {author} {\bibinfo {author} {\bibfnamefont {K.}~\bibnamefont {Ogata}}, \bibinfo {author} {\bibfnamefont {K.}~\bibnamefont {Yoshida}},\ and\ \bibinfo {author} {\bibfnamefont {K.}~\bibnamefont {Minomo}},\ }\bibfield  {title} {\bibinfo {title} {Asymmetry of the parallel momentum distribution of ($p,pn$) reaction residues},\ }\href {https://doi.org/10.1103/PhysRevC.92.034616} {\bibfield  {journal} {\bibinfo  {journal} {Phys. Rev. C}\ }\textbf {\bibinfo {volume} {92}},\ \bibinfo {pages} {034616} (\bibinfo {year} {2015})}\BibitemShut {NoStop}%
\bibitem [{\citenamefont {Bohr}\ and\ \citenamefont {Mottelson}(1998)}]{bohr-mottelson}%
  \BibitemOpen
  \bibfield  {author} {\bibinfo {author} {\bibfnamefont {A.}~\bibnamefont {Bohr}}\ and\ \bibinfo {author} {\bibfnamefont {B.~R.}\ \bibnamefont {Mottelson}},\ }\href {https://doi.org/10.1142/3530} {\emph {\bibinfo {title} {Nuclear Structure}}}\ (\bibinfo  {publisher} {World Scientific Publishing Company},\ \bibinfo {year} {1998})\ \Eprint {https://arxiv.org/abs/https://www.worldscientific.com/doi/pdf/10.1142/3530} {https://www.worldscientific.com/doi/pdf/10.1142/3530} \BibitemShut {NoStop}%
\end{thebibliography}%


\end{document}